\documentclass[acmconf]{acmart}

\AtBeginDocument{%
  }

\usepackage{subcaption}
\usepackage{graphicx}
\usepackage{lscape}
\usepackage{longtable}
\usepackage[normalem]{ulem}
\usepackage{xcolor}
\usepackage{soul}

\usepackage{tikz}
\usetikzlibrary{mindmap,trees}
\usepackage{verbatim}
\usepackage{subcaption}

\begin{document}

\title[Inflation of Interactivity?]{Inflation of Interactivity? Analyzing and Understanding Embodied Interaction in Interactive Art through a New Three-dimensional Model}
\author{Aven Le ZHOU}
\email{aven.le.zhou@gmail.com}
\orcid{0000-0002-8726-6797}
\affiliation{%
  \institution{The Hong Kong University of Science and Technology (Guangzhou)}
  \streetaddress{No.1 Du Xue Rd, Nansha District}
  \city{Guangzhou}
  \state{Guangdong}
  \country{P.R.China}
}

\renewcommand{\shortauthors}{Aven Le Zhou.}

\begin{abstract}

This insight paper examines embodied interaction in interactive art, focusing on body embodiment, bodily sensation (i.e., somaesthetic), and audience-artwork interaction. The authors propose a new three-dimensional descriptive model of interactive art based on literature and apply to analyze a curated corpus of 49 award-winning artworks from the Prix Ars Electronica between 2009 and 2023. The analysis reveals emergent patterns of interactive art that deepen the understanding of interactive art from an embodied perspective and prepare the ground for future research and art practices. This paper has discovered that embodied interaction remains under-explored in interactive art rather than an inflation of interactivity. Notable research gaps persist in exploring virtual embodiment within sociocultural contexts using immersive technologies. Furthermore, it also underscores the need to revisit the sociological and etymological roots of interaction to enhance interpersonality and relationality and advocates for a paradigm shift in future research and practice in interactive art.

\end{abstract}



\maketitle

\section{Introduction}


Interactive art traces its origins back to the 1950s but only gained widespread recognition as a formal discipline in 1990, when the Prix Ars Electronica introduced a dedicated category for it \cite[p.8]{kwastek_aesthetics_2013}. The Ars Electronica Festival, held in Linz, Austria, and renowned as the world's largest and oldest media art festival, embraced "Interactive Art" as a novel category in its international competition that year. Since then, the annual Prix Ars Electronica has consistently showcased a diverse array of artistic interfaces and interactive systems within this category. During the inaugural event, Roger Malina heralded this addition as "The Beginning of a New Art Form" in the festival catalog \cite[p.157]{leopoldseder_prix_1990}. 


The 2004 Prix Ars Electronica awarded the Golden Nica in Interactive Art to ``Listening Post'' by Mark Hansen and Ben Rubin, sparking widespread discussion and raising concerns about the state of interactive art. Erkki Huhtamo captured the ubiquity of interactive media in 2007, noting, ``Today interactive media is everywhere; its forms have become commonplace'' \cite{huhtamo_twintouchtestredux_2007}. Echoing this sentiment, Daniel Dieters observed that ``Interactivity is no longer an experiment in the media lab or an experience in a media art exhibition but part of everyday life in digital culture''\cite[p.56]{daniels_strategies_2008}. Extending this discourse, Guljajeva described these developments as ``a crisis of interactive art,'' arguing that while audiences have become ``less appreciative'' and ``much more knowledgeable'' about interactivity and technology in artworks, they have shifted their interest more towards the conceptual depth and are ``less hungry for entertainment''\cite[pp. 11-12]{guljajeva_interaction_2018}. This sentiment was reinforced by Ryszard Kluszczynski, who remarked, ``Long gone are the times of fascination just with the phenomenon of digital interactivity itself''\cite[p.27]{kluszczynski_strategies_2010}.


The focus on ``the crisis of interactive art'' and the discussion around the ``inflation of interactivity'' in interactive art forms the core of our investigation. In response, we introduce a new three-dimensional descriptive model of interactive art. This model is informed by philosophical and theoretical advancements related to body embodiment, somaesthetics (bodily sensation), and the dynamics of audience-artwork interaction. We apply this model to review and analyze artworks awarded at the Prix Ars Electronica over the last ten years (2009 - 2023). Through this analysis, our paper aims to uncover emergent patterns in interactive art that can deepen our understanding of this domain and set the stage for future research and artistic practices.


We outline this paper as follows: Sec.\ref{sec: ptaxonomy} reviews existing research on interactive art taxonomies and models. Sec.\ref{sec: literature} explores Don Ihde's philosophical perspectives on bodies in technology, Paul Dourish's definition of embodied interaction within human-computer interaction, and Brian Massumi's concepts of sensation and proprioceptive experiences in interactive art. In Sec.\ref{sec: model}, we introduce a new three-dimensional model of interactive art, drawing on these theoretical insights and the related literature. Sec.\ref{sec: describe} and Sec.\ref{sec: analysis} detail our corpus of interactive artworks from the Prix Ars Electronica and analyze them using our proposed model. Finally, Sec.\ref{sec: discuss} summarizes our findings and discusses future research directions.



\section{Previous Descriptive Model of Interactive Art} \label{sec: ptaxonomy}

Since the 1960s, active audience participation and interaction with artworks have garnered significant interest among artists and researchers alike \cite{schraffenberger_interaction_2012}. Roy Ascott emphasized the centrality of ``participation and interaction between audience and artwork'' in his seminal work, \textit{Behaviourist Art and the Cybernetic Vision} \cite{in_roy_nodate}. Similarly, Ernest Edmonds, a prominent figure in interactive art research, posited that the art experience is inherently interactive, involving the dynamic interplay among the environment, perception (or sensation), and the audience's generation of meaning \cite{muller_living_2006}. Edmonds subsequently developed various models to categorize interactive art, reflecting this comprehensive understanding.

\subsection{System-oriented Perspectives} \label{subsec: system}
In 1973, Edmonds and Cornock proposed a system perspective to describe artwork, categorizing relationships between artist, artifact, and audience as static, dynamic-passive, and dynamic-interactive \cite{cornock_creative_1973}. These categories, while broad, reflect interaction primarily in the dynamic-interactive system. Tab.\ref{tab:taxonomies1} includes developments of their four categories, adding ``dynamic-interactive (varying)'' \cite{edmonds_approaches_2004}, and Sparacino et al. expanded these into five classifications: scripted, responsive, behavioral, learning, and intentional \cite{sparacino_media_2000}.

Building on this system-oriented view of interactive art, Trifonova et al. delineated interaction types based on three properties: interaction rules (static and dynamic), triggering parameters (human presence, actions, and environment), and content origin (user input, predefined by the artist, and algorithmically generated content) \cite{trifonova_software_2008}. However, this classification focuses only on how the artwork is influenced by the audience and the environment, neglecting the artwork's impact on the audience \cite{schraffenberger_audience-artwork_2015}.

Conversely, Nardelli introduced a three-dimensional classification framework focusing on the content provider, the processing dynamics of the artwork, and the contributors to this processing \cite{nardelli_viewpoint_2014}. While similar to the categorization by Trifonova et al., Nardelli's framework, and others that view interactive art as merely software or hardware systems, tend to overlook the dynamic, bi-directional process between artwork and audience, focusing instead solely on the interaction's outcomes rather than the experience itself.

\subsection{Audience-experience and Interaction}

Focusing solely on the interactive system and its description has been critiqued by Nathaniel Stern, who argued that ``most writing on interactive art will explain that a given piece is interactive, and how it is interactive, but not how we interact'' \cite[p.3]{stern_implicit_2009}. In a parallel vein, Ernest Edmonds has introduced a ``critical language'' for describing, comparing, and discussing interactive art, drawing from ``experience design'' in human-computer interaction (HCI) to better understand user or audience experience \cite{edmonds_art_2010}. Collaboratively with Costello et al., they emphasized audience experience, examining its relation to movements, vocabulary, and behaviors, and employed the term ``embodied experience,'' categorizing it into response, control, contemplation, belonging, and disengagement \cite{costello_understanding_2005}. However, their model, being based on only one artwork, has not been widely validated for general applicability.

The artist duo Sommerer and Mignonneau have proposed a concept of non-linear or multi-layered interaction that is easy to grasp initially but allows audiences to ``continuously discover different levels of interactive experiences'' \cite{gwilt_towards_1997}. They connected evolutionary processes to the audience's actions, fostering unpredictable and open-ended artworks, and distinguishing between artworks that utilize pre-designed or pre-programmed ``paths of interaction.'' Their approach has become a significant reference in the study of audience-artwork interactions \cite{gwilt_towards_1997}. Similarly, Steve Dixon's taxonomy on ``concerning ascending levels and depth of interactivity'' identifies four interaction categories: navigation, participation, conversation, and collaboration \cite[p.597]{dixon_digital_2015}.

Building on the classification work originally developed by Apple Computer in \textit{Multimedia Demystified: A Guide to the World of Multimedia,} Hannington and Reed defined multiple levels of interactivity in multimedia applications: passive, where content is linear and users can only start or stop the content; interactive, allowing users to navigate through content; and adaptive, where users contribute content and control its usage \cite{hannington_towards_2002}. Although intended for multimedia, this framework has influenced categorizations in interactive art \cite{schraffenberger_audience-artwork_2015}. The V2\_ Institute for the Unstable Media further refined this idea in its taxonomy (details in Sec.\ref{v2_taxonomy}), delineating ``interaction levels'' as observational, navigational, participatory, co-authoring, and intercommunication. A comprehensive list of these models focusing on audience experience and interaction is presented in Tab.\ref{tab:taxonomies2} in the Appendix.

\subsection{Ars Electronica and V2\_ Institute for the Unstable Media} \label{v2_taxonomy}

The V2\_ Institute for the Unstable Media has focused on categorizing and archiving (interactive) media art, particularly in their research aimed at ``capturing unstable media'' \cite[p.19]{fauconnier_description_2003}. Their interactive art model specifies the context and conditions of interaction, detailed in Tab.\ref{tab:ars-v2} in the Appendix. This model includes time flexibility and interaction location to specify when and where interactions occur; user number to indicate permissible audience size; interaction level to describe the intensity and direction of communication, categorized as observational, navigational, participatory, co-authoring, and intercommunication; and sensory mode, which identifies which of the user's senses are engaged during the interaction.

In the context of the Prix Ars Electronica, collaborative efforts by Kwastek and the Ludwig Boltzmann Institute have characterized interactive art as ``an action-based aesthetic experience'' \cite{ludwig_boltzmann_institute_visualization_nodate}. They developed keywords based on analyses of Ars Electronica's archive, entries for the Prix Ars Electronica, expert feedback, and artists' descriptions of their submissions \cite{kwastek_research_2007, kwastek_research_2009}. Nine keyword categories, outlined in Tab.\ref{tab:ars-v2} in the Appendix, comprehensively describe interactive artworks. These primarily address interaction partners—the performer (visitor) and the work (project)—and their actions. Importantly, the category of interaction partners extends beyond just audience-artwork interactions to include various constellations, such as interactions among audience members.

While these descriptive models do not focus on an embodied perspective, they offer a robust framework of keywords and attributes that recognize the significance of both artwork and audience actions in modelling interactive art. However, they are limited in capturing certain aspects of the interactive art, such as the specifics (e.g., modality of sensations) of input and output in interactions and the communication's direction, which are crucial for understanding the reciprocal nature of interactions between the artwork and its audience.

\section{Embodiment, Embodied Interaction, and Somaesthetics in Interactive Art} \label{sec: literature}

This section builds upon previously reviewed descriptive models and taxonomies of interactive art, positioning our investigation within the context of embodied interaction. Our inquiry specifically focuses on embodied perspectives, which include body embodiment, bodily sensations, and audience-artwork interactions within interactive art experiences. Initially, we review Don Ihde's exploration of bodies in their phenomenal and sociocultural dimensions and their relationship to technology, establishing a philosophical foundation for our model. Subsequently, we delve into the concept of ``embodied interaction'' as defined in human-computer interaction by Paul Dourish. Finally, we examine interactive art theories that address embodiment and diverse sensational experiences.

\subsection{Bodies in Technology: Embodiment, Disembodiment, and Re-embodiment}
Don Ihde, a leading philosopher of technology, has explored the multifaceted ways we experience our bodies within a technologically textured world in his seminal work, \textit{Bodies in Technology} \cite{ihde_bodies_2002}. He draws upon Maurice Merleau-Ponty's concept of the ``body-subject,'' emphasizing that ``as subjects we are embodied, and to that extent, our being is inextricably entangled with the world. This fundamental relationship between body and world is the sole source of meaning.''

\subsubsection{Bodies in Technology}
Ihde elaborates on the primordial relationship between the body and the world by introducing additional dimensions. Beyond the phenomenal body, he identifies a second dimension: the socially and culturally constructed body, which emerges from the interactions of the phenomenal body with other ``body-subjects.'' A third dimension appears at the intersection of the phenomenal and social bodies with technology \cite{pepperell_where_2003}. Ihde suggests that experiencing the world through technology fosters a utopia that transcends human limitations and creates ``techno fantasies,'' extending the phenomenal body's capabilities and reach in the world \cite{koukal_bodies_2002}. Nappi further encapsulates this concept by describing it as an ``ontological triangulation of our experiences of our bodies in relation to technology,'' outlining three distinct embodying experiences: (1) as motile, perceptual, and emotive beings-in-the-world; (2) as entities shaped by social and cultural interactions; and (3) as entities that navigate through technology, linking the first two dimensions with technological mediations \cite{nappi_bodies_2004}. These can be succinctly described as the Phenomenal Body, the Sociocultural Body, and the Techno Phenomenal and Techno Sociocultural Body.

\subsubsection{Disembodiment}
Ihde explores the concept of disembodiment, where technology not only corrects or enhances the perceptual faculties of the body \cite{koukal_bodies_2002} but also projects and objectifies it into a ``virtual body.'' This form of body visualization, generated by various technologies, often results in a predominance of virtual bodies over situated bodies-in-the-world \cite{ihde_bodies_2002}. Ihde critiques the traditional use of scientific instruments in a visualist mode, which privileges these virtual representations \cite{ihde_bodies_2002}. This critique opens new avenues for interactive art practices and research to transcend visualist modes and engage more deeply with ``situated bodies-in-the-world.'' Motivated by the contemporary relevance of the ``virtual body'' in the realms of virtual, augmented, and mixed reality, we propose an expansion of Ihde’s original ``three bodies'' framework to include this additional ``virtual body.''

In his subsequent work, \textit{Embodied Technics} \cite{ihde_embodied_2010}, Ihde addresses the ``erroneous ascription of the dominance of the visual sense in humans'' \cite{schilhab_embodiment_2010}. Historically, from Plato to Descartes and Locke, the visualization of the relationship between humans and the world has over-emphasized vision. Despite humans, like most primates, having a well-developed visual system, Ihde argues that it should not overshadow other sensory modalities \cite[p.2]{ihde_embodied_2010}. The idea that our focus on visual input could circumvent most other bodily sensations is, according to Ihde, a significant oversight. He asserts that the considerable portion of our brain dedicated to the visual system should work in synergy with other senses, promoting a ``multidimensioned sense of body'' that integrates all sensory experiences \cite[p.4]{ihde_embodied_2010}.


\subsubsection{Re-embodiment} 
Ihde articulates that the multidimensionality of bodies is crucial for re-embodiment through technology. Frequent interactions with technologies diminish the unease and perceived distance between the body and the technological device, which Ihde refers to as an ``instrument.'' This process allows the human somatosensory system to seamlessly integrate technological objects into our embodied self-experience. Specifically, in the context of interactive art, the continuous detection of human behavior through technological device, while simultaneously overlooking moments that might emphasize the separation between user and the ``instruments,'' enhances the transparency of the interaction. This dynamic is particularly evident in how audience-artwork interactions in interactive art can transcend the mere physicality of the encounter. Overall, Ihde's post-phenomenological perspective incorporates both ``extended mind'' theories and bio-semiotic principles, which together highlight the evolutionary development of cognitive systems and underscore the complex interplay of bodily processes that shape our embodied identity.


\subsection{Embodied Interaction in Human-computer Interaction}
Paul Dourish has introduced the term ``embodied interaction'' in human-computer interaction in his ground-breaking work \cite{dourish_where_2001-1}, proposing ``a more coherent understanding of our interaction (intersection) with machines ... involve consideration of human mental activity as embedded in a wider physical context'' \cite{pepperell_where_2003}. Grounding his ideas in a rich philosophical foundation, Dourish embarks on a ``whirlwind tour'' \cite[p.124]{dourish_where_2001-1} through the phenomenological philosophies of Husserl, Heidegger, Merleau-Ponty, and Wittgenstein. He elucidates how these thinkers have collectively shaped a tradition in Western thought that views the human agent as intimately situated in—and dynamically interacting with—a pre-organized world filled with physical demands and possibilities.

Accordingly, Dourish posits that the mind is not separate but is embodied in the physical presence of the body; similarly, the body is deeply rooted in the ``real world''—a complex matrix of natural, social, and cultural objects and events. His concept of ``embodied interaction'' emerges from two foundational ideas: tangible and social computing, which underscore the integration of human bodily experience with the broader environmental context \cite{dourish_embodied_1999}. He argues that ``the history of interaction is a gradual expansion of the range of human skills and abilities that can be incorporated into interacting with computers'' \cite[p.17]{dourish_where_2001-1}.

``Tangible computing'' pushes human-computer interaction research beyond the traditional desktop metaphor towards more distributed and tactile environments, emphasizing the physical apparatus and the body's role in interaction. Conversely, ``social computing'' focuses on the sociological, cultural, and historical contexts within which technology is used. Dourish emphasizes that ``social action is embedded'' \cite[p.96]{dourish_where_2001-1}, grounding it in real-world practices rather than abstract theories. Drawing from the ethnomethodological work of Harold Garfinkel and Lucy Suchman who studied situated action, Dourish advocates for an ``action-oriented'' analysis that respects the mundane, often overlooked aspects of social life \cite[p.96]{dourish_where_2001-1}.


\subsection{Sensation, Proprioceptive Experience, and Affect in Interactive Art}
The examination of interactive art through an embodied perspective, particularly focusing on audience-artwork interactions, has been explored extensively. Nathaniel Stern critiques the dominant linguistic and visual signification paradigms in digital art discussions, arguing that they fail to fully address the complexities of interactive art. According to Stern, the essence of interactivity lies in the physical engagement of the audience, making embodied interaction central to the realization of artwork \cite[p.1]{stern_implicit_2009}. Caroline A. Jones, in her curatorial essays for the exhibition \textit{Sensorium}, challenges the ``reigning hegemony'' of the five senses, urging a reevaluation of digital art within an expanded sensorial framework \cite[p.8]{jones_sensorium_2006}. Similarly, the taxonomy developed by V2\_ (detailed in Tab.\ref{tab:ars-v2}) includes a ``sense mode'' dimension, which encompasses the visual, auditive, olfactory, tactile, and gustative senses, further emphasizing the multi-sensory experience of interactive art.

Brian Massumi advocates for deeper explorations into sensation, proprioceptive experience, and affect within both philosophy and art. He delineates three interrelated experiences: perception, affect, and proprioception, each contributing uniquely to our interaction with the world. For Massumi, perception is described as an ``object-oriented experience'' that pertains to the ``stasis-tending dimension of reality'' \cite[p.258]{massumi_parables_2002}, shaping our conscious understanding of sensory inputs. Affect, in contrast, represents a preconscious, autonomous bodily response that encompasses all sensory reactions. Proprioception, which includes both tactile sensibility (exteroceptive) and visceral sensibility (interoceptive) \cite[p.58]{massumi_parables_2002}, integrates tactile feedback and translates the physical interactions with objects into a muscular memory, enhancing relational understanding and responsiveness \cite[p.59]{massumi_parables_2002}.

Massumi posits that affect and proprioception together constitute ``sensation,'' which reveals a dimension of experience that is beyond conscious reflection or understanding of the body. He distinguishes between perception, which he terms ``exo-referential (extensive),'' and sensation (encompassing affect, proprioception, and other sensibilities), which he describes as ``endo-referential or self-referential (intensive)'' \cite[p.259]{massumi_parables_2002}. This distinction highlights that perception shapes sensation, altering how we sense in real-time and thereby transforming our engagement with the sensed world. Massumi argues that a focus on sensation enriches our understanding by reintegrating materiality and corporeality into cognitive processes, suggesting that an emphasis on sensation does not detract from but rather deepens perceptual experiences.



\section{Our Three-dimensional Model of Interactive Art from an Embodied Perspective} \label{sec: model}

This section presents our descriptive model of interactive art, conceptualized from an embodied perspective. The model is structured around three key dimensions: the various body embodiments of the audience, the sensations or sensibilities activated, applied, or experienced during interaction, and the interactions between audience and artwork. 

\begin{figure}[h]
\centering
\scalebox{0.5}{
\begin{tikzpicture}
  \path[mindmap,concept color=purple!50!red,text=white]{
            node[concept] {Embodiment}
                [clockwise from=-60]
                    child [concept color=purple!100!red]{ node[concept] {Sociocultural}
                        [clockwise from=80]
                        child { node[concept] {Techno Sociocultural (TS) Body} }
                        child { node[concept] {Virtual Sociocultural (VS) Body} }
                        child { node[concept] {Hyper Sociocultural (HS) Body} }
                    }
                    child [concept color=purple!100!red]{ node[concept] {Phenomenal} 
                        [clockwise from=-130]
                        child [concept color=purple!150!red]{ node[concept] {(Phenomenal (PB) Body)} }
                        child { node[concept] {Techno Phenomenal (TP) Body} }
                        child { node[concept] {Virtual (VB) Body} }
                    }
    }; 
\end{tikzpicture}
}
\caption{``Body Embodiment Dimension'' comprises two subcategories: Phenomenal Body and Sociocultural Body, further divided into five options. Recognizing that interactive art is inherently technology-mediated, the Phenomenal Body (PB) is not viable. The subcategory based on the Phenomenal Body includes the Techno Phenomenal Body and Virtual Body. The sociocultural subcategory features Techno Sociocultural Body, Virtual Sociocultural Body, and Hybrid Sociocultural Body.}
\label{fig: embodiment}
\end{figure}
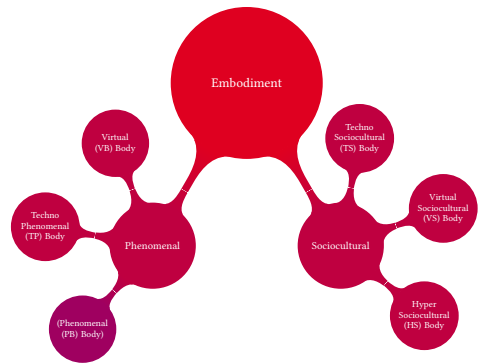



\subsection{Body Embodiment} \label{sec: embodiment}

Building on previously addressed philosophical perspectives (i.e., phenomenology) and technological perspectives (i.e., human-computer interaction) of embodiment and embodied interaction—particularly Ihde's concept of ``three bodies''—we conceptualize embodiment in interactive art as comprising ``five bodies'' of the audience, as illustrated in Fig.\ref{fig: embodiment}. We omit the Phenomenal Body since the experiences and interactions within Interactive Art are fundamentally technology-mediated. The five bodies include the Techno Phenomenal Body (TP; technology-mediated physical body), the Virtual Body (VB; virtually represented body), and three types of Sociocultural Bodies within the contexts of TP and VB. The sociocultural dimensions of interaction can occur among a group of Techno Phenomenal Bodies (Techno Sociocultural Bodies, TS), a group of Virtual Bodies (Virtual Sociocultural Bodies, VS), or a mixed group of both in a hybrid form (Hybrid Sociocultural Bodies, HS). For clarity, we use ``techno'' to refer to ``technology-mediated'' throughout this discussion.

\begin{figure}[h]
\centering
\scalebox{0.5}{
\begin{tikzpicture}
  \path[mindmap,concept color=blue,text=white]{
        node[concept] {Audience-artwork Interaction}
            [clockwise from=300]
                child [concept color=gray!40!blue]{ node[concept] {What the Audience Does [S]} }
                child [concept color=gray!40!blue]{ node[concept] {What the Audience Perceives [R]} }
    };
\end{tikzpicture}
}
\label{fig: audience-artwork}
\caption{Audience-artwork interaction model distinguishing ``what the audience does'' and ``what the audience perceives'' as ``Send'' and ``Receive'' from the audience's perspective and their bodily experience.}
\end{figure}
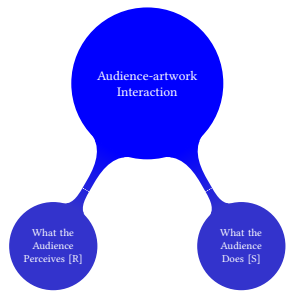

\subsection{Audience-artwork Interaction} \label{sec: audience-artwork}

``Audience-artwork Interaction'' is often conceptualized as ``Input versus Output'' in previous interactive models, particularly those with system-oriented perspectives discussed in Sec.\ref{subsec: system}. Following Kwastek's model for the Ars Electronica Archive, we adopt the framework of ``The performer (visitor) does'' versus ``The work (project) does,'' but with a specific focus on the audience's embodied experience. Thus, we describe ``what the audience does'' and ``what the audience receives, perceives, and experiences (or simply, what the audience perceives)'' as ``Send'' and ``Receive'' from the audience's perspective and their bodily experience. This is integrated with the concept of somaesthetics (i.e., exteroception and proprioception) discussed in Sec.\ref{sec: soma} to examine the bi-directional process of audience-artwork interaction and embodied experiences.

Applying Dourish's perspective of embodied interaction to interactive art, and incorporating Ihde's distinction between the phenomenal body and the sociocultural body, we recognize that the interaction between the audience and the artwork is not the only ``interaction'' occurring in an interactive art experience. We introduce the concept of ``audience-audience interaction'' to describe the sociocultural body embodiment of the audience, emphasizing the interpersonal interactions among audience members facilitated by or occurring within the artwork. This highlights the social dimensions of the interactive art experience. Since this concept primarily applies to artworks with TS embodiment, it is not developed as a separate dimension but is used to categorize relevant artworks into four subcategories in Sec.\ref{sec: TS}, where we discuss artworks involving TS embodiment.


\begin{figure}[h]
\centering
\scalebox{0.5}{
\begin{tikzpicture}
  \path[mindmap,concept color=purple!50!green,text=white]{ 
        node[concept] {Somaesthetic}        
            [clockwise from=210]
                child [concept color=black!75!green]{
                node[concept] {Exteroception \\(Five Senses)} 
                        [clockwise from= -15]
                        child { node[concept] {Visual (VI)} }
                        child { node[concept] {Auditive (AU)} }
                        child [concept color=purple!60!green]{ node[concept] {\sout{Olfactory (OL)}} }
                        child [concept color=purple!60!green]{ node[concept] {\sout{Gustative (GU)}} }
                        child { node[concept] {Tactile or Touch (TA)} }
                        % child { node[concept] {VP} }
                        % child { node[concept] {PA} }
                        % child { node[concept] {TE} }
                        }
                 child [concept color=black!75!green]{
                node[concept] {Exteroception \\(New)} 
                        [clockwise from= 180]
                        % child { node[concept] {VI} }
                        % child { node[concept] {AU} }
                        % child { node[concept] {OL} }
                        % child { node[concept] {GU} }
                        % child { node[concept] {TA} }
                        child { node[concept] {Vibratory (VP)} }
                        child { node[concept] {Pain (PA)} }
                        child { node[concept] {Temperature (TE)} }
                        }
                child [concept color=purple!60!green]{ 
                     node[concept] {\sout{Interoception}} }
                child [concept color=purple!60!green]{ 
                     node[concept] {\sout{Affect}} }
                child [concept color=black!75!green]{ 
                     node[concept] {Proprioception} 
                        [clockwise from=60]
                        child { node[concept] {Presence (PR)} }
                        child { node[concept] {Joint Position (JP)} }
                        child { node[concept] {Kinesthesia (KI)} }
                        }

    };    
\end{tikzpicture}}
\caption{``Somaesthetic'' with two primary subcategories: Exteroception and Proprioception. This review does not cover Interoception and Affect. Exteroception encompasses the traditional five senses along with three additional exteroceptive sensations. Proprioception is divided into three distinct aspects. Altogether, eleven options are explored within this dimension, but Olfactory (OL) and Gustative (GU) are not found in our search.}
\label{fig: somaesthetic}
\end{figure}
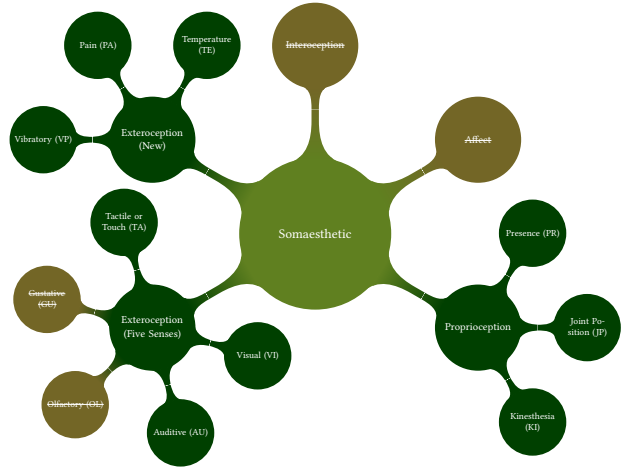



\subsection{Somaesthetics: Exteroception and Proprioception}\label{sec: soma}

We draw on medical and clinical definitions of the sensory system to establish the somaesthetic dimension of our model. Sensations are classified based on anatomical or functional criteria, providing individualized information relative to the environment \cite{bigley_sensation_1990}. Anatomically, sensory functions are divided into somatic and visceral components, each with general and special subgroups \cite{bigley_sensation_1990}. Functionally, sensory modalities are categorized into simple affective sensations (i.e., protopathic) and those that offer discriminative analysis of the environment (i.e., epicritic). Sherrington utilizes both anatomical criteria (i.e., types and locations of end organs) and functional criteria (i.e., types of stimuli measured by each modality) to differentiate exteroceptive and proprioceptive sensations. He also introduces a third sensory modality that interprets primary sensory information in a complex manner \cite{bigley_sensation_1990}.


Based on the clinical definition of exteroceptive sensation (also termed superficial sensation), we further expand the sense of tactile to encompass three distinct senses: tactile or touch sensation (thigmesthesia), pain sensation (algesia), and temperature sensation (thermesthesia). Conversely, proprioceptive sensations, although frequently present in many interactive artworks, are often underrecognized in existing models. To address this, we incorporate proprioceptive sensations in the somaesthetic dimension of our model, identifying three key sensations: joint position sense (arthresthesia), vibratory sensation (pallesthesia), and kinesthesia. By integrating both exteroceptive sensations (or exteroception) and proprioceptive sensations, we aim to present a comprehensive spectrum of bodily sensations potentially involved in interactive art experiences.


We have revised the categories and options for exteroception and proprioception based on the surveyed artworks (details in Appendix Tab.\ref{tab: corpus}) and our expertise. We have relocated the vibratory sensation from proprioception to exteroception, resulting in eight sensations, as it is more intuitive to understand it as an exteroceptive experience in interactive art. For proprioceptive experience, we further specify these sensations in the context of interactive art based on clinical definitions. We use joint position sensation (arthresthesia) to represent partial body involvement, such as the hand only, and kinesthesia for full-body engagement. Additionally, we introduce a ``presence'' option for this dimension, indicating that only the body's position matters for interaction. This also partially aligns with the dimension of ``triggering parameters'' in the taxonomy of Trifonova et al. \cite{trifonova_software_2008}.


As per the clinical categorization of sensations and Brian Massumi's arguments, interoception—particularly when connected to affect—is also crucial to consider as part of the embodied experience and embodiment. However, as with most model of interactive art, attempting to describe all aspects can lead to excessive complexity \cite{schraffenberger_audience-artwork_2015}. In our descriptive model, we have intentionally omitted these two subcategories of somaesthetic, thus focusing on exteroception and proprioception.

\begin{figure}[h]
\centering
\begin{tikzpicture}[scale=0.7, transform shape]
  \path[mindmap,concept color=purple,text=white]
    node[concept] {Body}
    [clockwise from=150] 
    child[concept color=purple!50!green] { 
        node[concept] {Somaesthetic}
            [clockwise from=210]
                child { node[concept] {Exteroception} 
                        [clockwise from=-15]
                            child { node[concept] {VI} }
                            child { node[concept] {AU} }
                            child { node[concept] {TA} }
                            child { node[concept] {VP} }
                            child { node[concept] {PA} }
                            child { node[concept] {TE} }
                        }
                child { node[concept] {Proprioception} 
                        [clockwise from=150]
                        child { node[concept] {PR} }
                        child { node[concept] {JP} }
                        child { node[concept] {KI} }
                        }
    }    
    child[concept color=purple!50!red] {
            node[concept] {Embodiment}
                [clockwise from=60]
                    child { node[concept] {Phenomenal}
                        [clockwise from=30]
                        child { node[concept] {TP} }
                        child { node[concept] {VB} }
                    }
                    child { node[concept] {Sociocultural} 
                        [clockwise from=15]
                        child { node[concept] {TS} }
                        child { node[concept] {VS} }
                        child { node[concept] {HS} }
                    }
    }
    child[concept color=white!100!white] {
    }
    child[concept color=blue] {
            node[concept] {Audience-artwork Interaction}
                [clockwise from=30]
                    child {node[concept] {Audience Perceives (R)}}
                    child {node[concept] {Audience Does (S)}} 
    };
    \node[draw,align=left,color=purple!50!green] at (-8.5, 9) {
        \textbf{Somaesthetic} (in Sec.\ref{sec: soma})
    };
    \node[align=left, color=purple!50!green] at (-9, 7.75){
        \textbf{PR:} Presence\\
        \textbf{JP:} Joint Position \\
        (Arthresthesia)\\
        \textbf{KI:} Kinesthesia
    };
    \node[align=left, color=purple!50!green] at (-9, 2.25) {
        \textbf{TE:} Temperature\\
        (Thermesthesia)\\
        \textbf{PA:} Pain (Algesia) \\
        \textbf{VP:} Vibratory \\
        (Pallesthesia) 
    };
    \node[align=left, color=purple!50!green] at (-8.75, -4) {
        \textbf{TA:} Tactile or Touch \\
        (Thigmesthesia)\\
        \textbf{AU:} Auditive\\
        \textbf{VI:} Visual
    };
    \node[draw,align=right, color=purple!50!red] at (8.75,9) {
        \textbf{Embodiment} (in Sec.\ref{sec: embodiment})
        };
    \node[align=right, color=purple!50!red] at (8.6, 5.5) {
        \textbf{TP:} Techno Phenomenal Body\\ 
        \textbf{VB:} Virtual Body\\ 
        \\
        \\
        \\
        \\
        \\
        \\
        \\
        \textbf{TS:} Techno Sociocultural Body\\
        \textbf{VS:} Virtual Sociocultural Body\\
        \textbf{HS:} Hybrid Sociocultural Body
    };
    \node[draw,align=right, color=blue] at (7.65,1.5) {
        \textbf{Audience-artwork Interaction} (in Sec.\ref{sec: audience-artwork})
    };
    \node[align=right, color=blue] at (8.6, 0.5) {
        \textbf{R:} What the Audience Perceives \\
        \textbf{S:} What the Audience Does\\
    };
\end{tikzpicture}
\caption{The three dimensions of our descriptive model of interactive art from the embodied perspective.}
\label{fig: three-dimension}
\end{figure}
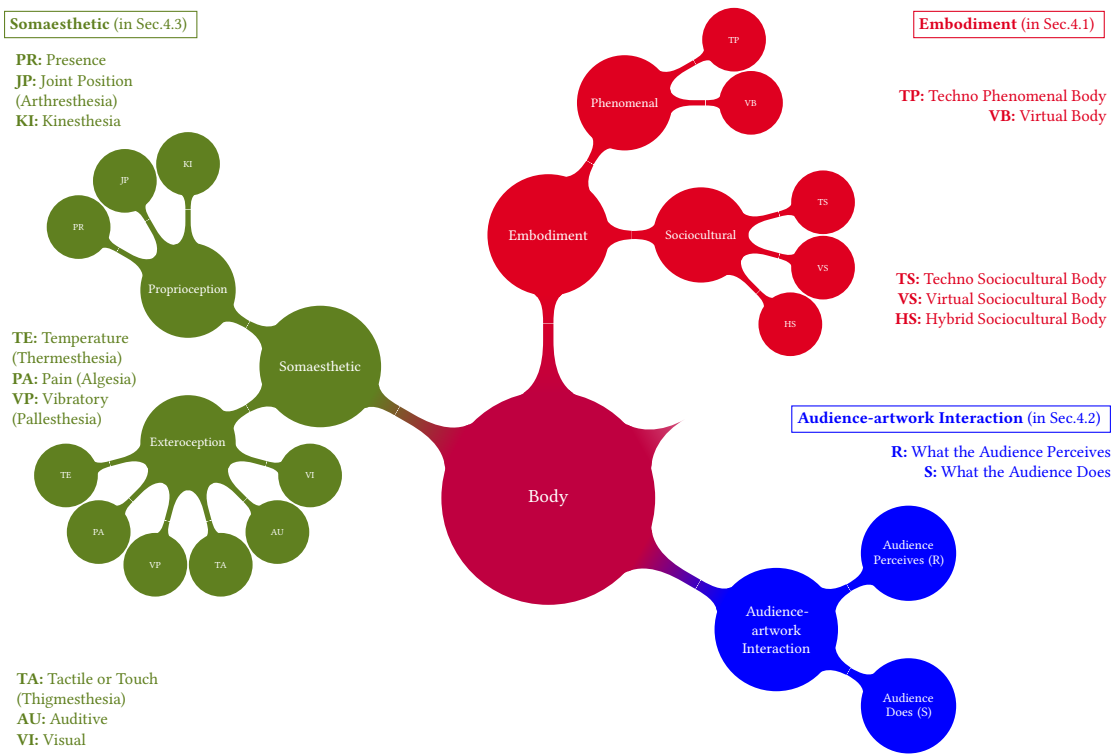

\begin{figure}[h]
    \centering
    \includegraphics[width=0.85\linewidth]{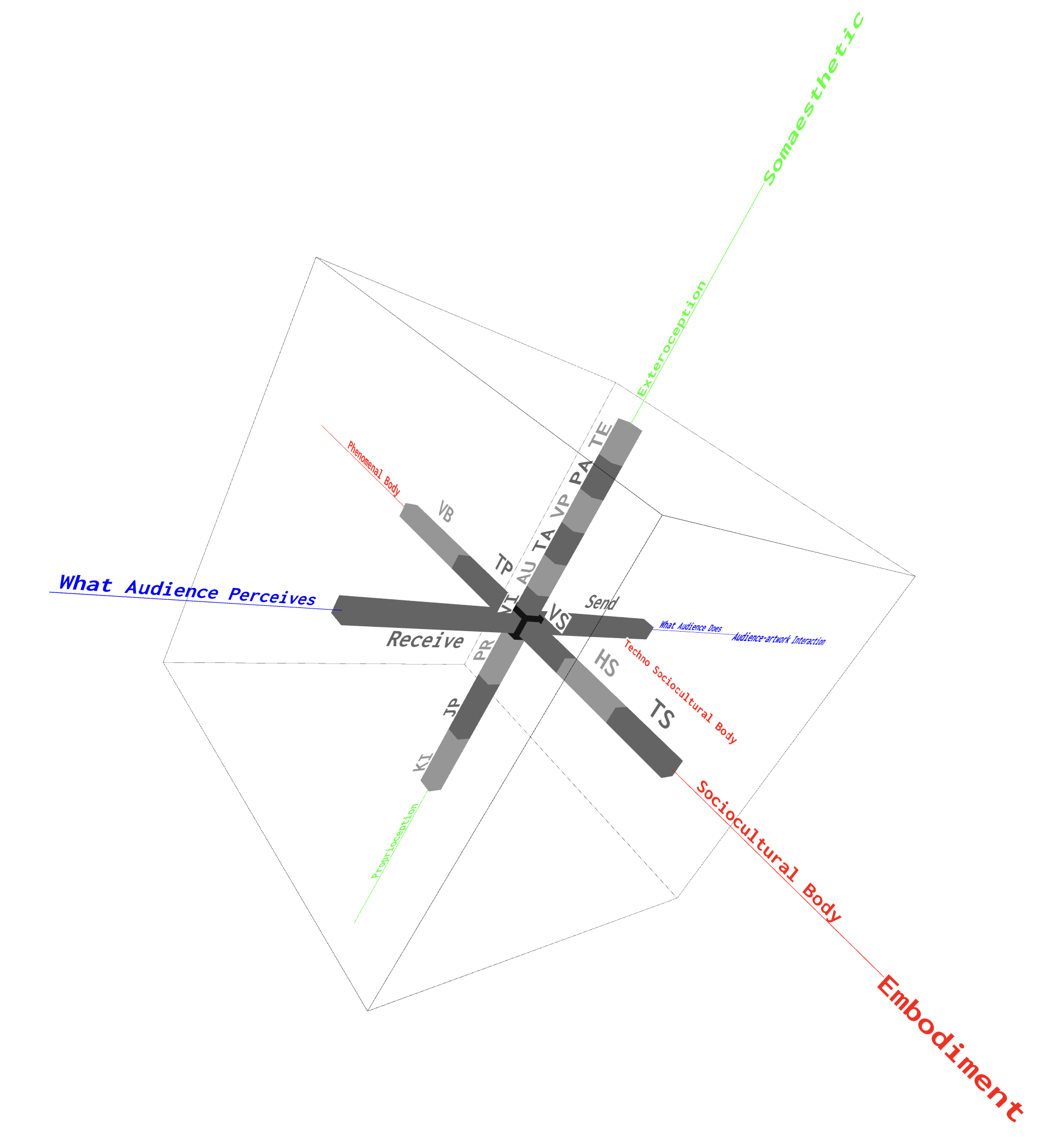}
    \caption{Visualization of our three-dimensional descriptive model.}
    \label{fig: model}
\end{figure}

\subsection{Three-dimensional Descriptive Model and Taxonomy}

Fig.\ref{fig: three-dimension} presents our descriptive model of interactive art from an embodied perspective. This model comprises three dimensions: (1) body embodiment, encompassing the five bodies; (2) audience-artwork interaction, detailing ``what the audience does'' and ``what the audience perceives''; and (3) somaesthetic, referring to the bodily sensations or sensibilities applied, activated, or experienced during the interactive art experience.


Combining the three aspects discussed, we have visualized the established model as shown in Fig.\ref{fig: model}. This model describes a given interactive artwork across the three dimensions—embodiment, audience-artwork interaction, and somaesthetic experience—corresponding to the axes in a Cartesian coordinate system. Each artwork is defined by a sets of points, referring its position on each axis as the tree dimensions. On the embodiment axis, each artwork is aligned with one of five possible categories: TP, VB, VS, HS, or TS, represented as ticks on the axis. The somaesthetic axis allows each artwork to incorporate multiple sensations from nine options. For each sensation, the audience-artwork interaction axis classifies it as ``S - the audience does'' and ``R - the audience perceives.'' 



\section{Body Embodiment and Embodied Audience-artwork Interaction} \label{sec: describe}

This section describes our corpus through the dimension of body embodiment and the audience-artwork interaction, including what audience does and perceives as well as artwork-mediated interpersonal interaction among audience members. The techno-phenomenal (TP) body embodiment focuses solely on audience-artwork interaction. Based on various behaviors of the audience—artwork interaction, we group the 21 artworks with TP embodiment into five clusters due to similar concepts and/or artistic mediums. The techno-sociocultural (TS) body embodiment features multi-audience interactions instead of individual ones. Sec.\ref{sec: TS} examines artworks in the corpus that support and/or require multiple audiences and discusses two types of interpersonal interaction (post and direct) among the audience. Finally, due to the limited number of artworks involving Virtual Bodies (VB), Virtual Sociocultural Bodies (VS), and Hybrid Sociocultural Bodies (HS), we group these together and discuss them in Sec.\ref{Sec: VBS}.



\subsection{Corpus Construction} \label{sec: corpus}

Dinkla, in defining the ``origin of interactive art,'' stated that ``interactive art are developed mainly outside traditional art institutions such as galleries and museums,'' and that ``interactive art are above all media art festivals'' \cite[p.279]{daniels_strategies_2008}. While this may be less true in today's context compared to when Dinkla wrote her essay in 2008, art festivals like the previously mentioned Prix Ars Electronica remain significant. The high-quality archives of such festivals serve as primary research materials that are widely recognized in the field. All artist submissions for the Prix since 1987 can be searched and viewed, with winning projects documented with extensive information and audio-visual media, available at \url{https://archive.aec.at/prix/}. Roger Malina described it as ``by establishing the prize for Interactive Art, the organizers of the Prix Ars Electronica have taken the lead in recognizing artworks in a newly emerging art form'' \cite[p.157]{leopoldseder_prix_1990}.




Due to its significant authority and longstanding reputation, we have chosen to review the Prix Ars Electronica archive over the years to establish our research corpus. The interactive art category has been awarded biennially since 2014 and was renamed ``Interactive Art +'' in 2018. Each year, fifteen submissions are awarded: one ``Golden Nica,'' two Awards of Distinction (``AUSZEICHNUNG''), and twelve Honorary Mentions (``ANERKENNUNG''). On average, there have been 3027 artwork submissions in the past five years, with approximately 1000 pieces in each category, such as Interactive Art +, Artificial Intelligence \& Life Art, etc. In other words, the ``Submission with Award'' list for the ``Interactive Art (+)'' category in each year's catalog and archive comprises fifteen interactive artworks selected from about one thousand submissions.



We review all ``Submissions with Award'' in the Prix Ars Electronica Archive - Interactive Art (IA) category over the past ten years and manually filtered artworks from our embodied perspective. Since the IA awards have been conducted biennially since 2013, we expanded the search timeframe to ``2009 - 2023'' to encompass ten years of awarded artworks. We retain only those artworks involving the audience's embodied interaction and filter out those not meeting the following criteria. First, the artwork must fall within our target timeframe, i.e., from 2009 to 2023, resulting in 150+ artworks. Second, the artwork must involve the audience's body and embodiment, which becomes the most ``influential'' criterion, filtering out nearly 80 artworks from 150. Forty-nine embodied interactive artworks are selected. In other words, nearly 10,000 artwork submissions are examined and selected by professional jury members, resulting in an awarded list of over 150 artworks through the years. Using our selection criteria, we further narrowed the corpus to 49 artworks for detailed analysis.

In the corpus, almost half artworks belong to the Techno Phenomenal Body category (23 out of 49), and the other half to the Techno Sociocultural Bodies category (20 out of 49). Only a few artworks fall into the Hybrid Sociocultural Bodies category (4 out of 49), with one artwork each in the Virtual Body and Virtual Sociocultural Bodies categories. This indicates a significant disparity, with far fewer VB-involved artworks (6 out of 49) compared to TP-relevant artworks (43 out of 49). 

For the somaesthetic dimension, the artworks are distributed across various options, which are further analyzed in Sec.\ref{sec: analysis}. Tab.\ref{tab: corpus} details all the selected artworks, including the artwork title, artist name(s), awarded year, and their categorization within the three dimensions of our descriptive model. Following common practice, we chose not to create a reference entry for each artwork, as they are not published academic resources. However, all relevant information is clearly listed in Tab.\ref{tab: corpus} and in the text. Notably, no artworks within our scope and corpus involved Olfactory (OL) or Gustative (GU) sensations, leading to their omission from the ``Exteroception'' subcategory in the table.



\subsection{Techno Phenomenal (TP) Body} \label{sec: TP}

The audience interacting with artwork embodying a Techno Phenomenal (TP) Body typically engages solely with the artwork or, at most, takes turns individually. This subsection reviews those artworks with a focus on their audience-artwork interaction, specifically what the audience does and perceives. Some artworks involve both TP and Techno Sociocultural (TS) Bodies, supporting both individual and group participation modes. Depending on the significant differences between these two modes, we describe those whose TS interaction and experience are almost identical to TP in this section (e.g., RAIN ROOM and IDEOGENETIC MACHINE); otherwise, they are discussed in Sec.\ref{sec: TS} Techno Sociocultural (TS) Bodies.

Based on the behaviors and perceptions of the audience, we cluster the 23 selected TP artworks into five distinct groups: (1) Audience Sending Bodily Movements as Image, where the audience's movements are captured and transformed into visual representations; (2) Artworks Innovating the ``Screen-display Experience,'' which re-imagines traditional screen-based interactions; (3) Audience Performing more than Dancing or Posing, Artwork not Merely Displaying Screen-based Contents, involving complex performances beyond simple gestures; (4) Audience Engaging with (Sports-like) Entertaining Activities, focusing on interactive artworks that incorporate physical, sports-like activities; and (5) Other Interactive Experiences for Social and Cultural Caring, which address social and cultural themes through interactive engagement. 


\subsubsection{Audience Sending Bodily Movements as Image}

In our definition, we consider performers, such as dancers, working with real-time interactive choreographic systems as the audience in interactive artwork. Examples include ``MORPHECHORE, 2022'' by Daito Manabe, ``THE ZIZI SHOW, 2022'' by Jake Elwes, ``PATHFINDER, 2016'' by Christian Loclair and Onformative, and ``PERFUME GLOBAL SITE PROJECT, 2013'' by Daito Manabe et al. These artworks typically translate the audience's real-time movements into coherent visual forms in image data format. From the perspective of proprioception, these works primarily involve kinesthesia, as the full body of the audience is engaged. This category also includes works like ``IDEOGENETIC MACHINE, 2012'' by Nova Jiang and ``SHADOW STALKER, 2020'' by Lynn Hershman Leeson, where real-time portraits of the audience are translated into customized visual styles.

\begin{figure}[h]
  \centering
    \hfill
   \begin{subfigure}{0.245\textwidth}
    \centering
    \includegraphics[height=.5\textwidth,keepaspectratio]{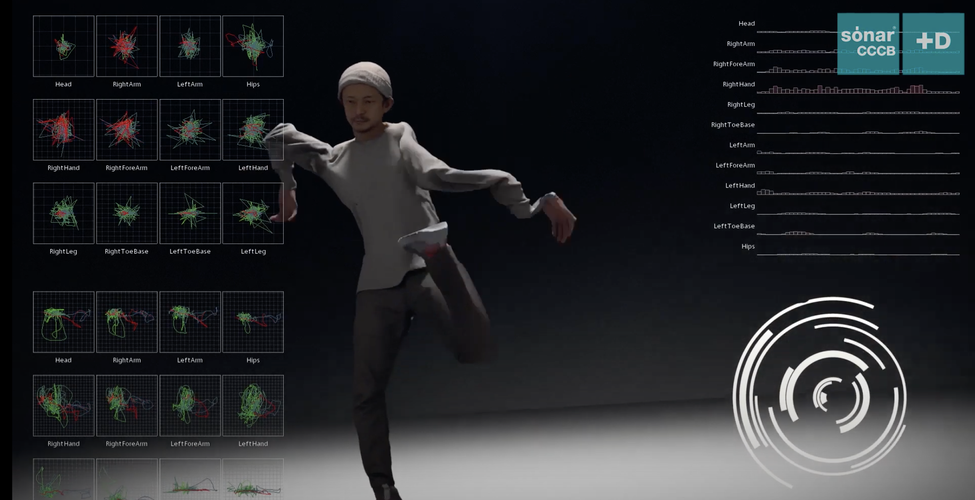}
      \caption{MORPHECHORE, 2022.\\Daito Manabe.}
      \label{fig: MORPHECHORE}
  \end{subfigure}
    \hfill
  \begin{subfigure}{0.245\textwidth}
        \centering
      \includegraphics[height=.5\textwidth,keepaspectratio]{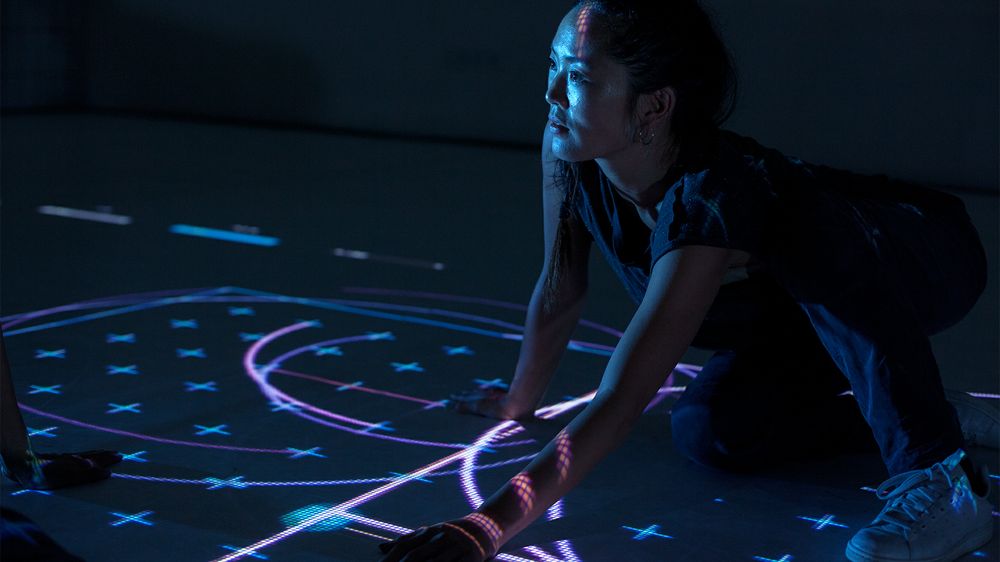}
      \caption{PATHFINDER, 2016.\\Christian Loclair.}
      \label{fig: PATHFINDER}
  \end{subfigure}
  \hfill
  \begin{subfigure}{0.245\textwidth}
        \centering
        \includegraphics[height=.5\textwidth,keepaspectratio]{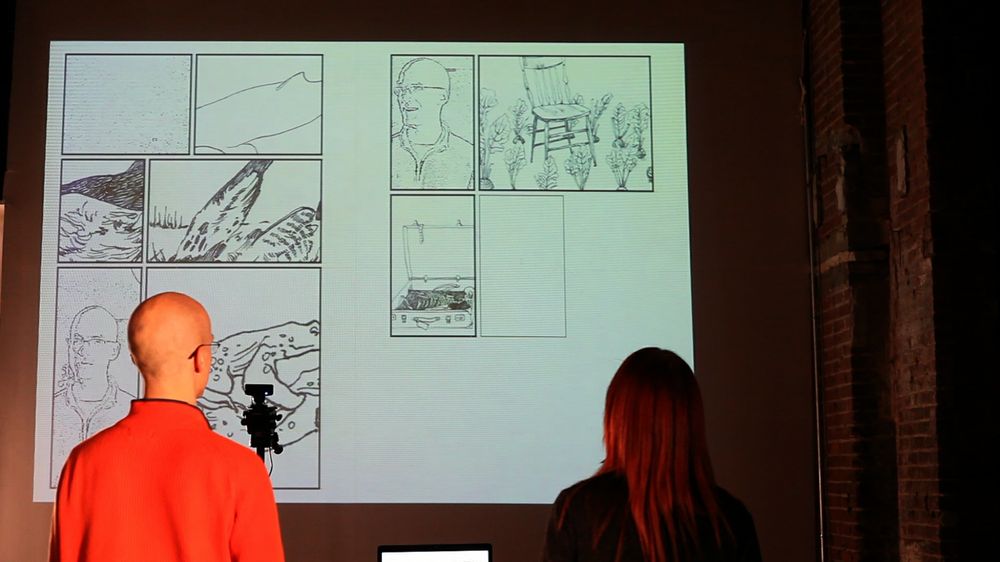}
      \caption{IDEOGENETIC MACHINE,\\2012. Nova Jiang}
      \label{fig: IDEOGENETIC}
  \end{subfigure}
          \hfill
    \begin{subfigure}{0.245\textwidth}
        \centering
        \includegraphics[height=.5\textwidth,keepaspectratio]{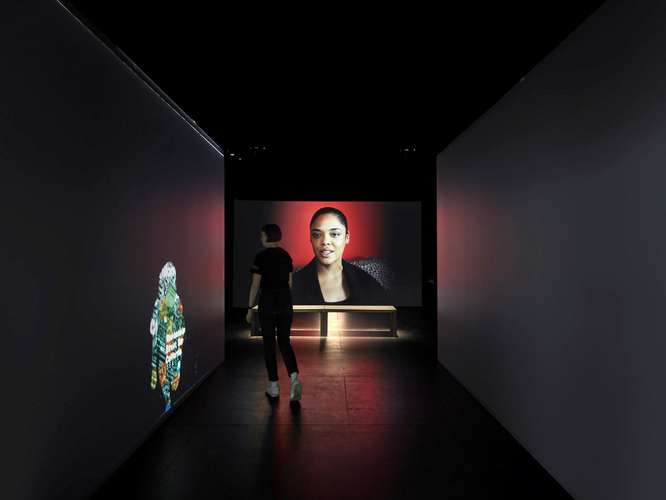}
      \caption{SHADOW STALKER, 2020.\\Lynn Hershman Leeson}
      \label{fig: STALKER}
  \end{subfigure}
  \caption{Artwork translating the audience's real-time portrait and bodily movement to coherent visual forms.}
  \label{fig: 1}
  \end{figure}
  

The only differences lie in the evolution of the ``translation'' functions over time. In earlier works, such as those by Jiang and Manabe, the translation relied on ``traditional'' algorithms, including various filtering effects in computer vision or customized procedural-generative algorithms. In contrast, more recent works, like Manabe's ``MORPHECHORE'' in Fig.\ref{fig: MORPHECHORE} utilize AI techniques to achieve conditional generative functions. In this type of artwork, the audience (or performer) engages their full body with the artwork through a vision-tracking system. The algorithmic generative components within the artwork then translate their bodily movements into digital visual (or audio-visual) content. The real-time generated visuals are subsequently displayed through computer or projection screens, providing the audience with aesthetic or conceptual experiences.
  


\subsubsection{Artworks Innovating the ``Screen-display Experience''}

Many early interactive artworks, particularly those from around 2009, focused on innovating the ``screen experience'' by pixelating image content and presenting it with arrays of physical units. For example, ``ANGLES MIRROR, 2013'' by Daniel Rozin uses triangular units at different rotating angles to simulate the illuminating light of an LED screen, continuously ``displaying'' a live portrait of the audience. Similarly, ``IN THE LINE OF SIGHT, 2009'' by Sauter and Winkler utilizes 100 computer-controlled tactical flashlights to project low-resolution video footage of suspicious human motion. These artworks exemplify the creative manipulation of physical and digital elements to transform traditional screen-based interactions.

\begin{figure}[h]
  \centering
      \begin{subfigure}{0.245\textwidth}
        \centering
        \includegraphics[height=.65\textwidth,keepaspectratio]{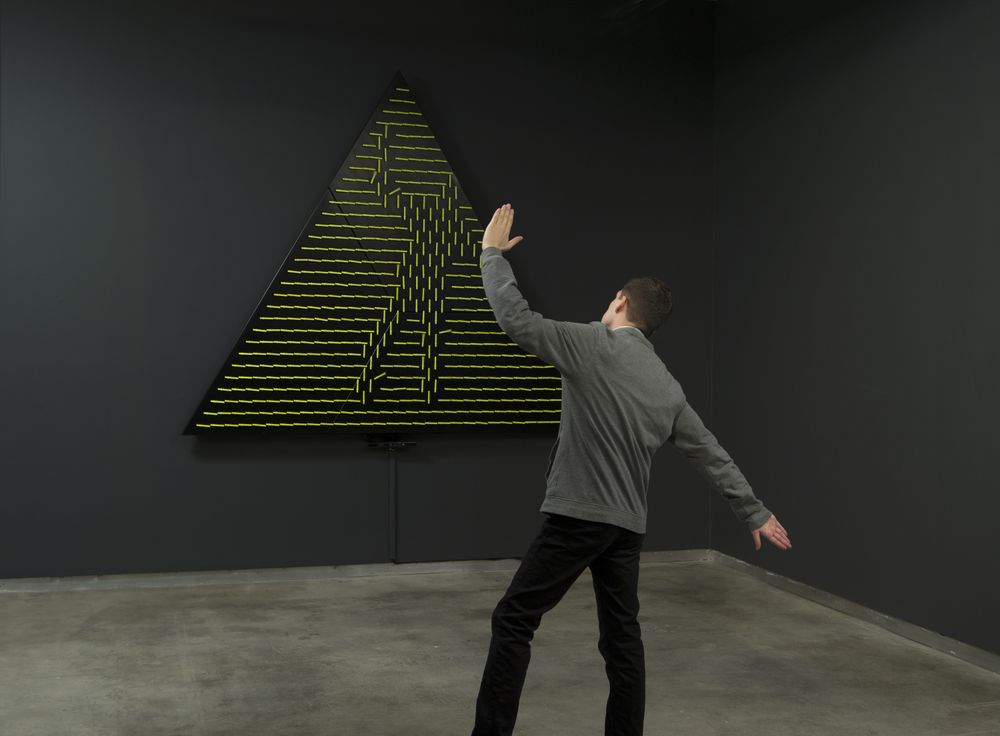}
      \caption{ANGLES MIRROR, 2013.\\Daniel Rozin}
      \label{fig: ANGELS}
  \end{subfigure}
      \hfill
  \begin{subfigure}{0.245\textwidth}
    \centering
      \includegraphics[height=.65\textwidth,keepaspectratio]{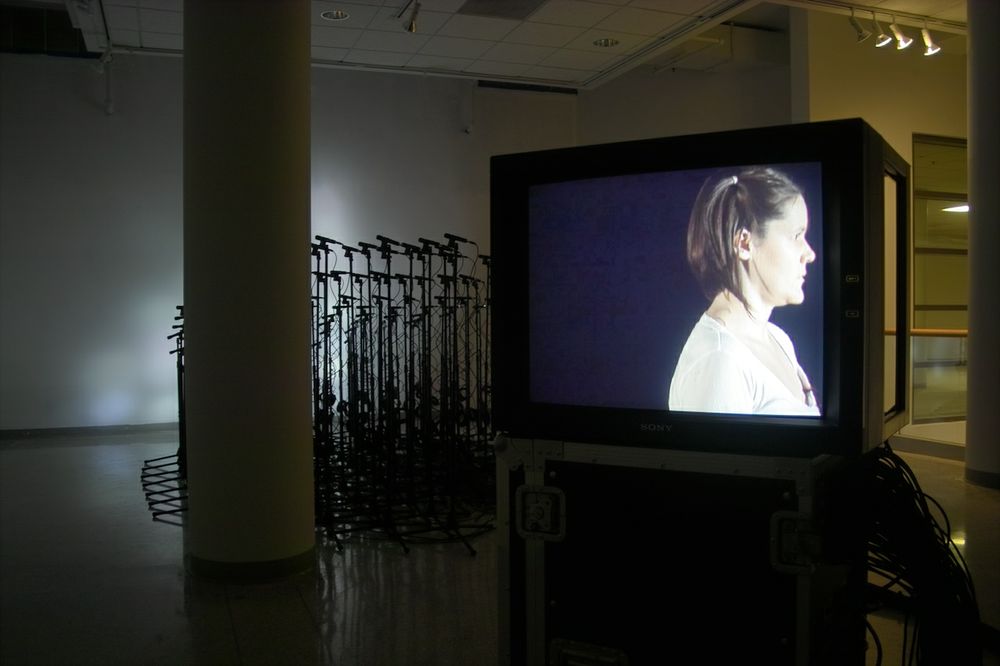}
      \caption{IN THE LINE OF SIGHT, 2009.\\Sauter and Winkler.}
      \label{fig: SIGHT}
  \end{subfigure}
      \hfill
  \begin{subfigure}{0.245\textwidth}
      \centering
      \includegraphics[height=.65\textwidth,keepaspectratio]{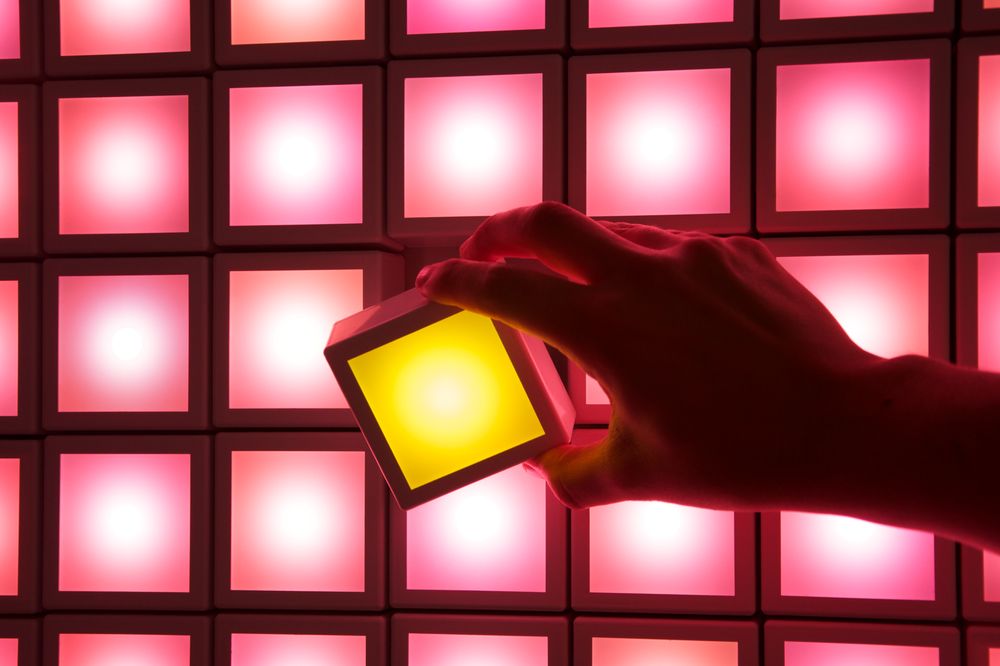}
      \caption{SIX-FORTY BY EIGHTY, 2011.\\Coelho and Zigelbaum.}
      \label{fig: SIX-FORTY}
  \end{subfigure}
      \hfill
    \begin{subfigure}{0.245\textwidth}
      \centering
      \includegraphics[height=.65\textwidth,keepaspectratio]{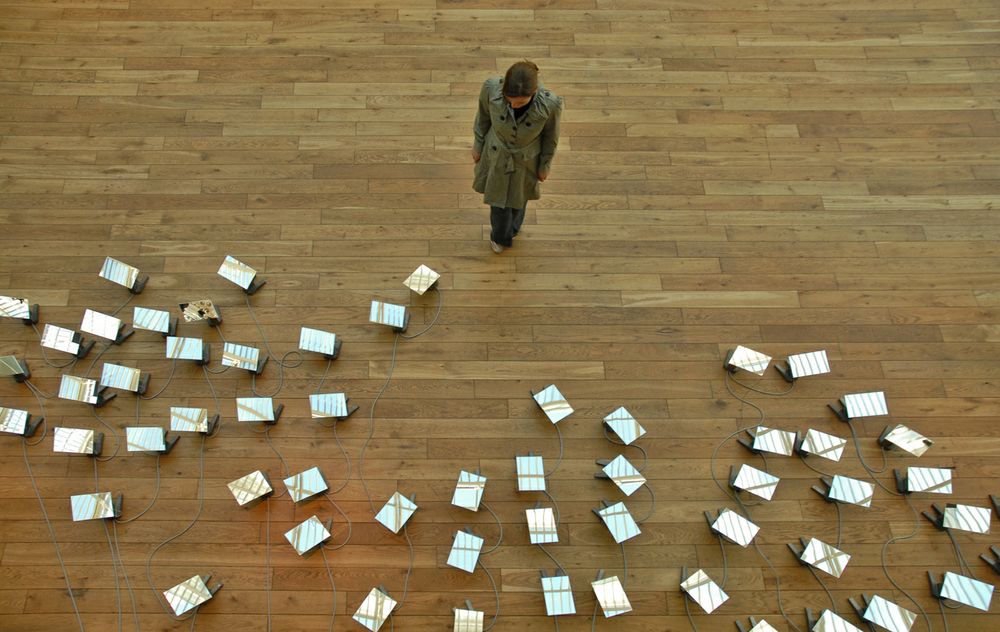}
      \caption{AUDIENCE, 2009.\\Hannes Koch et al.}
      \label{fig: AUDIENCE}
  \end{subfigure}
  \caption{Early interactive artworks focusing on innovating the ``screen experience.''}
  \label{fig: 2}
  \end{figure}
  

Fig.\ref{fig: SIX-FORTY} illustrates ``SIX-FORTY BY FOUR-EIGHTY, 2011'' by Coelho and Zigelbaum, which ``transposes the pixel from the confines of the screen and into the physical world,'' using the human body as a conduit for digital information. In ``AUDIENCE, 2009'' by Hannes Koch et al., the artists employ ``a large crowd of head-size mirror objects'' that rotate towards the audience's presence (i.e., real-time position), conceptually ``reversing the roles of the viewer and the viewed'' in the interaction. These artworks push the boundaries of traditional screen-based experiences, creating immersive and interactive installations that engage audiences in new ways.


\subsubsection{Audience Performing More than Dancing or Posing, Artwork not Merely Displaying Screen-based Contents}

\begin{figure}[h]
  \centering
      \begin{subfigure}{0.245\textwidth}
        \centering
        \includegraphics[height=.65\textwidth,keepaspectratio]{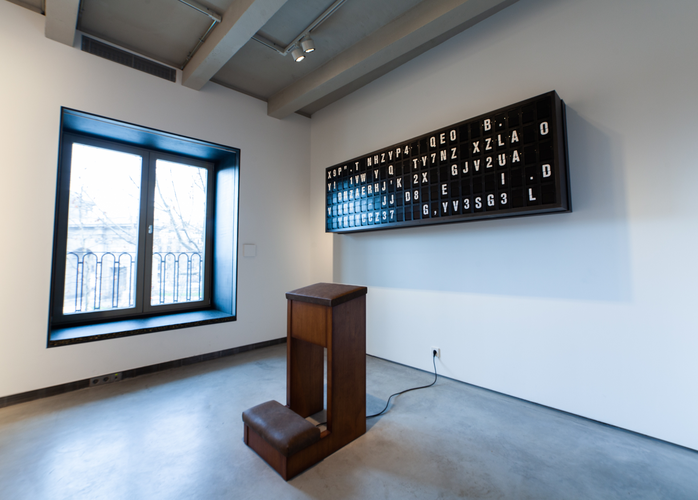}
      \caption{APPROPRIATE RESPONSE,\\2020. Mario Klingemann}
      \label{fig: APPROPRIATE}
  \end{subfigure}
      \hfill
  \begin{subfigure}{0.245\textwidth}
    \centering
      \includegraphics[height=.65\textwidth,keepaspectratio]{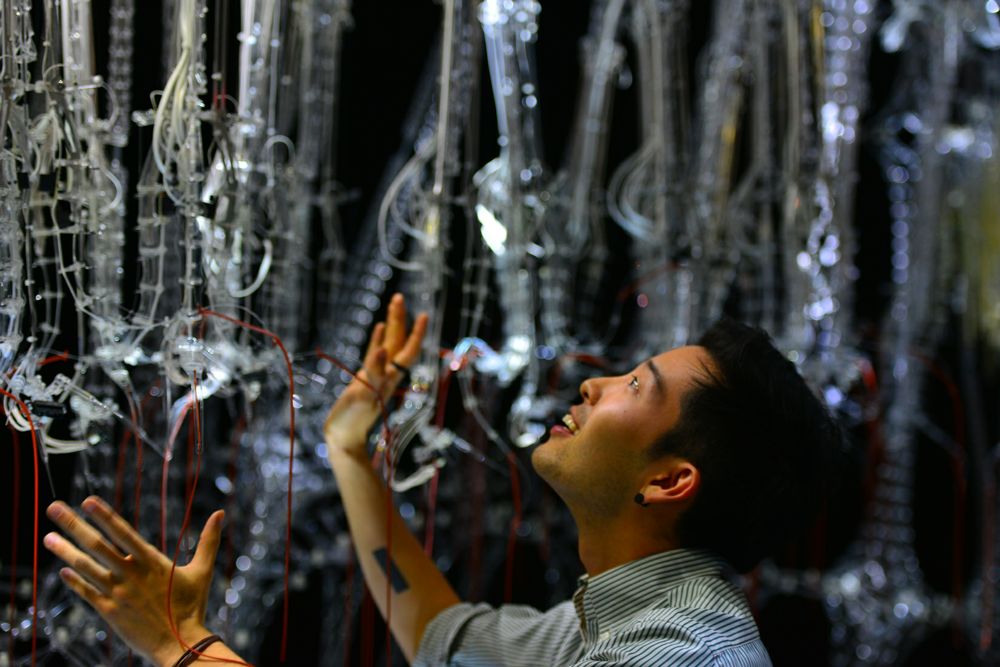}
      \caption{EPIPHYTE CHAMBER, 2014.\\Philip Beesley.}
      \label{fig: EPIPHYTE}
  \end{subfigure}
    \hfill
  \begin{subfigure}{0.245\textwidth}
        \centering
      \includegraphics[height=.65\textwidth,keepaspectratio]{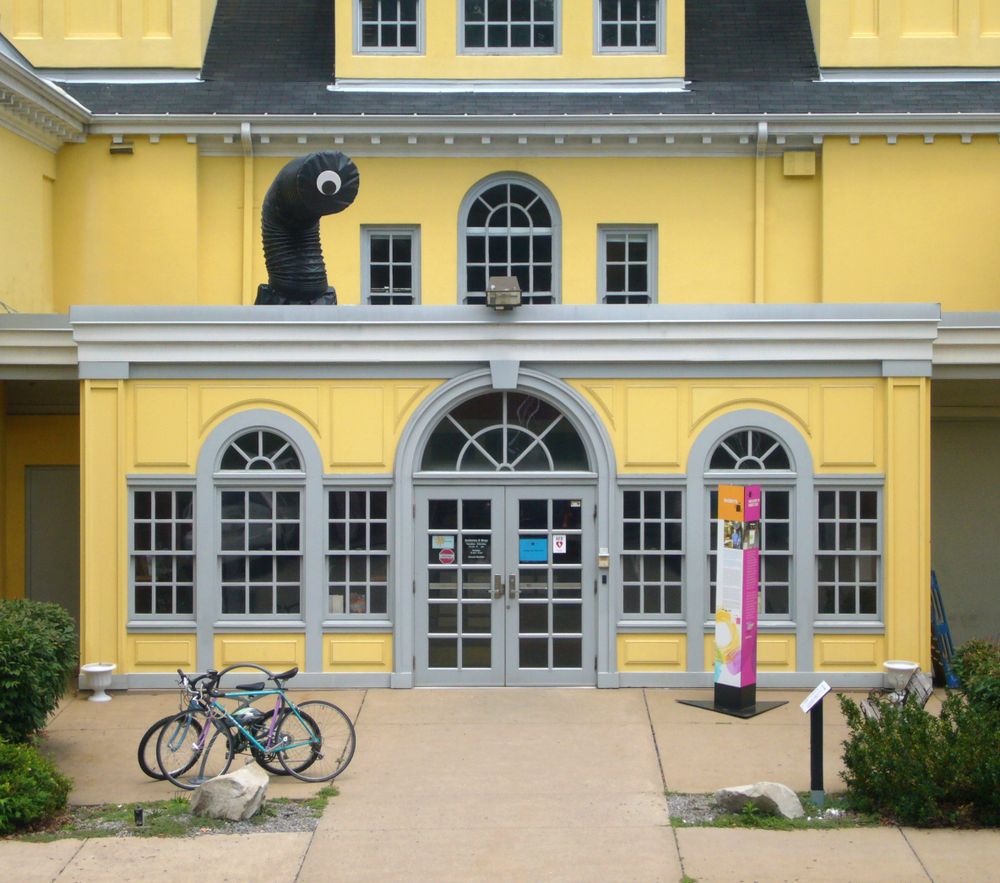}
      \caption{DOUBLE-TAKER, 2009.\\Steven Benders et al.}
      \label{fig: DOUBLE}
  \end{subfigure}
    \hfill
    \begin{subfigure}{0.245\textwidth}
        \centering
        \includegraphics[height=.65\textwidth,keepaspectratio]{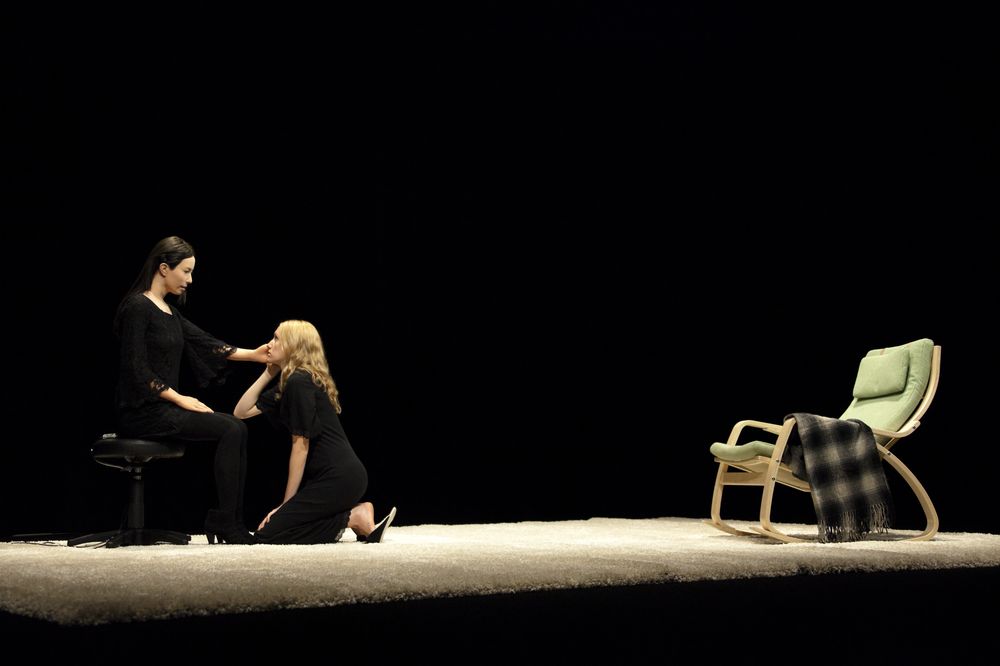}
      \caption{ANDROID THEATER, 2011.\\Oriza Hirata}
      \label{fig: ANDROID}
  \end{subfigure}
  \caption{The audience performs more than posing, dancing, and artwork does not merely display screen-based content.}
  \label{fig: more_than_posing}
  \end{figure}

Some artworks engage the audience in more complex performances beyond merely posing or dancing. These artworks respond to the audience through various methods rather than simply displaying screen-based audio or visual content. For example, Fig.\ref{fig: APPROPRIATE} illustrates ``APPROPRIATE RESPONSE, 2020'' by Mario Klingemann. In this piece, the audience (or as Klingemann terms them, ``kneelers'') kneel in front of the artwork, transforming the interaction into a ritual-like experience. The audience participates by kneeling and contemplating the GPT-2 generated short and coherent aphorisms on a split-flap display, navigating a balance between hope and fear in relation to AI.

Other artworks in Fig.\ref{fig: more_than_posing}, such as ``EPIPHYTE CHAMBER, 2014'' by Philip Beesley, ``DOUBLE-TAKER (SNOUT), 2009'' by Benders et al., and ``ANDROID THEATER, 2011'' by Hirata and Ishiguro, invite the audience to navigate through spaces, move around, and even act in a play on stage. These artworks function like creatures that react, ``stare,'' or even ``talk'' to the audience. While the audience primarily uses their visual and auditory senses, they ``receive'' far more than screen-based audio-visual content. 



\subsubsection{Audience Engaging with (Sports-like) Entertaining Activities}

In ``DOWN WITH WRESTLERS WITH SYSTEMS AND MENTAL NONADAPTERS! 2013,'' the audience experiences a complex blend of empowerment and insignificance as they activate various mechanical components by running on a simulator. Similarly, ``VOICES OF ALIVENESS, 2013'' transforms participants' cycling behaviors and shouts into a digital sculpture composed of their collective memory. In both cases, the audience engages with the artwork through sports-like entertaining activities, with their interactions further captured and reflected in socially relevant concepts.

\begin{figure}[h]
  \centering
    \hfill
  \begin{subfigure}{0.245\textwidth}
        \centering
      \includegraphics[height=.6\textwidth,keepaspectratio]{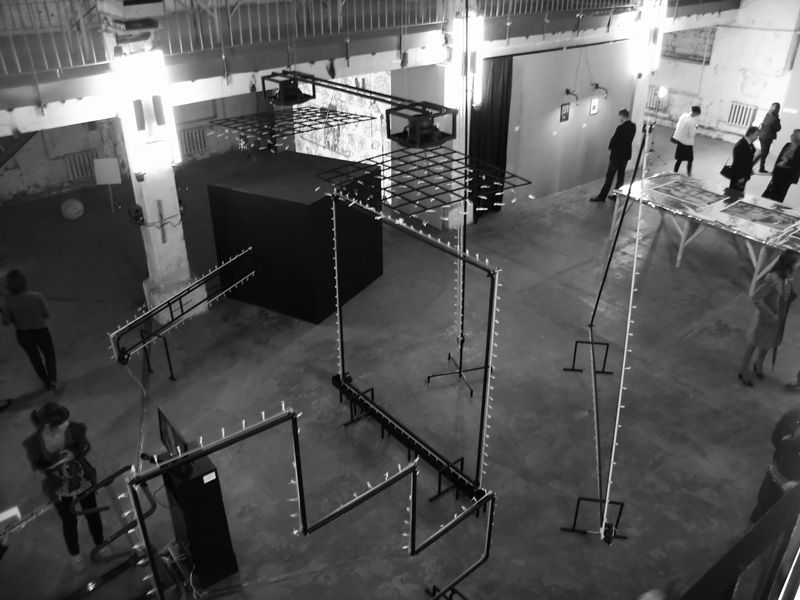}
      \caption{DOWN WITH WRESTLERS,\\2013. Kawarga and Kawarga.}
      \label{fig:DOWN}
  \end{subfigure}
  \begin{subfigure}{0.245\textwidth}
    \centering
      \includegraphics[height=.6\textwidth,keepaspectratio]{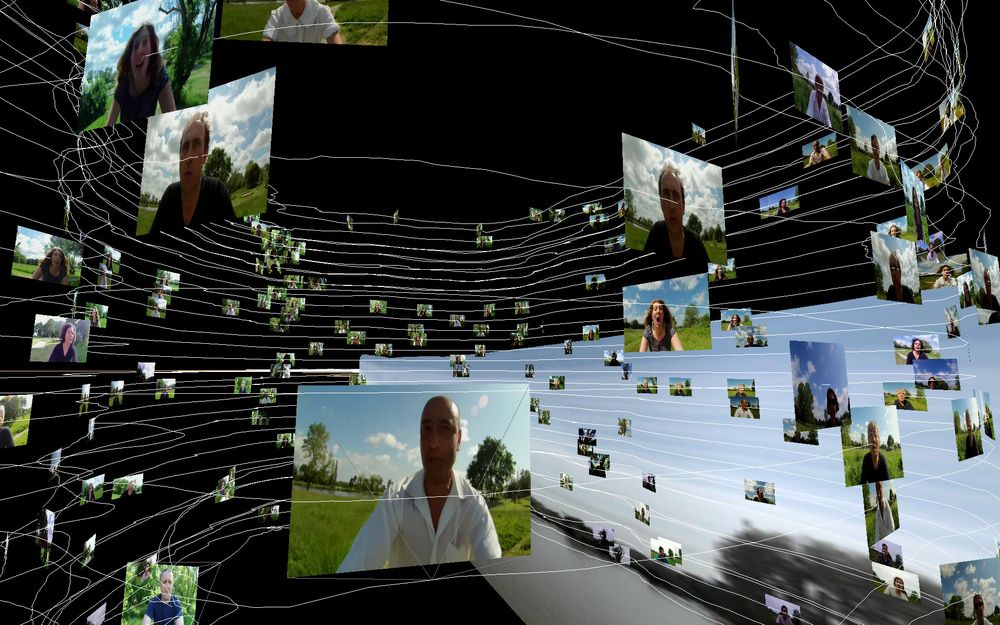}
      \caption{VOICES OF ALIVENESS,\\2013. Masaki Fujihata.}
      \label{fig:VOICE}
  \end{subfigure}
        \hfill
    \begin{subfigure}{0.245\textwidth}
    \centering
    \includegraphics[height=.6\textwidth,keepaspectratio]{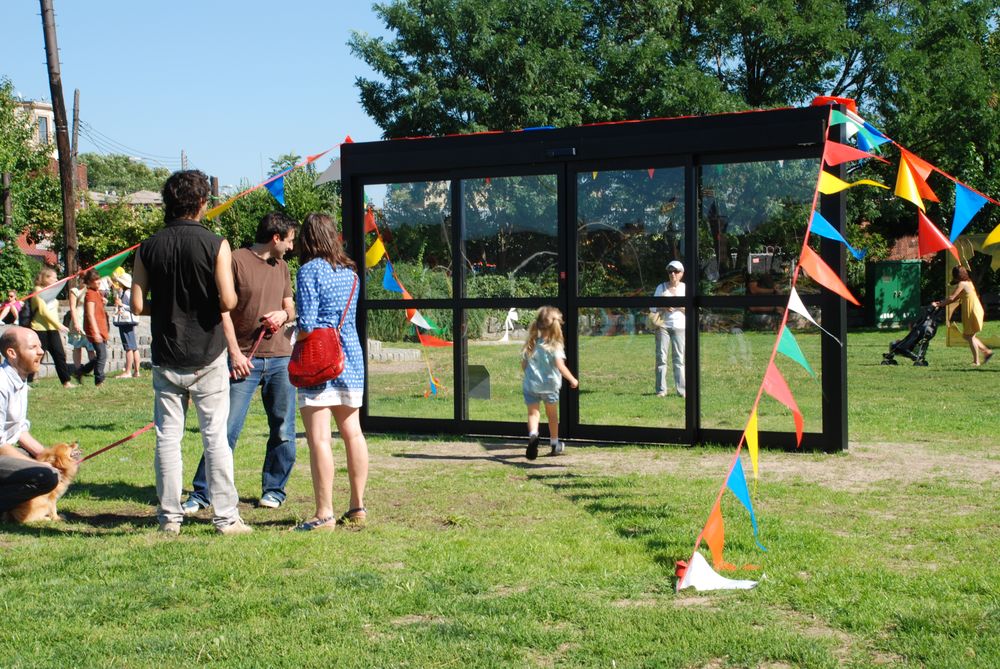}
      \caption{WHEN LAUGHTER TRIPS,\\2009. Beck and Khan.}
      \label{fig:DOOR}
    \end{subfigure}
        \hfill
    \begin{subfigure}{0.245\textwidth}
    \centering
    \includegraphics[height=.6\textwidth,keepaspectratio]{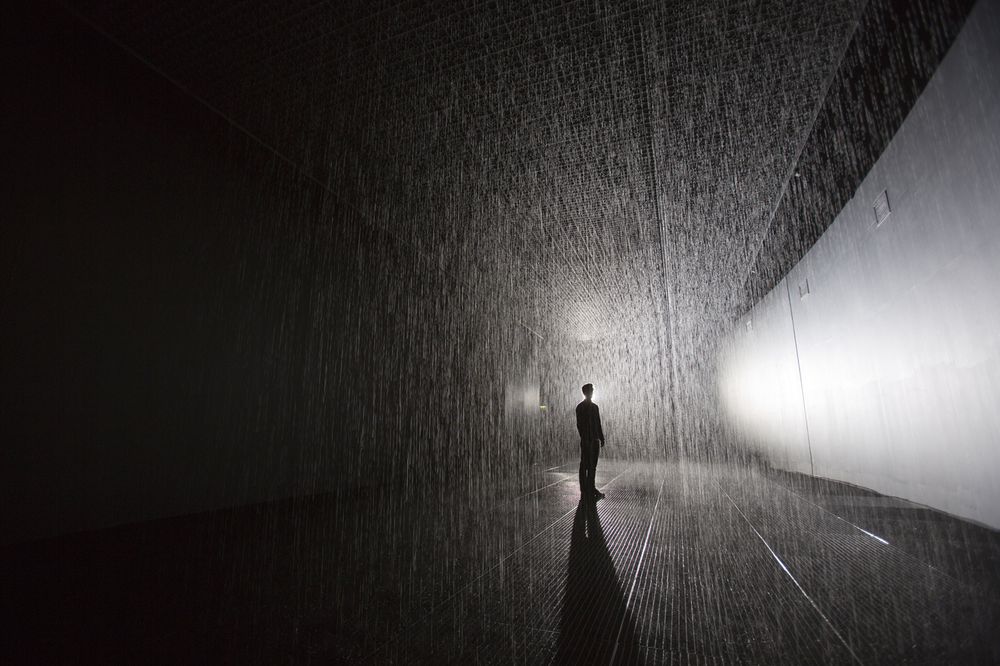}
      \caption{RAIN ROOM, 2013.\\rAndom International}
      \label{fig: RAIN}
  \end{subfigure}
  \caption{Sports-like Embodied Engagement and Entertainment.}
  \label{fig: sports-like}
  \end{figure}


``WHEN LAUGHTER TRIPS AT THE THRESHOLD OF THE DIVINE, 2009'' by Beck and Khan reimagines a public space by placing an automatic sliding door in a park, transforming an ordinary door into a playful experience that invites visitors to physically engage with it. Similarly, ``RAIN ROOM, 2013'' by rAndom International (Fig.\ref{fig: RAIN}) presents a 100-square-meter field of falling water, allowing the audience to walk through the exhibition space without getting wet, as the downpour carefully choreographs itself in response to their movements and presence. Both works engage the audience in spatial navigation and exploration experiences within the artwork.


\subsubsection{Other Interactive Experiences for Social and Cultural Caring}


Addressing issues relevant to social and cultural realms is a common practice in contemporary art, including interactive art. For example, in ``GOOGLE MAPS HACKS, 2020'' by Simon Weckert, the artist creates a virtual traffic jam on Google Maps by dragging a handcart loaded with 99 secondhand smartphones, turning a green-lit street into a red alert. This intervention underscores the tangible impact of digital inputs on physical reality, highlighting the pervasive influence of technology on everyday life. Another example is ``THE EYEWRITER, 2010'' by Zach Lieberman et al., which functions as an instrument to empower individuals suffering from neuromuscular disorders and injuries, enabling artists to draw using only their eyes.

\subsection{Techno Sociocultural (TS) Bodies}\label{sec: TS}

Audience embodiment as TS in interactive art experiences emphasizes the social and cultural aspects of the body, particularly the interpersonal interactions among audience members when engaging with the artwork. Similar to Sec.\ref{sec: TP}, the audience-artwork interaction describes what the audience does to the artwork and perceives from it, but with less focus. This does not diminish the importance of audience-artwork interaction in TS embodiment; rather, we prioritize our analysis of audience-audience interaction as a key aspect of the overall interactive experience. Our definition of Techno Sociocultural (TS) bodies is primarily drawn from Don Ihde's philosophy of bodies in technology. In other contexts, this has been termed ``co-located interaction,'' referring to interactions occurring among two or more co-located audiences and the artwork \cite{xu_describing_2023}.

This ``co-located interaction'' or interpersonal interaction focuses on how other audience members influence an individual's engagement with the artwork or how the artwork mediates interactions among them. We further divide this interaction into ``multi-audience'' and ``interpersonal interaction.'' The ``multi-audience'' category includes scenarios where the artwork supports engagement from multiple audience members without requiring direct interaction between them, or where the artwork necessitates the presence of multiple audience members to function. The ``interpersonal interaction'' category is subdivided into ``post-interpersonal interaction'' and ``direct-interpersonal interaction.'' ``Post-interpersonal interaction'' refers to situations where audiences do not directly interact with each other but engage in collective, competitive, or other non-real-time interactions. ``Direct-interpersonal interaction,'' on the other hand, involves real-time interactions among audience members.

\begin{figure}[h]
  \centering
    \begin{subfigure}{0.245\textwidth}
    \centering
    \includegraphics[height=.55\textwidth,keepaspectratio]{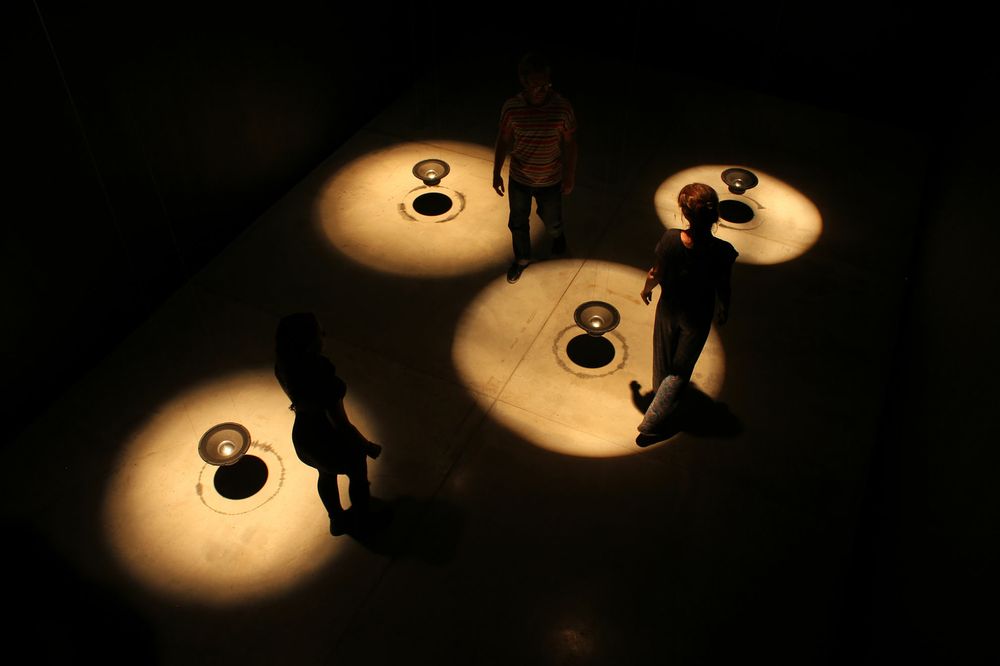}
      \caption{AHORA, 2013.\\Kerllenevich and Alsina.}
      \label{fig: AHORA}
    \end{subfigure}
        \hfill
    \begin{subfigure}{0.245\textwidth}
    \centering
    \includegraphics[height=.55\textwidth,keepaspectratio]{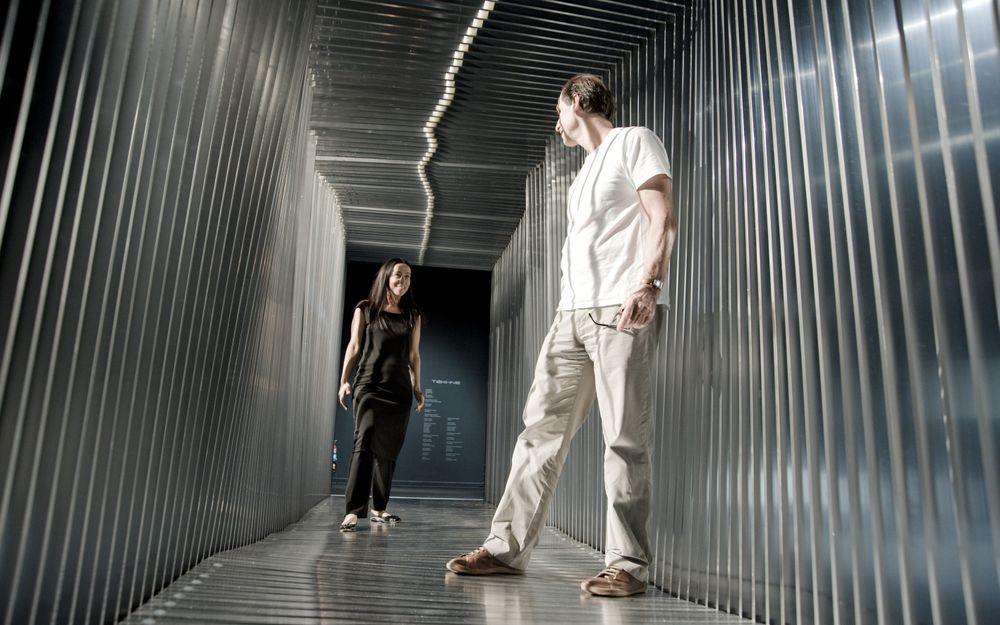}
      \caption{TUNNEL, 2011.\\Cantoni and Crescenti.}
      \label{fig: TUNNEL}
    \end{subfigure}
        \hfill
    \begin{subfigure}{0.245\textwidth}
    \centering
      \includegraphics[height=.55\textwidth,keepaspectratio]{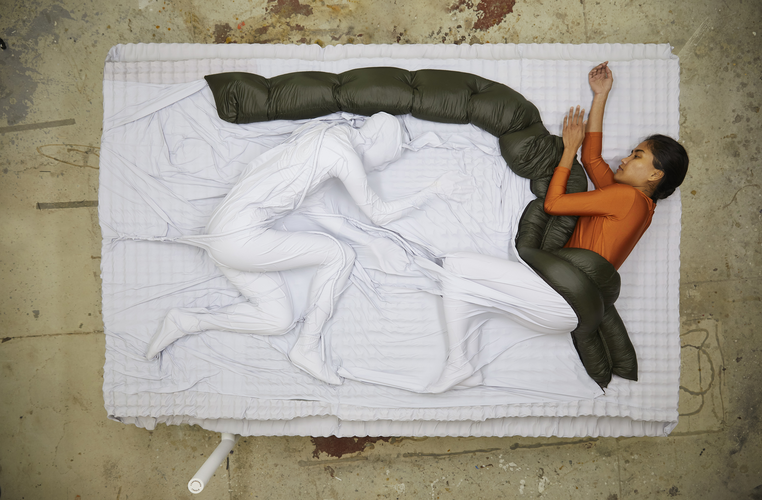}
      \caption{COMPRESSION CRADLE,\\2020. Lucy McRae.}
      \label{fig: TURNSTILE}
    \end{subfigure}
        \hfill
  \begin{subfigure}{0.245\textwidth}
        \centering
      \includegraphics[height=.55\textwidth,keepaspectratio]{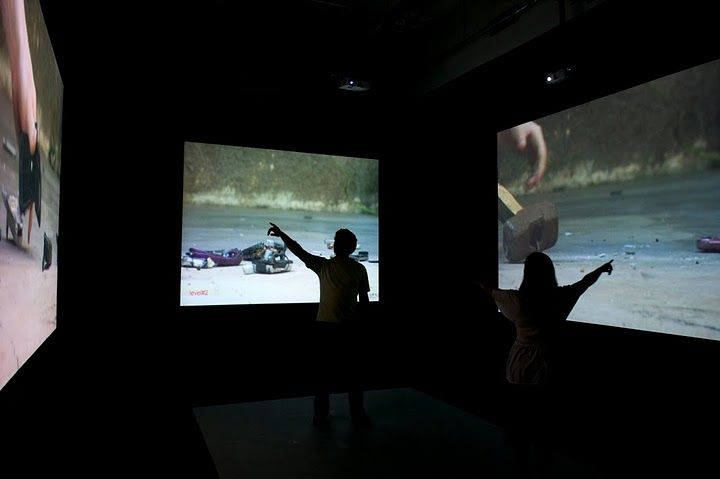}
      \caption{MOBILE CRASH, 2010.\\Lucas Bambozzi}
      \label{fig: CRASH}
  \end{subfigure}
    \caption{Artwork supporting multi-user participation and interaction.}
    \label{fig: supported}
\end{figure}

\subsubsection{Supported Multi-audience}

Many artworks support both single and multiple audiences participating simultaneously; those without obvious differences in interactive experience across these modes have been previously analyzed in Sec.\ref{sec: TP}. In ``AHORA, 2013'' by Kerllenevich and Alsina, the audience triggers and alters sounds mapped to spatial coordinates through their movement, creating a personal auditory experience. In ``TUNNEL, 2011'' by Cantoni and Crescenti, the arrangement of 92 porticos changes based on the audience's location and body weight. Movement to one side causes the floor to tilt slightly, prompting nearby porticos to rotate and generating waves across the installation, creating dynamic optical effects visible to external observers. ``COMPRESSION CRADLE, 2020'' by Lucy McRae simulates human touch using aerated volumes that gently squeeze the body, mimicking the sensation of being held. ``MOBILE CRASH, 2010'' by Lucas Bambozzi allows participants to trigger audiovisual sequences by pointing to any of the four projection screens.

\begin{figure}[h]
  \centering
    \begin{subfigure}{0.245\textwidth}
    \centering
    \includegraphics[height=.55\textwidth,keepaspectratio]{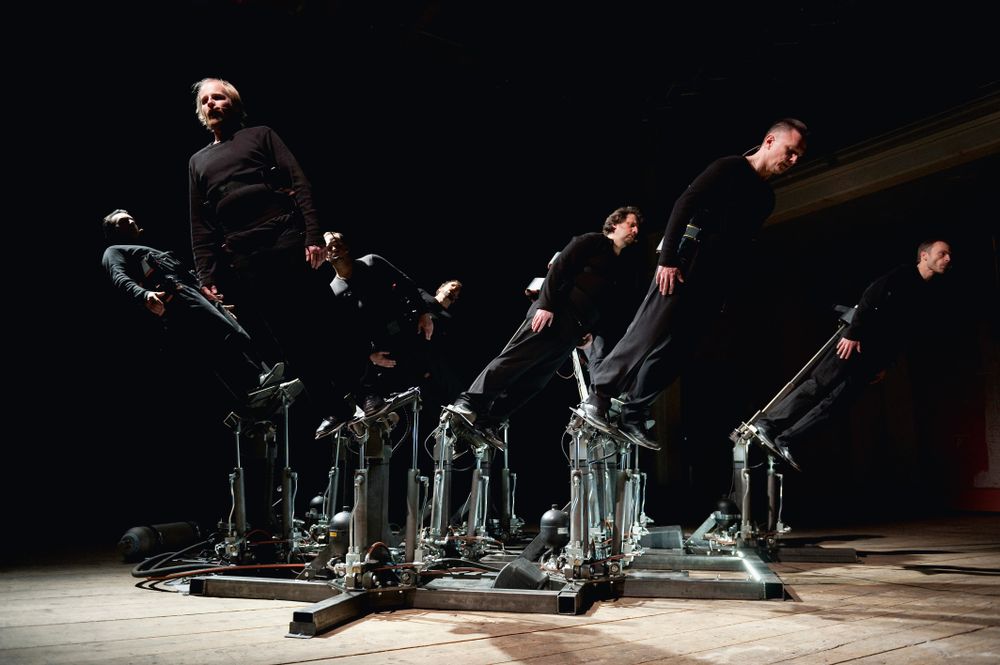}
      \caption{PENDULUM CHOIR, 2013.\\Decosterd and Decosterd}
      \label{fig: CHOIR}
    \end{subfigure}
        \hfill
    \begin{subfigure}{0.245\textwidth}
    \centering
    \includegraphics[height=.55\textwidth,keepaspectratio]{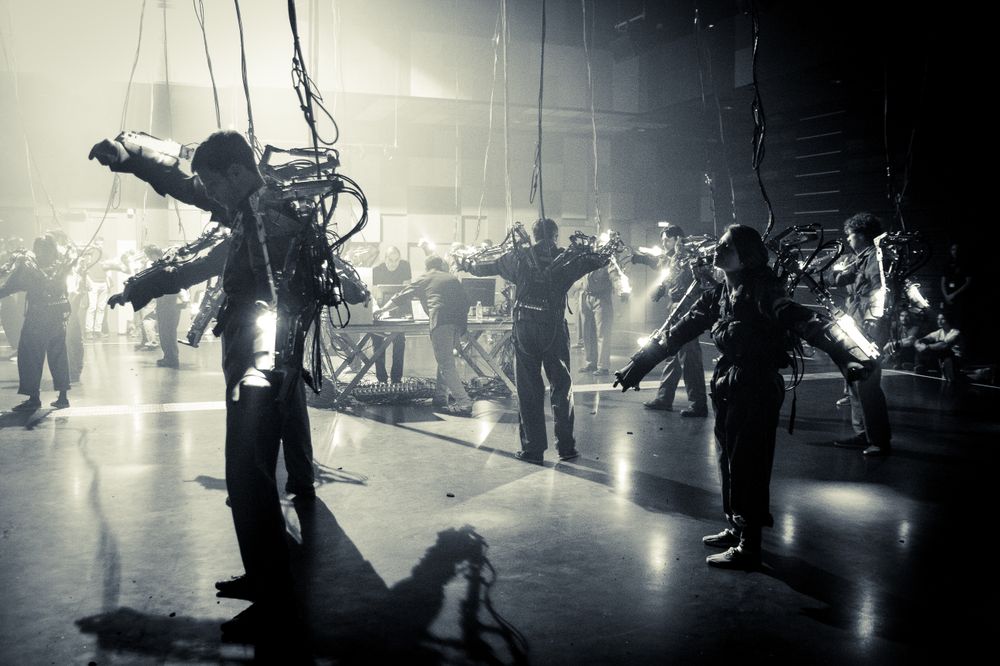}
      \caption{INFERNO, 2016.\\Demers and Vorn.}
      \label{fig: INFERNO}
    \end{subfigure}
        \hfill
    \begin{subfigure}{0.245\textwidth}
    \centering
      \includegraphics[height=.55\textwidth,keepaspectratio]{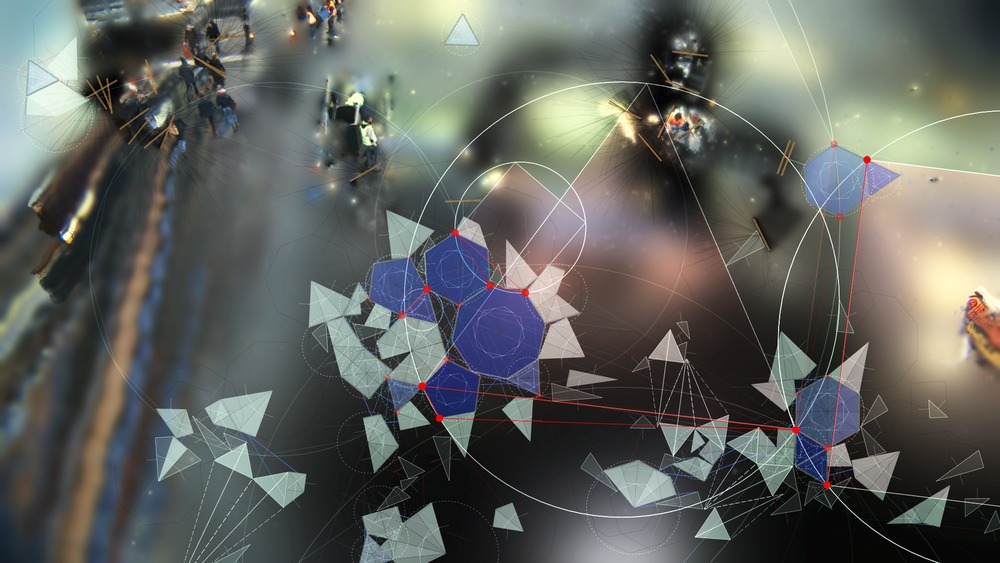}
      \caption{TURNSTILE, 2018.\\Ursula Damm.}
      \label{fig: TURNSTILE}
    \end{subfigure}
        \hfill
  \begin{subfigure}{0.245\textwidth}
        \centering
      \includegraphics[height=.55\textwidth,keepaspectratio]{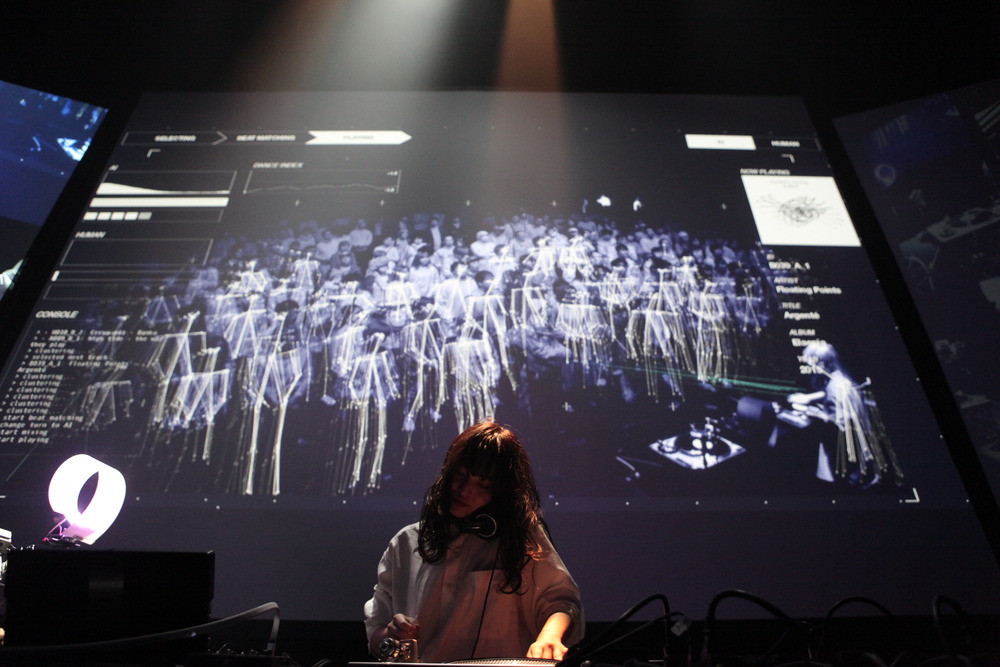}
      \caption{AI DJ PROJECT, 2018.\\Shoya Dozono et al.}
      \label{fig: DJ}
  \end{subfigure}
    \caption{Artwork requiring multi-user participation and interaction.}
    \label{fig: required-multi}
\end{figure}

\subsubsection{Required Multi-audience}

This type of artwork requires more than one audience member to function. ``PENDULUM CHOIR, 2013'' by Decosterd and Decosterd features nine a cappella singers on tilting platforms controlled by 18 hydraulic jacks, where the collective vocal expressions and movements are influenced by the platform's tilting. In ``INFERNO, 2016'' by Demers and Vorn, the audience is fitted with robotic exoskeletons, becoming active participants in a performance that explores the dynamics of control and human-machine interaction. ``TURNSTILE, 2018'' by Ursula Damm transforms pedestrian movements into dynamic visual patterns on an LED wall, responding to the flow of people in the traffic station. ``AI DJ PROJECT, 2018'' by Shoya Dozono et al. features a live performance influenced by the audience's movements, analyzed through motion tracking.

\begin{figure}[h]
  \centering
    \begin{subfigure}{0.245\textwidth}
    \centering
    \includegraphics[height=.6\textwidth,keepaspectratio]{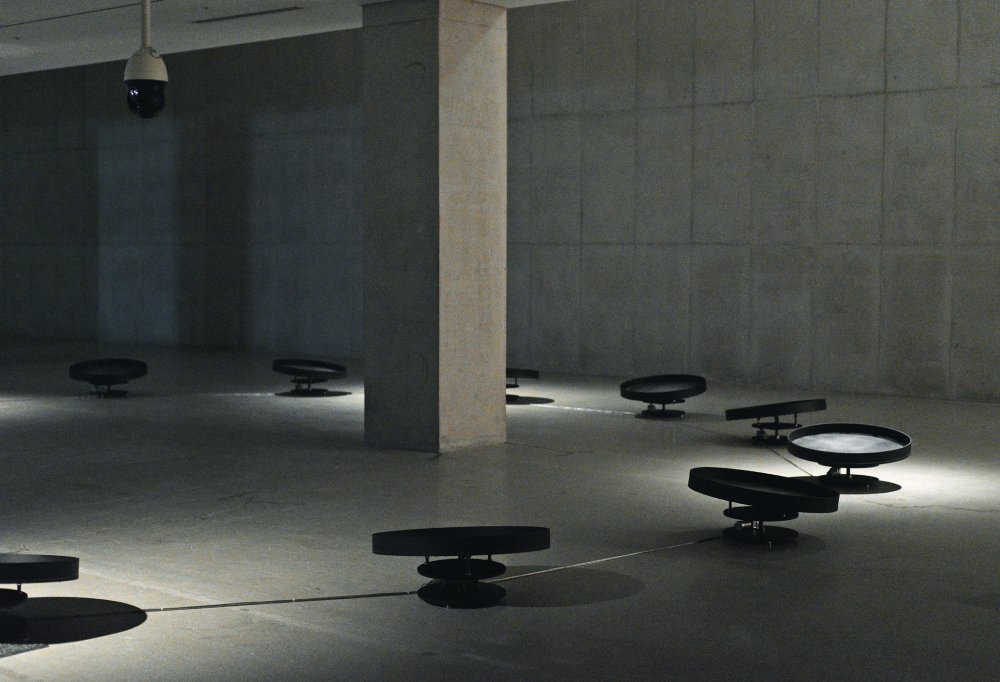}
      \caption{MIND, 2020.\\Shinseung Kimyonghun}
      \label{fig: MIND}
    \end{subfigure}
        \hfill
    \begin{subfigure}{0.245\textwidth}
    \centering
    \includegraphics[height=.6\textwidth,keepaspectratio]{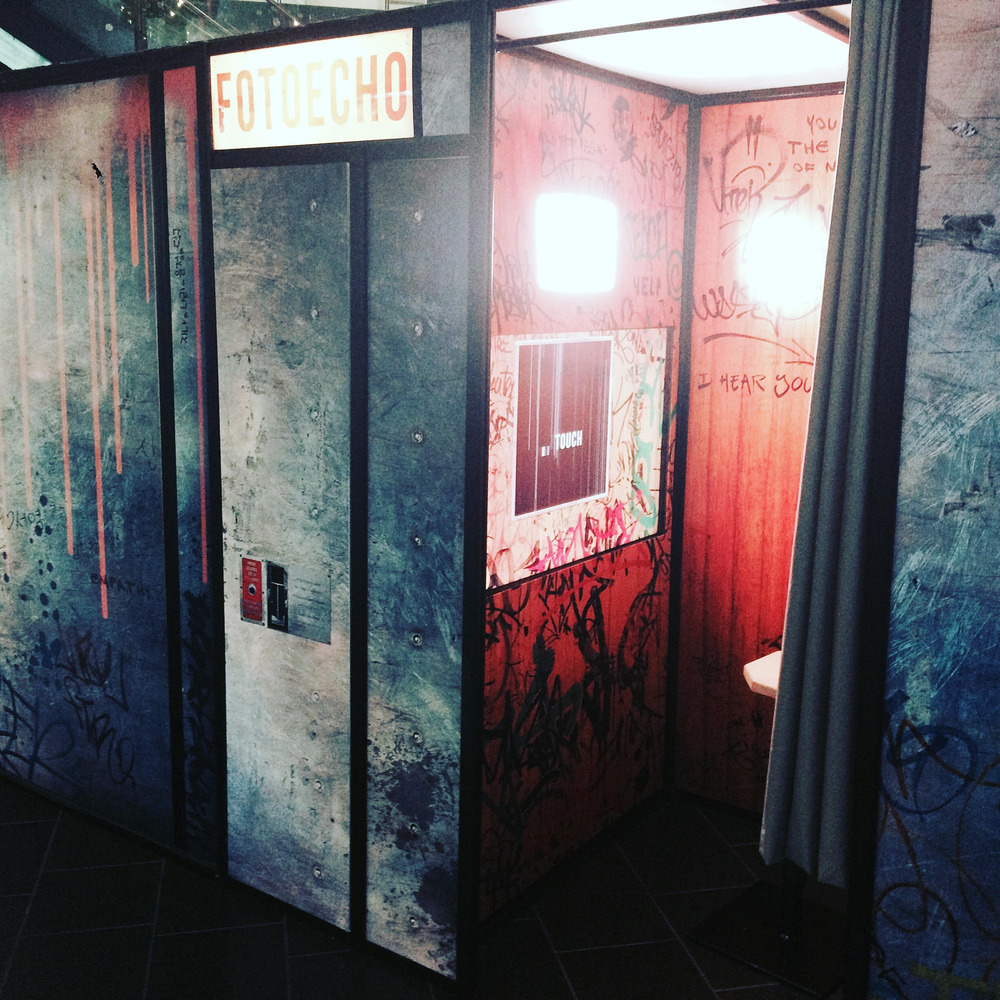}
      \caption{ECHO, 2018.\\Pinn and Rossi.}
      \label{fig: EHCO}
    \end{subfigure}
        \hfill
    \begin{subfigure}{0.245\textwidth}
    \centering
      \includegraphics[height=.6\textwidth,keepaspectratio]{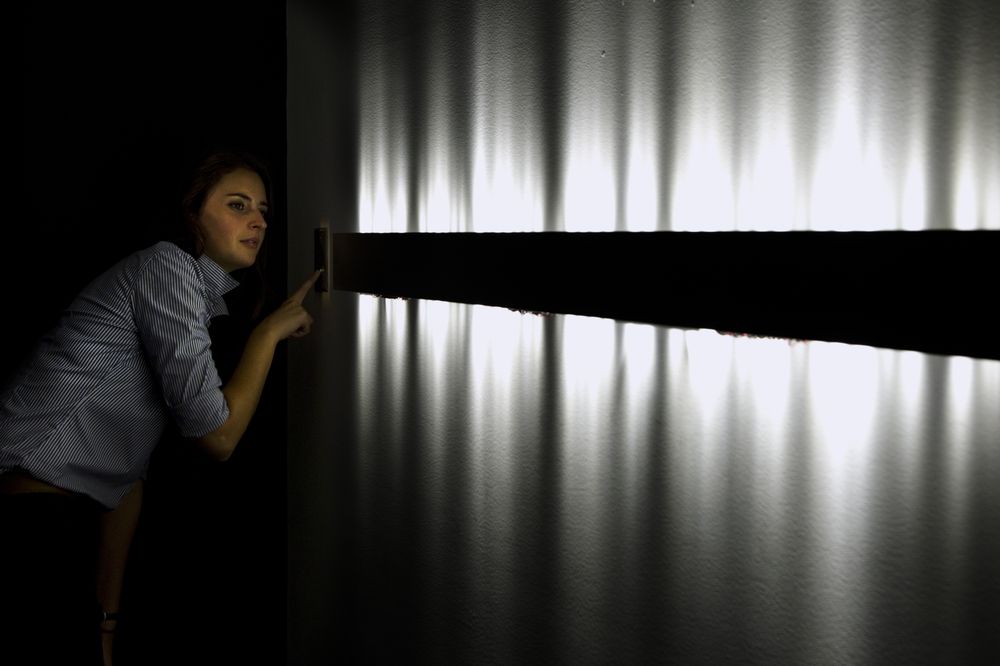}
      \caption{VOICE ARRAY, 2013.\\Rafael Lozano-Hemmer.}
      \label{fig: ARRAY}
    \end{subfigure}
        \hfill
  \begin{subfigure}{0.245\textwidth}
        \centering
      \includegraphics[height=.6\textwidth,keepaspectratio]{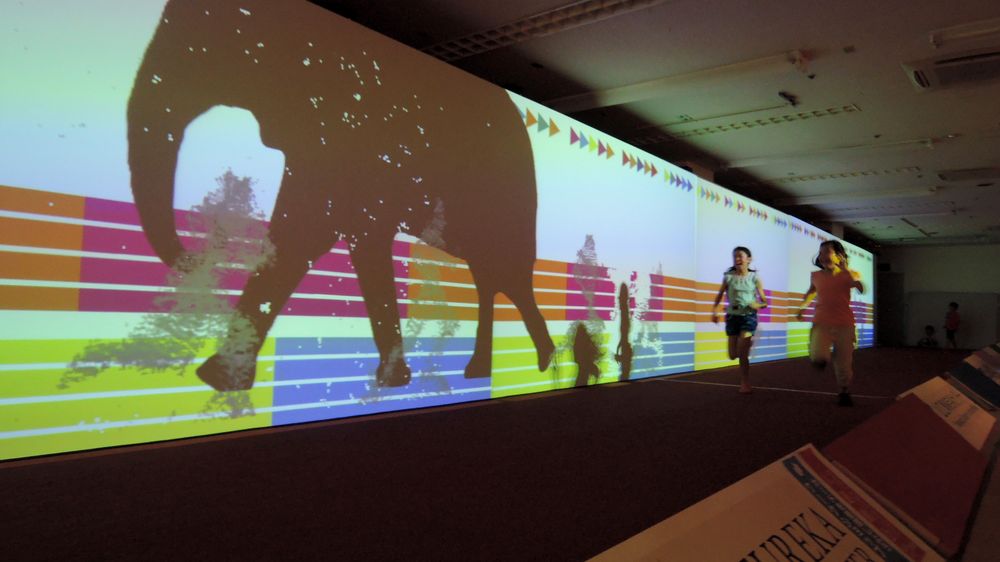}
      \caption{SPORTS TIME MACHINE,\\2014. Ando and Inukai.}
      \label{fig: SPORTS}
  \end{subfigure}
    \caption{Post Interpersonal Interaction.}
    \label{fig: post}
\end{figure}

\subsubsection{Post Interpersonal Interaction}

In this subcategory, the artworks facilitate ``post'' rather than real-time interaction among audiences. ``MIND, 2020'' by artist duo Shinseungback Kimyonghun involves a central rotating camera that captures participants' facial expressions to modulate sea-like sounds from mechanical ocean drums, reflecting the collective emotions of the last 100 visitors. ``ECHO, 2018'' by Pinn and Rossi uses facial capture technology in a booth where participants interact with a virtual mirror, choosing another person’s face to hear their life story as interaction with the previous audience. ``VOICE ARRAY, 2013'' by Rafael Lozano-Hemmer translates participants' voices into unique flashing light patterns via an intercom. Each new voice recording is stored as a looping light pattern in the first light of an array, pushing older recordings from the earlier audience down the line. ``SPORTS TIME MACHINE, 2014'' by Ando and Inukai allows runners to race against past performances projected onto a wall.

\begin{figure}[h]
  \centering
    \begin{subfigure}{0.245\textwidth}
    \centering
    \includegraphics[height=.6\textwidth,keepaspectratio]{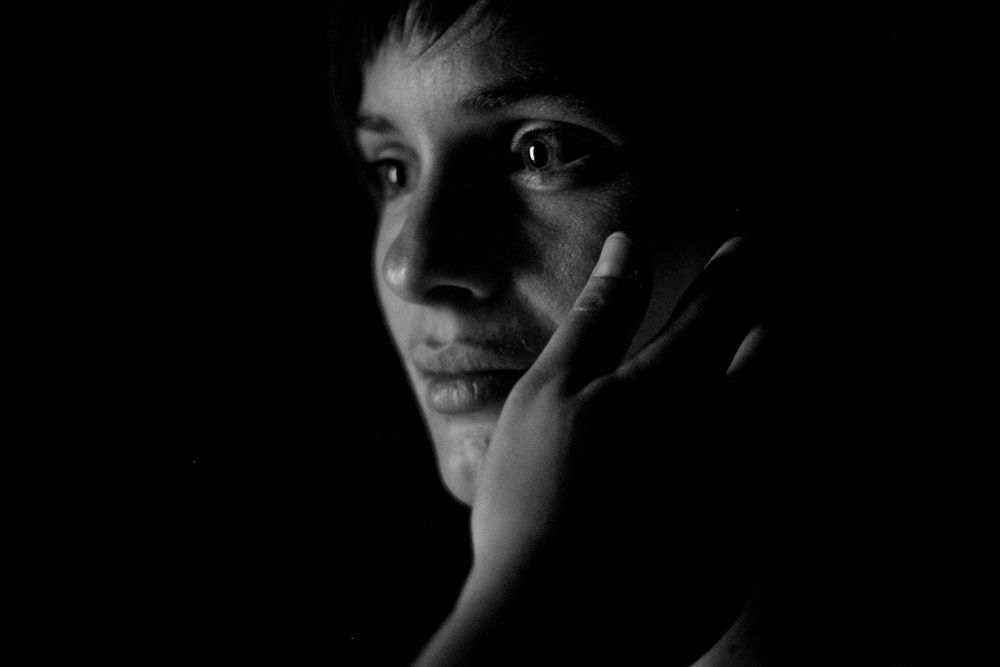}
      \caption{SENSITIVE TO PLEASURE,\\2011. Sonia Cillari}
      \label{fig: SENSITIVE}
    \end{subfigure}
        \hfill
    \begin{subfigure}{0.245\textwidth}
    \centering
    \includegraphics[height=.6\textwidth,keepaspectratio]{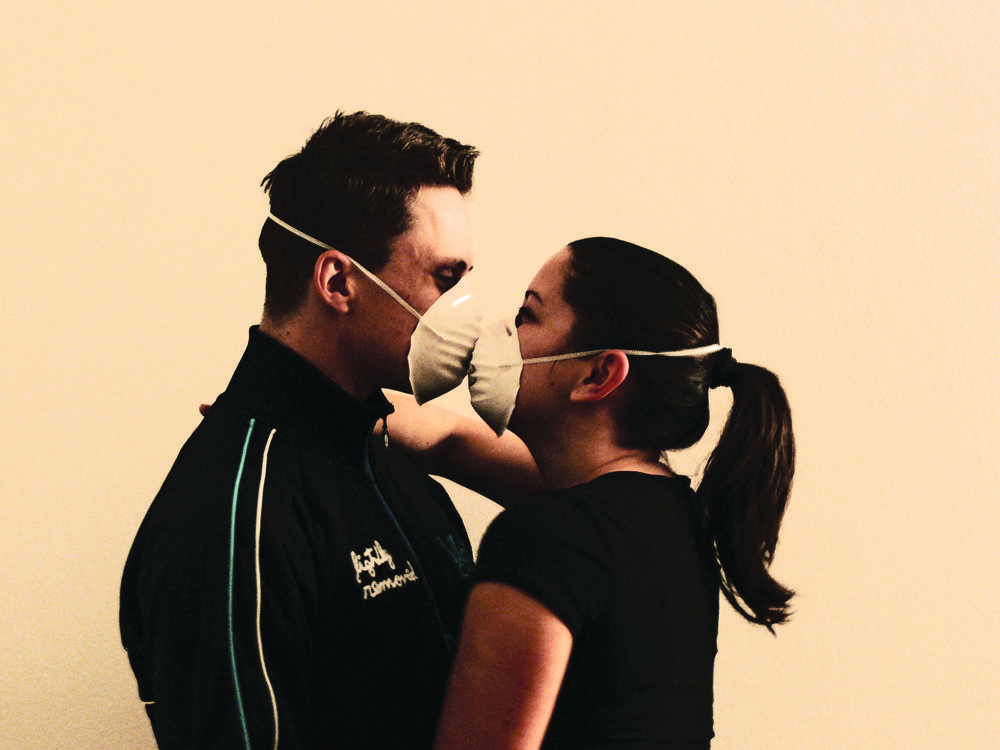}
      \caption{FUTURE KISS, 2009.\\Lenka Klimesova.}
      \label{fig: KISS}
    \end{subfigure}
        \hfill
    \begin{subfigure}{0.245\textwidth}
    \centering
      \includegraphics[height=.6\textwidth,keepaspectratio]{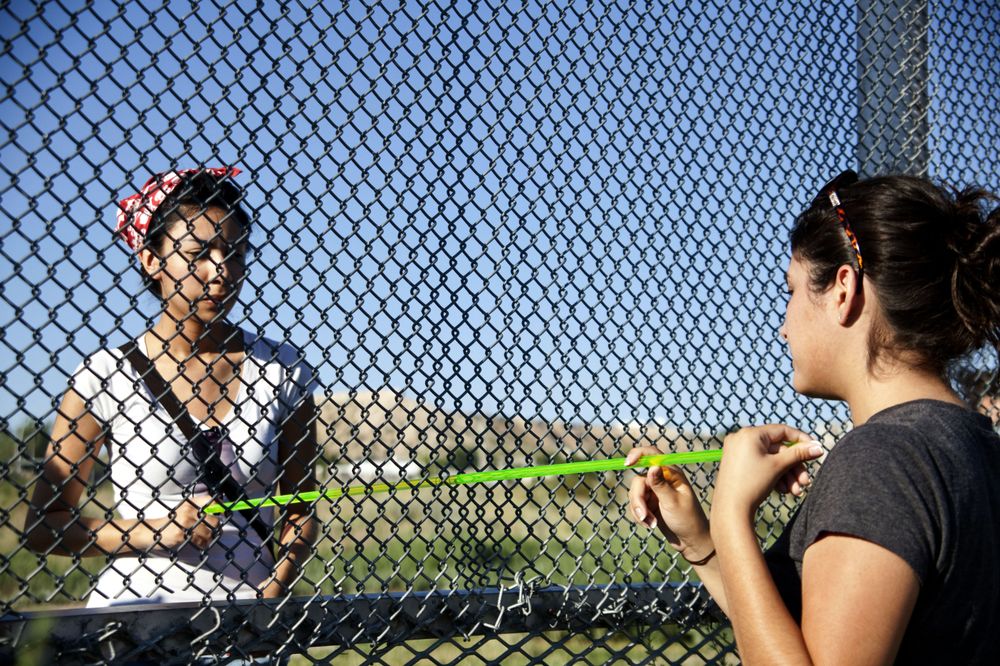}
      \caption{CROSS COORDINATES,\\2012. Iván Abreu.}
      \label{fig: CROSS}
    \end{subfigure}
        \hfill
  \begin{subfigure}{0.245\textwidth}
        \centering
      \includegraphics[height=.6\textwidth,keepaspectratio]{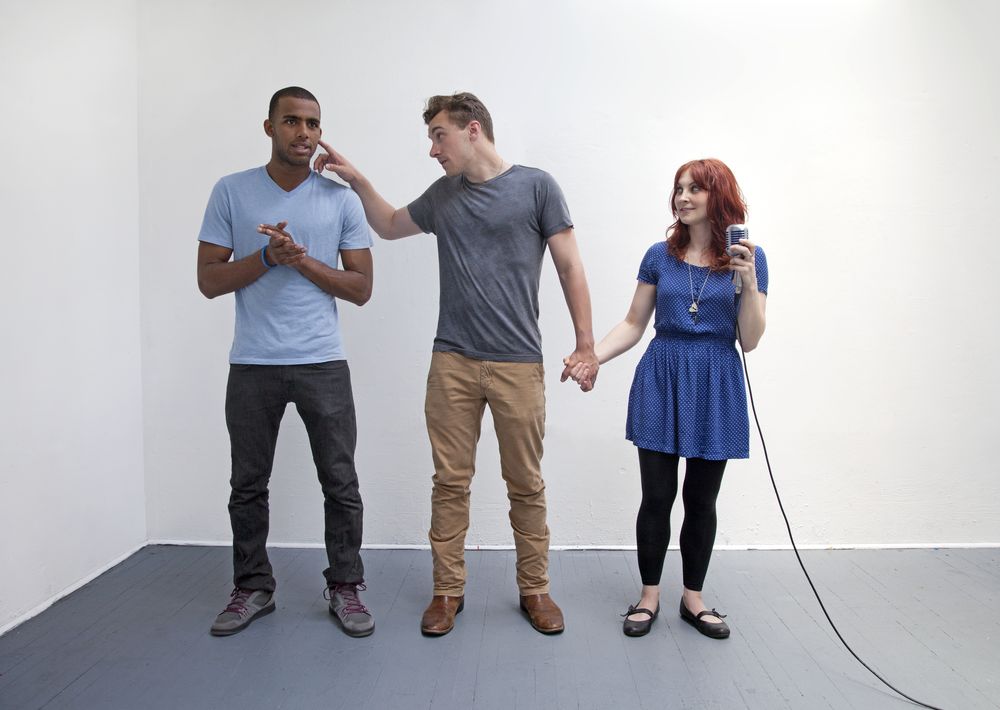}
      \caption{ISHIN-DEN-SHIN, 2013.\\Bau et al.}
      \label{fig: ISHIN}
  \end{subfigure}
    \caption{Direct (Duo) Interpersonal Interaction.}
    \label{fig: indirect}
\end{figure}

\subsubsection{Direct Interpersonal Interaction}

This type of work involves real-time interaction among audiences interacting with the artwork in pairs or groups. In ``SENSITIVE TO PLEASURE, 2011,'' the audiences engage with a naked human within an ambisonic cube, experiencing physical sensations that echo the artist's controversial relationship with the work. In ``FUTURE KISS, 2009'' by Lenka Klimesova, participants wear masks with vibration chips and kiss detectors, seeking connections by finding and kissing their ``other half.'' ``CROSS COORDINATES (MX-US), 2012'' by Iván Abreu involves participants from the US-Mexico border area in a cooperative game to balance a carpenter's level, with efforts logged as meters aiming to exceed the length of the border. ``ISHIN-DEN-SHIN, 2013'' by Olivier Bau et al. uses the body to transmit sound, where recorded sounds become audible when touching someone's ear, allowing intimate messages to be shared directly.

\begin{figure}[h]
  \centering
    \begin{subfigure}{0.245\textwidth}
    \centering
    \includegraphics[height=.75\textwidth,keepaspectratio]{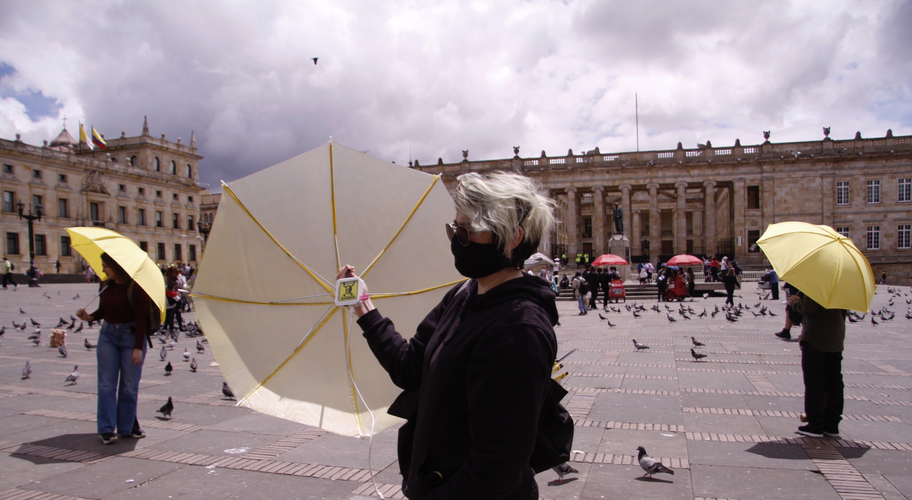}
      \caption{BI0FILM.NET, 2022.\\Hsu and Rivera}
      \label{fig: biofilm}
    \end{subfigure}
        \hfill
    \begin{subfigure}{0.245\textwidth}
    \centering
    \includegraphics[height=.75\textwidth,keepaspectratio]{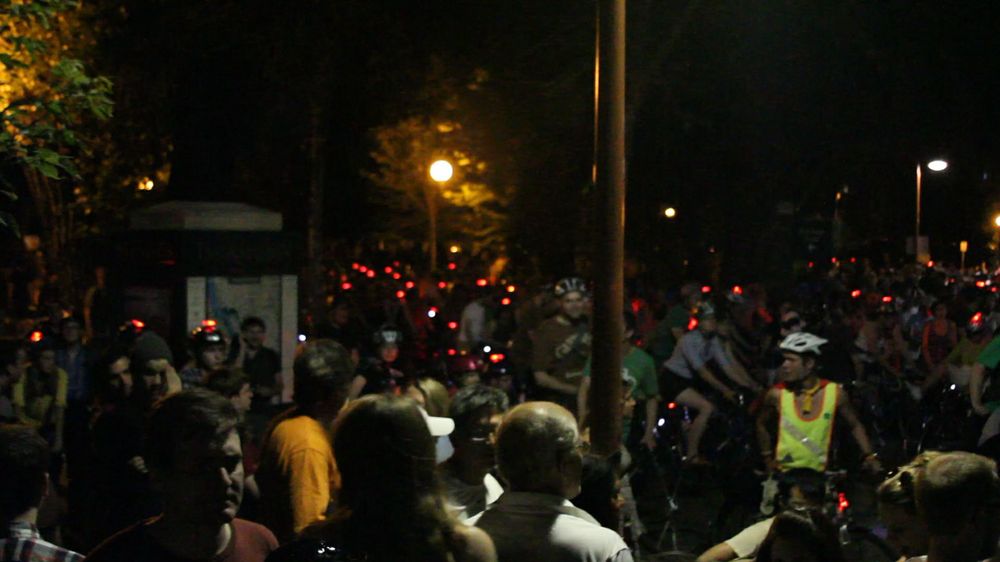}
      \caption{THE KURAMOTO MODEL,\\2013. David Allan Rueter}
      \label{fig: MODEL}
    \end{subfigure}
        \hfill
    \begin{subfigure}{0.245\textwidth}
    \centering
      \includegraphics[height=.75\textwidth,keepaspectratio]{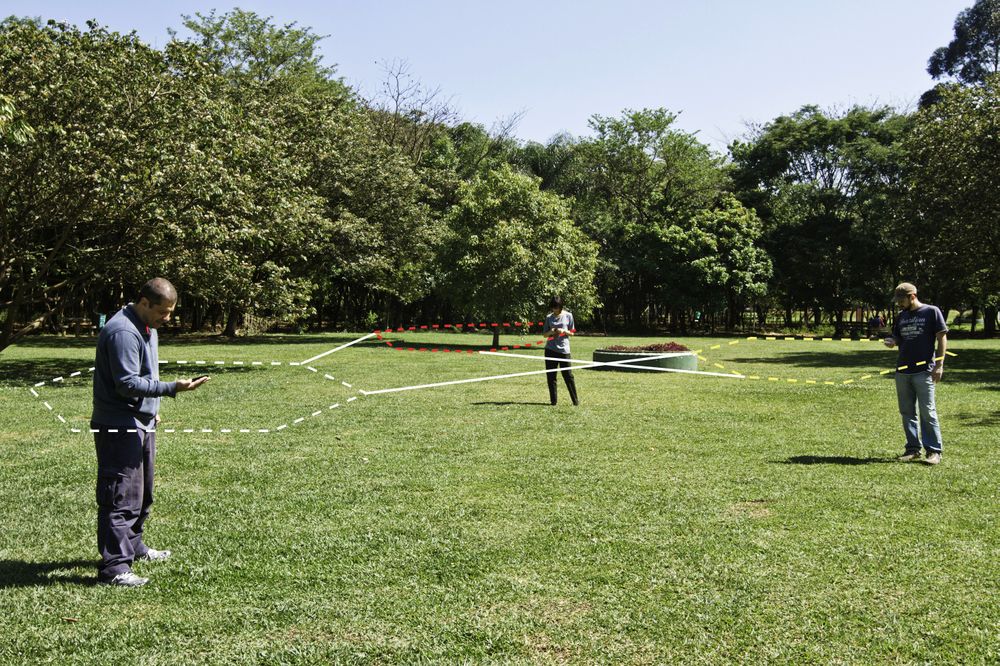}
      \caption{WEARABLE NETS, 2011.\\Claudio Bueno.}
      \label{fig: WEARABLE}
    \end{subfigure}
    \caption{Direct (Crowd) Interpersonal Interaction.}
    \label{fig: direct}
\end{figure}


 ``BI0FILM.NET: RESIST LIKE BACTERIA, 2022'' by Hsu and Rivera transforms yellow umbrellas into parabolic WiFi antennas, enhancing communication for demonstrators by creating a decentralized network for chatting, file sharing, and storage. ``THE KURAMOTO MODEL (1,000 FIREFLIES), 2013'' by David Allan Rueter features 1,000 custom bike lights equipped with radio transceivers that synchronize their blinking patterns with nearby devices, fostering a citywide self-organizing system among cyclists. ``REDES VESTÍVEIS / WEARABLE NETS, 2011'' by Claudio Bueno is a collective performance using a virtual elastic net visualized on mobile phones through a geo-localized app. Both works utilize technology to create dynamic, participatory relationships that emphasize movement and collaboration among physical bodies.

\subsection{Virtual (VB) Body, Virtual Sociocultural (VS) Bodies and Hybrid Sociocultural (HS) Bodies} \label{Sec: VBS}
There are, in general, very few artworks in the selected corpus (6 out of 51) that involve the Virtual Body and only one for each of the Virtual (VB) Body and Virtual Sociocultural (VS) Bodies categories. VB features the audience interacting with the artwork or other audience through their virtual embodiment. 


\begin{figure}[h]
  \centering
    \begin{subfigure}{0.35\textwidth}
    \centering
    \includegraphics[height=.5\textwidth,keepaspectratio]{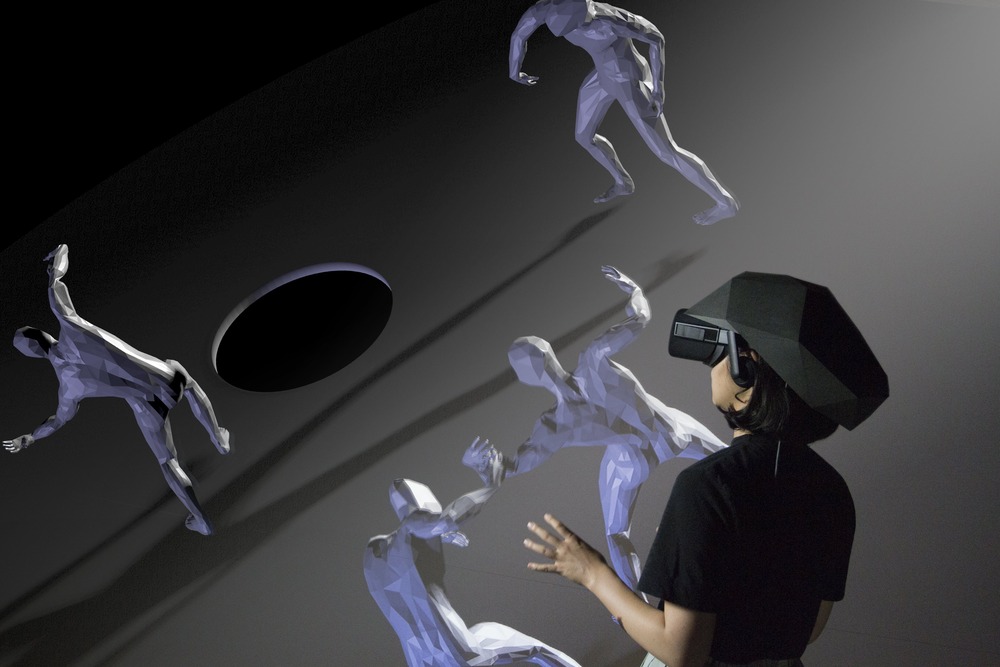}
      \caption{THE OTHER IN YOU, 2018.\\Owaki and YCAM.}
      \label{fig: OTHER}
    \end{subfigure}
    \begin{subfigure}{0.35\textwidth}
    \centering
    \includegraphics[height=.5\textwidth,keepaspectratio]{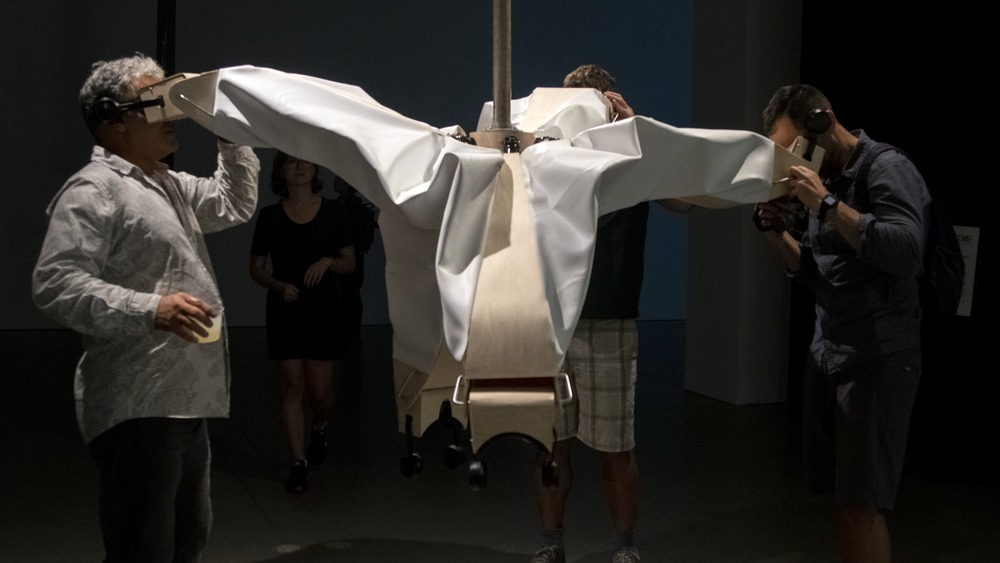}
      \caption{CONSPIRACY, 2018.\\Kristin McWharter.}
      \label{fig: CONSPIRACY}
    \end{subfigure}
    \caption{Virtual Body and Virtual Sociocultural Bodies in Interactive Art.}
    \label{fig: vb+vs}
\end{figure}

``THE OTHER IN YOU, 2018'' by Richi Owaki and YCAM uses Virtual Reality (VR) and haptic feedback technology to transform traditional dance viewing experiences. Audience members can move freely around virtual dancers and see themselves in the performance through real-time 3D capturing cameras. ``CONSPIRACY: CONJOINING THE VIRTUAL, 2018'' by Kristin McWhorter combines VR with a sculptural form, allowing five audiences to participate in a collective decision-making game of capturing the flag.

\begin{figure}[h]
  \centering
    \begin{subfigure}{0.32\textwidth}
    \centering
    \includegraphics[height=.6\textwidth,keepaspectratio]{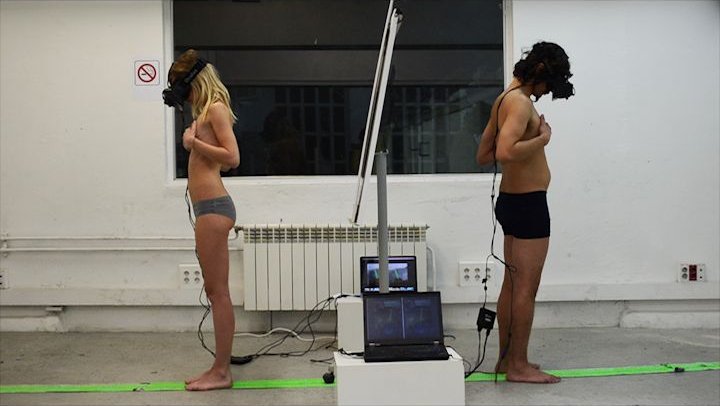}
      \caption{THE MACHINE TO BE ANOTHER, 2014.\\BeAnotherLab.}
      \label{fig: ANOTHER}
    \end{subfigure}
        \hfill
        \begin{subfigure}{0.32\textwidth}
    \centering
    \includegraphics[height=.6\textwidth,keepaspectratio]{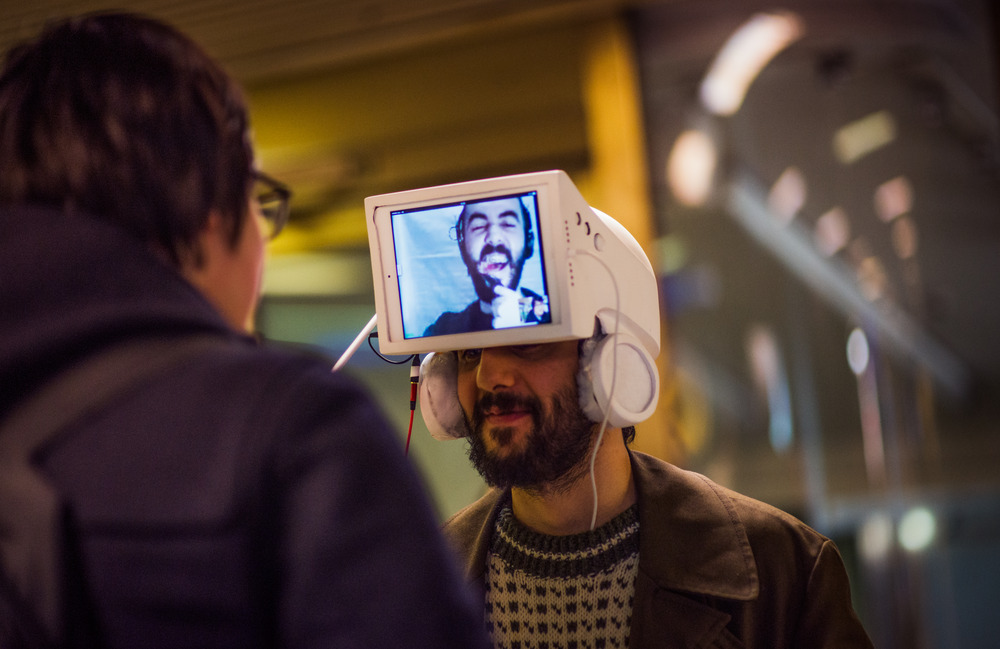}
      \caption{MONITOR MAN, 2018.\\Yassine Khaled.}
      \label{fig: MONITOR}
    \end{subfigure}
    \hfill
    \begin{subfigure}{0.32\textwidth}
    \centering
    \includegraphics[height=.6\textwidth,keepaspectratio]{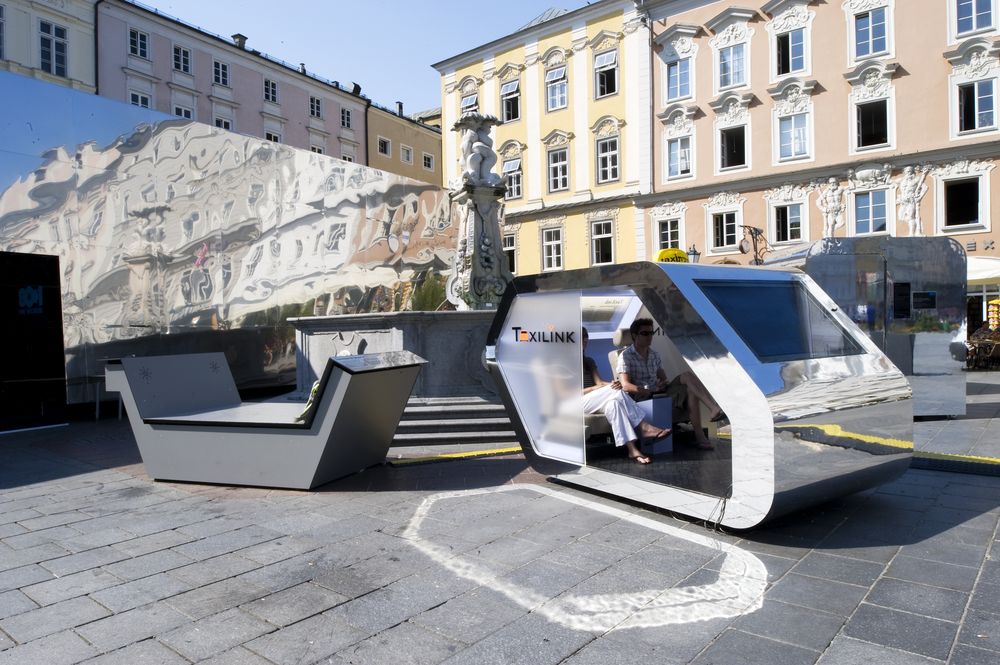}
      \caption{TAXILINK, 2010.\\Tal Chalozin et al.}
      \label{fig: TAXILINK}
    \end{subfigure}
    \caption{Hybird Sociocultural Bodies in Interactive Art.}
    \label{fig: vb+vs}
\end{figure}

There are four artworks that work with the audience's Hybrid Sociocultural (HS) Bodies. ``THE MACHINE TO BE ANOTHER, 2014'' by BeAnotherLab features an embodied interaction system combining performance, neuroscience, and telepresence, allowing the audience to experience inhabiting another person's body. ``MONITOR MAN, 2018'' by Yassine Khaled features a helmet equipped with an iPad that connects individuals outside of Europe and the Western world in real-time. ``TAXILINK, 2010'' by Tal Chalozin et al. offers a simulated virtual taxi ride in Jerusalem to audiences worldwide, engaging in conversations with the driver and listening to local radio broadcasts.

\section{Analysis} \label{sec: analysis}

\begin{figure}[h]
    \centering
    \includegraphics[width=\linewidth]{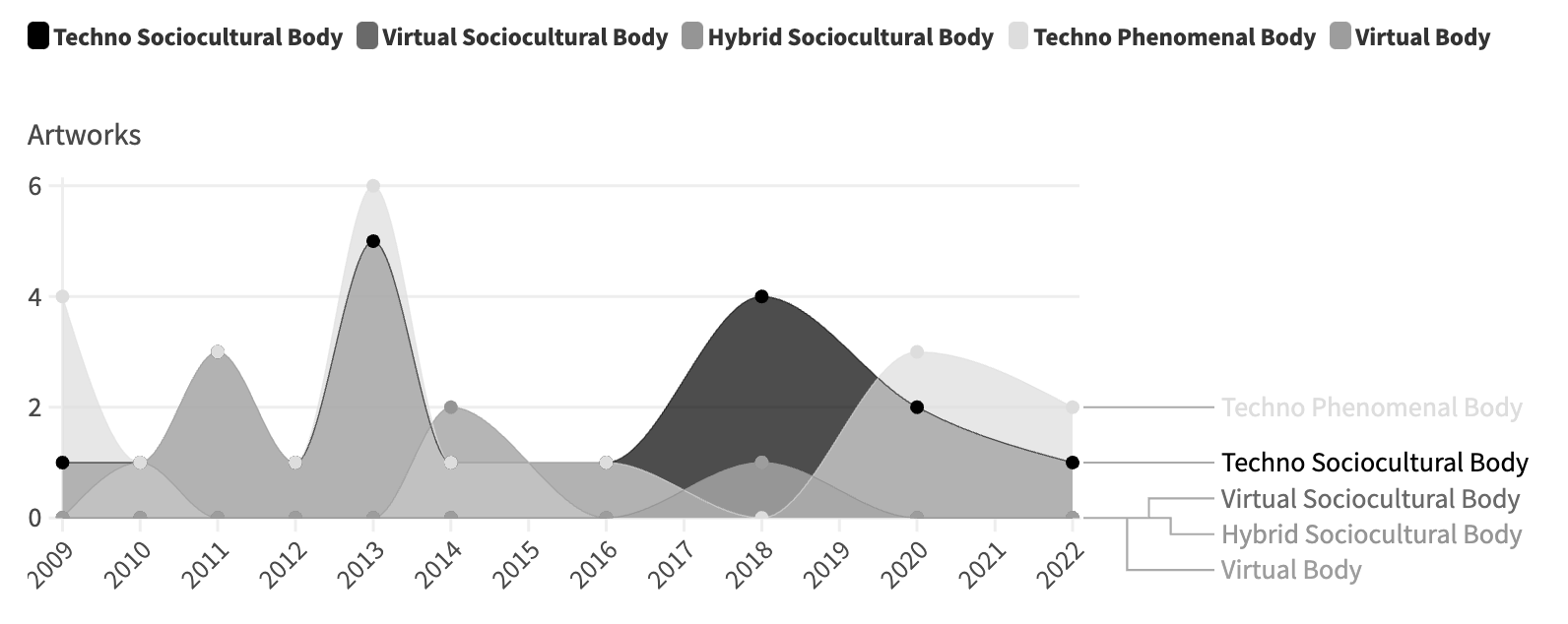}
    \caption{A chronological perspective: TP artworks have been less frequent in the recent ten years (except for the recent two award years), constituting only half the number of TS pieces, identifying a trend favoring TS artworks.}
    \label{fig:yearly-TPVB}
\end{figure}



\subsection{Sociocultural versus Phenomenal Body} \label{sec: analysis-SB-PB}

Fig.\ref{fig:yearly-TPVB} shows the body embodiment in selected artworks chronologically. In the two most recent award years (2020 and 2022), there was a higher occurrence of TP artworks (5 versus 3 for TS), likely influenced by the outbreak of AI and its experimental applications, which emphasize medium exploration. However, the 2022 Golden Nica was awarded to a TS artwork. Overall, TP artworks are less frequent, constituting only half the number of TS pieces. Extending our analysis to the last decade (2014 to 2020), we observe a recent trend favoring TS artworks. This trend reflects the field's interest shift, especially at Ars Electronica and in Europe, from focusing on medium and interactivity to more socially, politically, and critically engaged art perspectives.

\subsection{Interpersonal Interaction with Sociocultural Body Embodiment}


Alongside the growing interest in sociocultural body involvement, more than half of the 20 Techno Sociocultural (TS) artworks analyzed focus on interpersonal relationships in diverse forms. These artworks act as mediators among audiences, demonstrating that the interest in interactive art extends beyond mere human-machine interactions to embrace interpersonal interaction, aligning with the etymological and sociological origins of interaction.

Some artworks facilitate direct sensation transmission between people, such as tactile sensations in ``Sensitive to Pleasure, 2011'' and auditory sensations in ``Ishin-Den-Shin, 2013.'' Others visualize interconnectedness, as seen in ``The Kuramoto Model (1,000 Fireflies), 2011'' and ``Redes Vestíveis / Wearable Nets, 2011.'' ``Future Kiss, 2009'' uses the act of kissing to identify ``the other half,'' illustrating how these works bridge interpersonal gaps. Additionally, some pieces connect people across geographical, social, and political divides in challenging scenarios, such as ``Cross Coordinates (MX-US), 2012,'' ``Monitor Man, 2018,'' and ``BI0FILM.NET: RESIST LIKE BACTERIA, 2022.''

\begin{figure}[h!]
    \centering
    \includegraphics[width=0.5\linewidth]{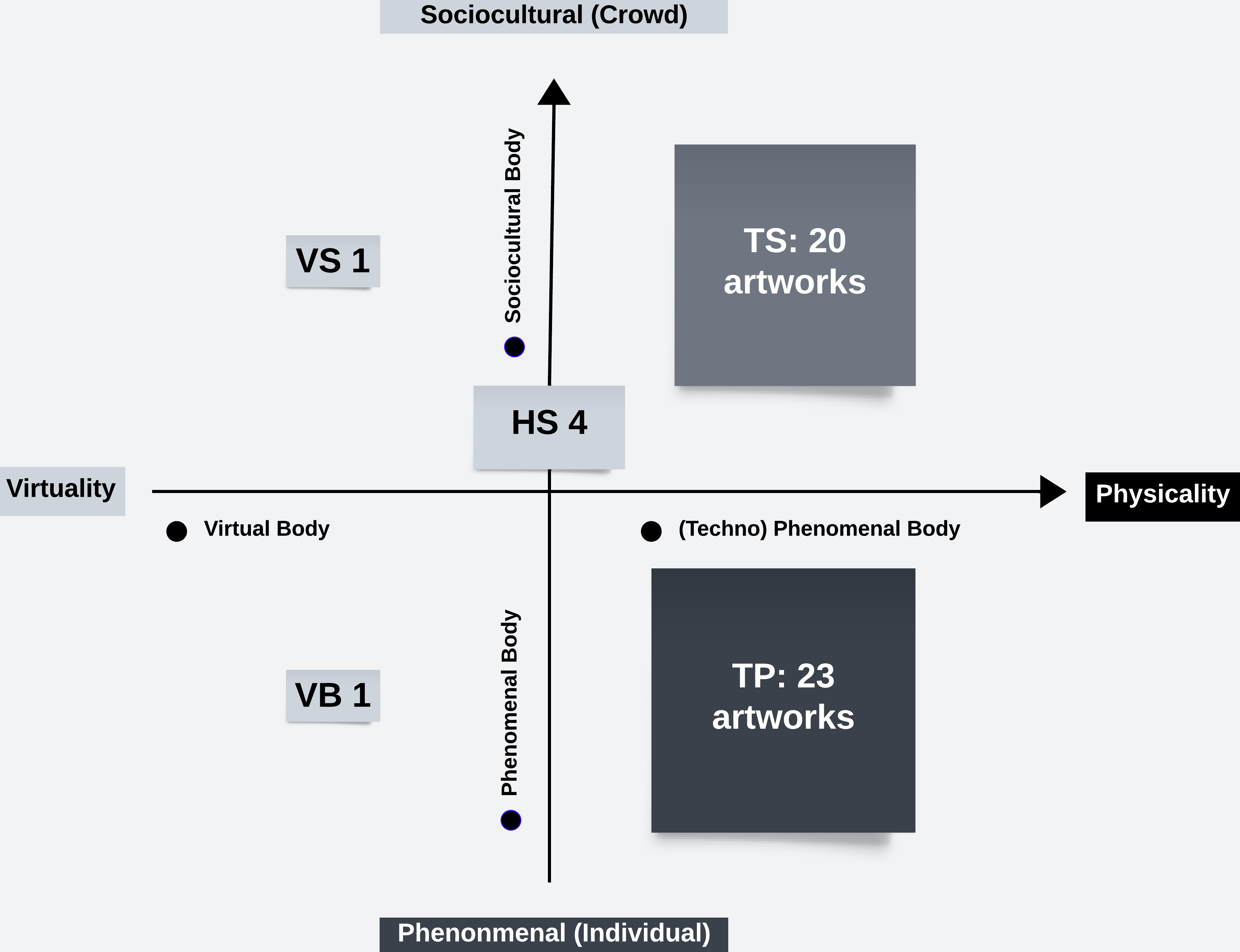}
    \caption{Techno Phenomenal Body has been thoroughly explored over the years, with both TP and TS artworks receiving significant artistic attention. Artworks related to the Virtual Body, including VB, VS, and HS categories, appear underrepresented.}
    \label{fig:VBvTS}
\end{figure}

\subsection{Phenomenal Body versus Virtual Body}

If examining the physicality and virtuality of body embodiment, the artworks in the corpus (as detailed in Tab.\ref{tab: corpus}) predominantly relate to the Techno Phenomenal Body, with only a few pieces relevant to the Virtual Body. Fig.\ref{fig:VBvTS} demonstrates that there are 23 TP artworks in the corpus compared to 20 TS artworks and only 6 VB-related artworks. This observation leads to a dual perspective: on the one hand, the Techno Phenomenal Body has been thoroughly explored over the years, with both TP and TS artworks receiving significant artistic attention. On the other hand, artworks related to the Virtual Body, including VB, VS, and HS categories, appear underrepresented in our analysis, suggesting a potential research gap in current interactive art, especially considering the advancements in immersive technologies and growing interest in virtual and mixed realities. This raises questions about whether or how soon Virtual Body artworks will gain mainstream recognition in the field.

The virtual body introduces a new context to these interactions; it can manifest as VS, where virtual sociocultural bodies interact, or HS, where interactions span virtual and phenomenal bodies. The current VB-related artwork, ``The Other in You, 2018,'' allows the audience to engage in an immersive experience from an out-of-body perspective with virtual dancers, although the dancers are pre-recorded and programmed; otherwise, it could be an HS piece. Artistic concepts like telepresence can be seen as a classical form of HS, facilitating interaction across vast distances, even across realities. ``Taxilink, 2010'' enables experiences that transcend national borders, while ``Monitor Man, 2018'' addresses geographical, political, and social barriers. The virtual body can transcend the limitations of our physical form, fostering a move toward greater social, political, and critical engagement and paving the way for advancements in immersive technologies within artistic exploration.

\subsection{Somaesthetic and Audience-artwork Interaction}

This subsection analyses somaesthetic and audience-artwork interaction. An artwork may engage multiple sensations; thus, to tally the frequency of all sensations involved in both ``what the audience does'' and ``what the audience perceives'' across the 49 artworks, the former is recorded 52 times, while the latter appears 74 times. Generally, it is observed that the audience ``perceives'' more than they ``does,'' indicating the audience ``interacts'' (slightly) more passively than actively. Even though ``interactive,'' these interactive artworks tend toward unidirectional, rather than evenly bidirectional, interactions.  

\begin{figure}[h]    
    \centering
    \includegraphics[width=0.5\linewidth]{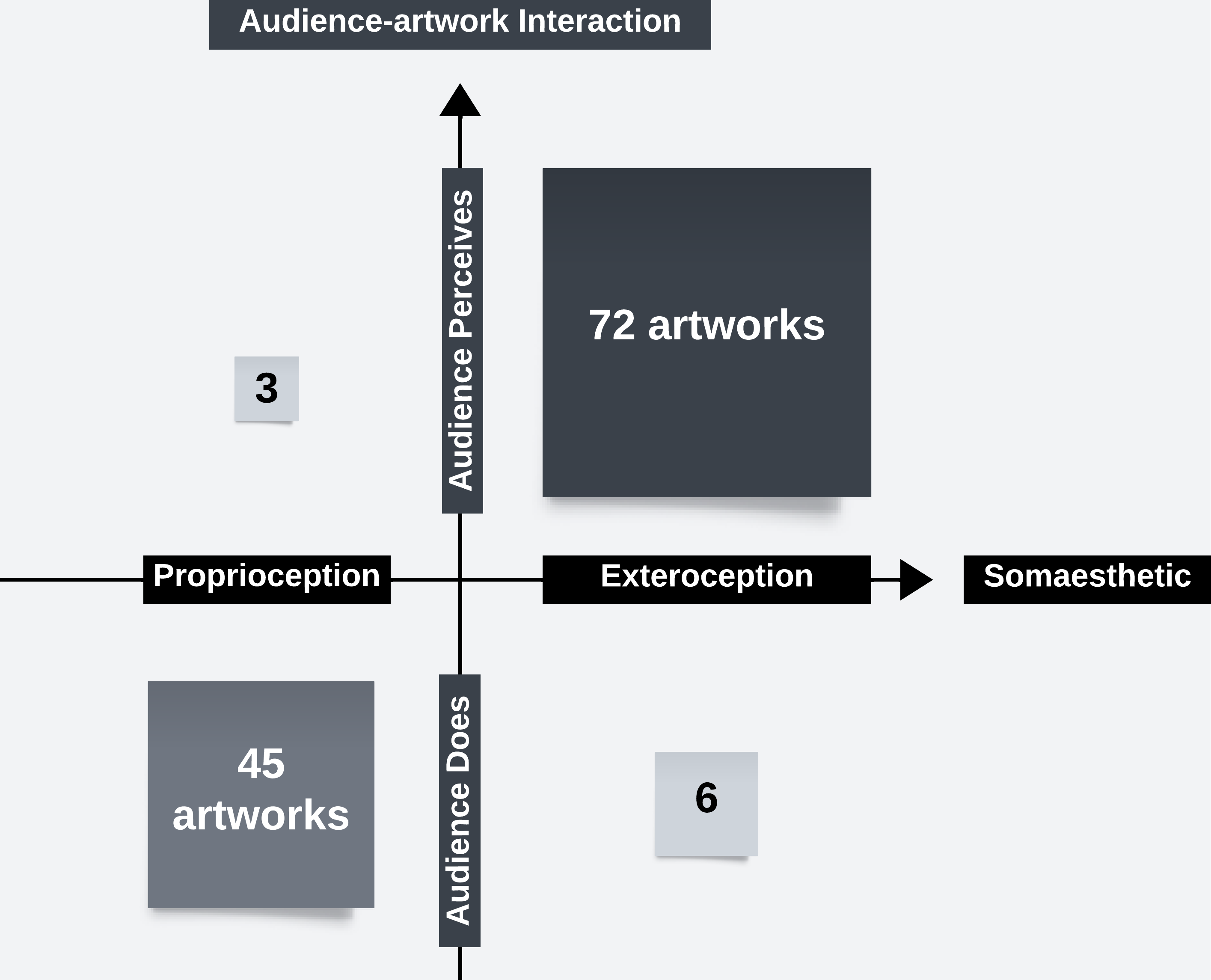}
    \caption{Audience participates (slightly) more passively than actively. ``what the audience does'' is primarily driven by proprioception, whereas ``what the audience perceives'' mostly involves exteroception, specifically through ``Visual (VI)'' and ``Auditive (AU)'' sensations.}
    \label{fig:soma-aai}
\end{figure}

This trend becomes more pronounced when examining the subcategories of exteroception and proprioception. Fig.\ref{fig:soma-aai} shows there are 72 instances of artworks which the ``audience perceives'' through exteroception, and only in 6 pieces, the ``audience does.'' At the same time, proprioception shows a reverse trend, with only in 3 instances ``audience perceives,'' and 45 ``audience does.'' In the current audience-artwork interaction, ``what the audience does'' is primarily driven by proprioception, whereas ``what the audience perceives'' mostly involves exteroception, specifically through ``Visual (VI)'' and ``Auditive (AU)'' sensations.


\begin{figure}[h]
    \centering
    \includegraphics[width=\linewidth]{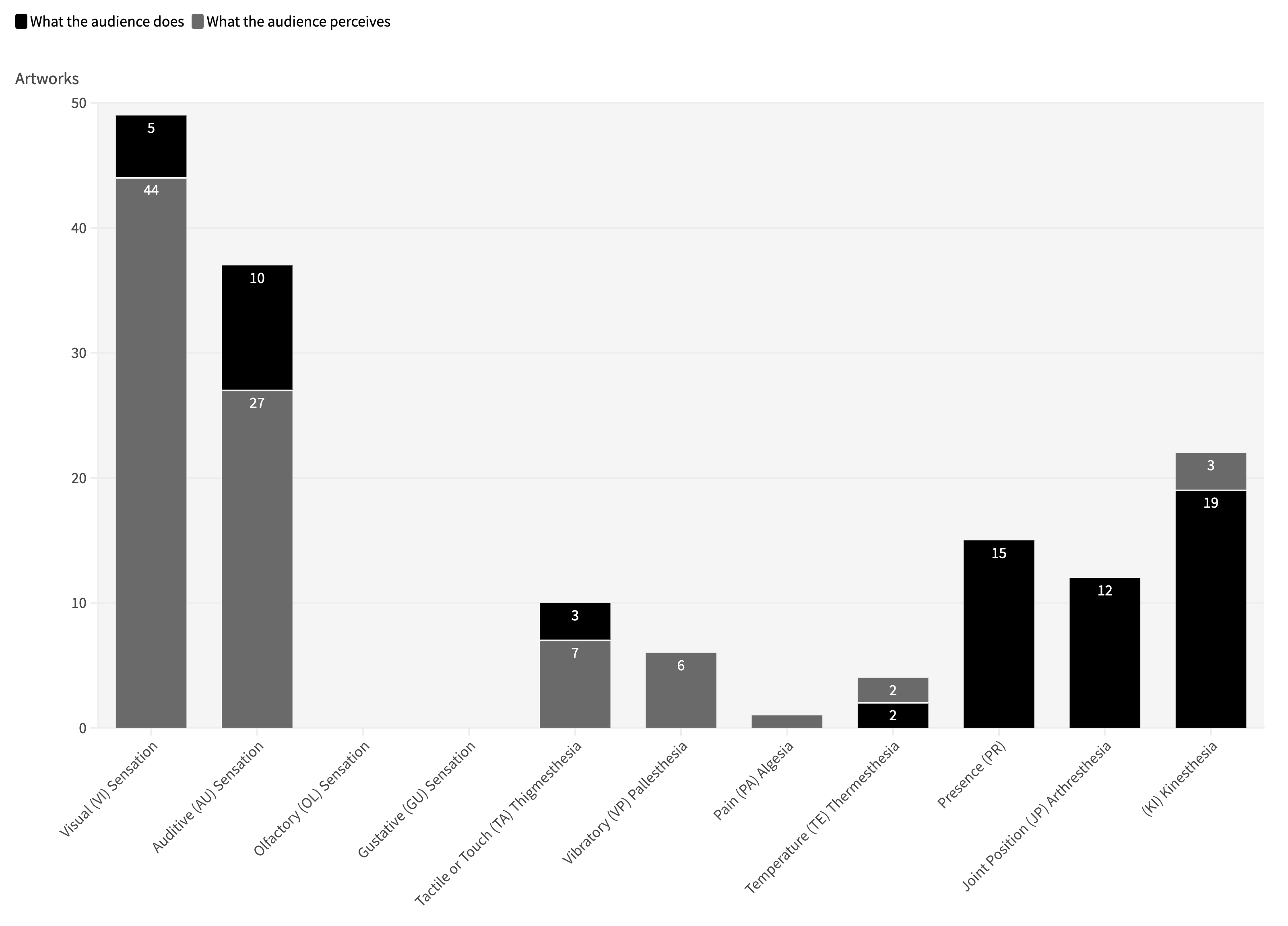}
    \caption{Employing proprioception to enhance ``what the audience does'' has become mature, commonplace, and perhaps even conventional within interactive art. The dominance of visual sensation in interactive art underlines, superisely, this critique, suggesting that the field still overly relies on visual-based engagement. It's evident that interactive art has not fully tapped the extensive spectrum of somaesthetic or bodily sensations.}
    \label{fig:aa-soma}
\end{figure}

When we look into details of the somaesthetic dimension and through each sensation separately, Fig.\ref{fig:aa-soma} shows the three sensations in the proprioception subcategory appear to be thoroughly explored and evenly distributed, but only from the perspective of ``what the audience does.'' As previously discussed, employing proprioception to enhance ``what the audience does'' has become mature, commonplace, and perhaps even conventional within interactive art. However, there are only two artworks where kinesthetic (KI) sensation is used for the audience to perceive, involving an exoskeleton that controls the audience's body movements rather than the audience controlling it. This highlights a distinct approach within the interaction dynamics, where the typical roles of control and agency are reversed.

Regarding the exteroception, no artwork within our scope involves Olfactory (OL) or Gustative (GU) sensations in either ``What Audience Does: Send'' or ``What Audience Perceives: Receive.'' There are 7 artworks in which ``the audience perceives'' Tactile or Touch Sensation (TA), and in 3 artworks, ``the audience does.'' Similarly, six artworks engage the audience's Vibratory Sensation (VP), but here, ``the audience perceives'' exclusively. For instance, in ``PENDULUM CHOIR, 2013,'' the performer-audience perceives TA and VP from their exoskeleton. In ``SENSITIVE TO PLEASURE, 2011'' and ``ISHIN-DEN-SHIN, 2013,'' TA is activated in terms of both ``what the audience does and perceives.''

When involving Temperature Sensation (TE), there are two artworks where ``the audience does'' and another two where ``the audience perceives.'' For example, in ``INSTITUTE OF HUMAN OBSOLESCENCE, 2018,'' the audience utilizes their body heat to interact with the artwork and generate bio-electricity for bitcoin mining. Despite this, TE is seldom utilized in terms of ``what the audience perceives.'' One notable exception is ``Rain Room, 2013,'' which manipulates moisture and temperature as the audience navigates through the space. Regarding Pain Sensation (PA), ``Sensitive to Pleasure, 2011'' is the only artwork where PA is employed, specifically for ``the audience perceives,'' with the artist participant experiencing PA transmitted from another ``creature.''

On the other hand, most of the artworks (45 out of 49) engage the visual sensation, typically ``the audience perceives.'' Only in one artwork does the audience interact through eye movement; in the other four works, concepts akin to telepresence enable the audience to share their real-time visual perceptions with others. Don Ihde's critique, originally aimed at scientific instrumentation, of the dominance of visual or visualist sensations appears to be equally, if not more, relevant here. The dominance of visual sensation in interactive art underlines, superisely, this critique, suggesting that the field still overly relies on visual-based engagement.

Based on the statistics of the application and involvement of somaesthetic sensations in the corpus, it's evident that interactive art has not fully tapped the extensive spectrum of somaesthetic or bodily sensations. Specifically, a wide range of non-visual and non-audio sensations, such as Olfactory (OL), Gustative (GU), Temperature (TE), and Pain (PA) Sensations, remain underexplored, contrary to what might be commonly assumed. Though thoroughly studied, those sensations, e.g., proprioception and visual and auditive sensations, are used conventionally unidirectional; in other words, the other half is not explored.

\section{Discussion and Outlook} \label{sec: discuss}

Several conclusions can be delineated based on the previous comprehensive review and analysis. Firstly, it is evident that interaction and interactivity are still insufficiently explored within the realm of interactive art, rather than ``so ubiquitous in our daily lives that it’s almost vanished into the background, become ordinary, even banal \cite[p.144]{leopoldseder_prix_2009},'' particularly when assessed from an embodied perspective. Secondly, there appears to be a trend indicating that interaction (i.e., interpersonal interaction or audience-audience interaction) in interactive art is returning to, or should return to, its etymological and sociological origin. Thirdly, notable research gaps exist in three key areas: the insufficiently explored somaesthetic sensations, the dynamics of interpersonal interactions, and the development of the virtual body within sociocultural forms. 

\subsection{Inflation of Interactivity? Neither True nor Crisis!} 

The ``crisis of interactive art'' discussion, ignited nearly 20 years ago by the Gold Nica Awardee ``Listening Post, 2004'' by Mark Hansen and Ben Rubin, sparked a debate about the ubiquity of interactivity and diminishing audience appreciation due to familiarity with interactivity. These discussions suggest, perhaps misleadingly, that the domain of interactivity and interaction within interactive art is ``exhaustively'' explored. While this may hold true in the realm of body tracking (kinesthetic, KI) \footnote{Daniel Dieters stated that ``Today interactivity is no longer an experiment ... but part of everyday life ... Thus, no one is surprised any longer by ... body tracking \cite[p.56]{daniels_strategies_2008}.''}, which corresponds to the observation that most ``what the audience does'' involves proprioception with minimal exteroception, or it can also be true if we talk about projections (Visual Sensation, VI), but it overlooked broader possibilities. Due to the analysis from our embodied perspective, the somaesthetic in interactive art, opposite to common assumptions, is underexplored. The current interactive art yet relies too heavily on visual and auditive sensations. Further exploration of the extensive spectrum of somaesthetic deserves more investigation.

\subsection{Trending Relationality Instead of Interactivity} 

The analytic results in Sec.\ref{sec: analysis-SB-PB} suggest a discernible trend in TS artworks, corroborating our earlier statements about a shift in interest within interactive art from human-machine (or human-artwork) interactions towards interpersonal interactions. In the broader scope of contemporary art, the fascination with the relationship between audience and artwork and interpersonal ones is not exclusively in interactive art. Claire Bishop, in her book \textit{Participation}, looks at audience participation from the social dimension \cite{bishop_participation_2010}. According to Bishop, participation represents a social mode of audience engagement, distinct from interactivity, which involves direct human-artwork connections. It's noteworthy that, as Christian Paul points out, Bishop, although he applies similar terminology as interactive art, is ignorant of new media and interactive art \cite{paul_new_2011}. Similarly, Nicolas Bourriaud introduced the concept of ``relational aesthetic'' to highlight the significance of interpersonal human relations and their social contexts within contemporary art \cite{bourriaud_relational_1998}. The shift in interactive art from focusing on human-artwork interaction to interpersonal interaction could be characterized as a transition from interactivity to relationality.

\begin{figure}[h]
    \centering
    \includegraphics[width=0.65\linewidth]{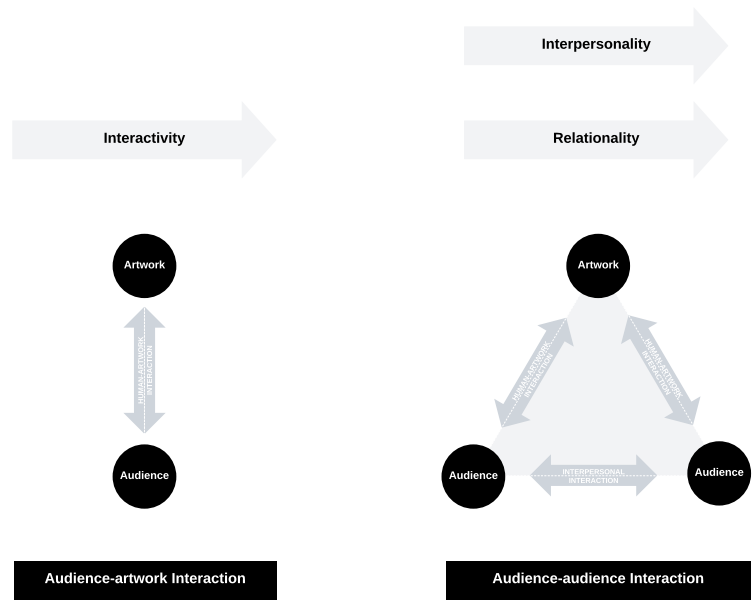}
    \caption{Interactivity vs. Relationality (new triangle)}
    \label{fig:enter-label}
\end{figure}

This observed pattern or shift of paradigm aligns with the evolving idea of interactive art at the Prix Ars Electronica. When commenting on ``Pendulum Choir'' by Michel and Andre Decoster, the Golden Nica award in the interactive art category in 2013, the jury stated that ``...interactivity should no longer be defined by a specific formal language prescribed by the vocabulary of the 1990s. Today, it is a set of strategies that can be situated in multiple contexts, within the art world and outside of it'' \cite[P.156]{leopoldseder_cyberarts_2014}. This aligns with the fact that Ars Electronica renamed the Interactive Art category to ``Interactive Art +'' to include a broader range of artworks in 2016. The jury stated that ``the language of interaction in the realm of art has matured to the extent that it can now occupy spaces outside the artists' laboratory and beyond the confines of the screen'', and ``in the hyper-connected era, interaction is the most pervasive medium that allows art to create a dialogue with every form of human activity'' \cite{de_jaeger_transcending_2016}. 

We share the opinion of those jury members at the Prix Ars Electronica, and regard the expanding, yet increasingly undefined, scope of interactive art as a sign of its maturity rather than a crisis. Just as ``digital'' is no longer a necessary adjective in front of ``media art,'' the term ``interactive'' as human-machine or technology-mediated interaction could also fade into the background. Interaction should not only be constrained within the dualism of human versus artwork (or system and machine) but can be widely expanded to include interpersonal interaction, returning to its etymological and sociological origins.

As interaction and interactivity mature as mediums, we observe interactive art integrating into the wider category of ``contemporary art.'' In this case, it adopts contemporary art's shared characteristics and interests, with interaction and interactivity as well as its inevitable technical approach, serving as artistic mediums - comparable to any other established and prevalent forms. Despite Christian Paul's criticism that mainstream contemporary art frequently neglects the inclusion of technology-based art fields, such as interactive art, within its domain \cite{paul_new_2011}, it is arguably time to recognize, understand, and embrace the emerging trend where interactive art shifts its emphasis towards relationality and other common themes of ``contemporary art'' \cite{paul_new_2011}. The stance that technology-based artworks fall outside the realm of contemporary art may be antiquated or, at the very least, overlooks the recent evolution of interactive art. 

\subsection{Concept Matters More than Ever} 

Christa Sommerer and Laurent Mignonneau, in an interview with Pau Waelder in 2013, have already summarised the importance of the concept in their interactive art as ``... the main quality of an interactive artwork must be its concept. Technology should only be a means to achieve the realization of the artistic concept in the best way.'' As previously discussed, it might be a good chance to return to its etymological and sociological origins when interactivity fades out in the background. Interaction does not indicate a technological medium but a conceptual framework, which will lead interactive art to socially, politically, and critically engaged concepts. Undoubtedly, on some level, this is also interconnected with the trends of sociocultural body involvement and the trending relationality. 

The Prix Ars Electronica, when addressing the ``Interactive Art +'' category, even though dedicated to all forms and formats from installations to performances, unveiled their priorities. They included (1) artistic quality in the interaction; (2) a harmonious dialog between the content and interaction, i.e., the inherent principles of interaction and the interfaces; (3) particular interest with an explicit sociopolitical agenda; (4) noteworthy technological or scientific achievements; (5) strong innate potential to expand the scope of human action and participation. More or less, it's pointing in a direction similar to our discussion; long gone are the medium-specific perspectives, and a paradigm shift toward art concepts about interpersonality and relatinality, if not occurred, is emerging. 

\subsection{Insight and Conclusion}


For future interactive art practices, we suggest adopting both medium-specific and concept-oriented perspectives. The former indicates further exploration of non-visual and non-auditory sensations in exteroception is needed; frequently-used sensations, such as visual and auditive exteroception and proprioception, deserve further investigation in their unconventional formats. Additionally, the virtual body within sociocultural forms and immersive technologies should be given more attention. From a concept-oriented perspective, we encourage a re-evaluation of the etymological and sociological origins of interaction and promote a shift toward socially, politically, and critically engaged art concepts. 


\bibliographystyle{ACM-Reference-Format}
\bibliography{references}


\begin{thebibliography}{43}


\ifx \showCODEN    \undefined \def \showCODEN     #1{\unskip}     \fi
\ifx \showDOI      \undefined \def \showDOI       #1{#1}\fi
\ifx \showISBNx    \undefined \def \showISBNx     #1{\unskip}     \fi
\ifx \showISBNxiii \undefined \def \showISBNxiii  #1{\unskip}     \fi
\ifx \showISSN     \undefined \def \showISSN      #1{\unskip}     \fi
\ifx \showLCCN     \undefined \def \showLCCN      #1{\unskip}     \fi
\ifx \shownote     \undefined \def \shownote      #1{#1}          \fi
\ifx \showarticletitle \undefined \def \showarticletitle #1{#1}   \fi
\ifx \showURL      \undefined \def \showURL       {\relax}        \fi
\providecommand\bibfield[2]{#2}
\providecommand\bibinfo[2]{#2}
\providecommand\natexlab[1]{#1}
\providecommand\showeprint[2][]{arXiv:#2}

\bibitem[Bigley(1990)]%
        {bigley_sensation_1990}
\bibfield{author}{\bibinfo{person}{G.~Kim Bigley}.} \bibinfo{year}{1990}\natexlab{}.
\newblock \showarticletitle{Sensation}.
\newblock In \bibinfo{booktitle}{\emph{Clinical {Methods}: {The} {History}, {Physical}, and {Laboratory} {Examinations}} (\bibinfo{edition}{3rd} ed.)}, \bibfield{editor}{\bibinfo{person}{H.~Kenneth Walker}, \bibinfo{person}{W.~Dallas Hall}, {and} \bibinfo{person}{J.~Willis Hurst}} (Eds.). \bibinfo{publisher}{Butterworths}, \bibinfo{address}{Boston}.
\newblock
\showISBNx{978-0-409-90077-4}
\urldef\tempurl%
\url{http://www.ncbi.nlm.nih.gov/books/NBK390/}
\showURL{%
\tempurl}


\bibitem[Bishop(2010)]%
        {bishop_participation_2010}
\bibfield{editor}{\bibinfo{person}{Claire Bishop}} (Ed.). \bibinfo{year}{2010}\natexlab{}.
\newblock \bibinfo{booktitle}{\emph{Participation} (\bibinfo{edition}{3. pr} ed.)}.
\newblock \bibinfo{publisher}{Whitechapel}, \bibinfo{address}{London}.
\newblock
\showISBNx{978-0-85488-147-5 978-0-262-52464-3}


\bibitem[Bourriaud(1998)]%
        {bourriaud_relational_1998}
\bibfield{author}{\bibinfo{person}{Nicolas Bourriaud}.} \bibinfo{year}{1998}\natexlab{}.
\newblock \bibinfo{booktitle}{\emph{Relational {Aesthetics}} (\bibinfo{edition}{les presses du reel edition} ed.)}.
\newblock \bibinfo{publisher}{Les Presse Du Reel,Franc}, \bibinfo{address}{Dijon}.
\newblock
\showISBNx{978-2-84066-060-6}


\bibitem[Cameron(2009)]%
        {leopoldseder_prix_2009}
\bibfield{author}{\bibinfo{person}{A Cameron}.} \bibinfo{year}{2009}\natexlab{}.
\newblock \bibinfo{booktitle}{\emph{Prix ars electronica: {CyberArts} 2009: international compendium {Prix} {Ars} {Electronica}.}}
\newblock


\bibitem[Cornock and Edmonds(1973)]%
        {cornock_creative_1973}
\bibfield{author}{\bibinfo{person}{Stroud Cornock} {and} \bibinfo{person}{Ernest Edmonds}.} \bibinfo{year}{1973}\natexlab{}.
\newblock \showarticletitle{The {Creative} {Process} {Where} the {Artist} is {Amplified} or {Superseded} by the {Computer}}.
\newblock \bibinfo{journal}{\emph{Leonardo}} \bibinfo{volume}{6}, \bibinfo{number}{1} (\bibinfo{year}{1973}), \bibinfo{pages}{11--16}.
\newblock
\showISSN{1530-9282}
\urldef\tempurl%
\url{https://muse.jhu.edu/pub/6/article/597993}
\showURL{%
\tempurl}
\newblock
\shownote{Publisher: The MIT Press}.


\bibitem[Costello et~al\mbox{.}(2005)]%
        {costello_understanding_2005}
\bibfield{author}{\bibinfo{person}{Brigid Costello}, \bibinfo{person}{Lizzie Muller}, \bibinfo{person}{Shigeki Amitani}, {and} \bibinfo{person}{Ernest Edmonds}.} \bibinfo{year}{2005}\natexlab{}.
\newblock \showarticletitle{Understanding the experience of interactive art: {Iamascope} in {Beta}\_space}.
\newblock  (\bibinfo{date}{Nov.} \bibinfo{year}{2005}), \bibinfo{pages}{49--56}.
\newblock
\showISSN{0-9751533-2-3}


\bibitem[Daniels(2008)]%
        {daniels_strategies_2008}
\bibfield{author}{\bibinfo{person}{Dieter Daniels}.} \bibinfo{year}{2008}\natexlab{}.
\newblock \showarticletitle{Strategies of {Interactivity}}.
\newblock In \bibinfo{booktitle}{\emph{The {Art} and {Science} of {Interface} and {Interaction} {Design}}}, \bibfield{editor}{\bibinfo{person}{Christa Sommerer}, \bibinfo{person}{Lakhmi~C. Jain}, {and} \bibinfo{person}{Laurent Mignonneau}} (Eds.). \bibinfo{publisher}{Springer}, \bibinfo{address}{Berlin, Heidelberg}, \bibinfo{pages}{27--62}.
\newblock
\showISBNx{978-3-540-79870-5}
\urldef\tempurl%
\url{https://doi.org/10.1007/978-3-540-79870-5_3}
\showDOI{\tempurl}


\bibitem[De~Jaeger(2016)]%
        {de_jaeger_transcending_2016}
\bibfield{author}{\bibinfo{person}{C De~Jaeger}.} \bibinfo{year}{2016}\natexlab{}.
\newblock \showarticletitle{Transcending the {Medium}}.
\newblock In \bibinfo{booktitle}{\emph{{CyberArts} 2016}}, \bibfield{editor}{\bibinfo{person}{Hannes Leopoldseder}, \bibinfo{person}{Christine Schöpf}, {and} \bibinfo{person}{Gerfried Stocker}} (Eds.).
\newblock
\urldef\tempurl%
\url{https://www.artbook.com/https:/www.artbook.com/9783775741941.html}
\showURL{%
\tempurl}


\bibitem[Dixon(2015)]%
        {dixon_digital_2015}
\bibfield{author}{\bibinfo{person}{Steve Dixon}.} \bibinfo{year}{2015}\natexlab{}.
\newblock \bibinfo{booktitle}{\emph{Digital {Performance}: {A} {History} of {New} {Media} in {Theater}, {Dance}, {Performance} {Art}, and {Installation}}}.
\newblock \bibinfo{publisher}{MIT Press}.
\newblock
\showISBNx{978-0-262-52752-1}
\newblock
\shownote{Google-Books-ID: yL34DwAAQBAJ}.


\bibitem[Dourish(1999)]%
        {dourish_embodied_1999}
\bibfield{author}{\bibinfo{person}{Paul Dourish}.} \bibinfo{year}{1999}\natexlab{}.
\newblock \bibinfo{title}{Embodied {Interaction}: {Exploring} the {Foundations} of a {New} {Approach} to {HCI}}.  (\bibinfo{year}{1999}).
\newblock


\bibitem[Dourish(2001)]%
        {dourish_where_2001-1}
\bibfield{author}{\bibinfo{person}{Paul Dourish}.} \bibinfo{year}{2001}\natexlab{}.
\newblock \bibinfo{booktitle}{\emph{Where the {Action} {Is}: {The} {Foundations} of {Embodied} {Interaction}}}.
\newblock \bibinfo{publisher}{The MIT Press}.
\newblock
\showISBNx{978-0-262-25605-6}
\urldef\tempurl%
\url{https://doi.org/10.7551/mitpress/7221.001.0001}
\showDOI{\tempurl}


\bibitem[Edmonds(2010)]%
        {edmonds_art_2010}
\bibfield{author}{\bibinfo{person}{Ernest Edmonds}.} \bibinfo{year}{2010}\natexlab{}.
\newblock \showarticletitle{The art of interaction}.
\newblock \bibinfo{journal}{\emph{Digital Creativity}} \bibinfo{volume}{21}, \bibinfo{number}{4} (\bibinfo{date}{Dec.} \bibinfo{year}{2010}), \bibinfo{pages}{257--264}.
\newblock
\showISSN{1462-6268}
\urldef\tempurl%
\url{https://doi.org/10.1080/14626268.2010.556347}
\showDOI{\tempurl}
\newblock
\shownote{Publisher: Routledge \_eprint: https://doi.org/10.1080/14626268.2010.556347}.


\bibitem[Edmonds et~al\mbox{.}(2004)]%
        {edmonds_approaches_2004}
\bibfield{author}{\bibinfo{person}{Ernest Edmonds}, \bibinfo{person}{Greg Turner}, {and} \bibinfo{person}{Linda Candy}.} \bibinfo{year}{2004}\natexlab{}.
\newblock \showarticletitle{Approaches to interactive art systems}. In \bibinfo{booktitle}{\emph{Proceedings of the 2nd international conference on {Computer} graphics and interactive techniques in {Australasia} and {South} {East} {Asia}}} \emph{(\bibinfo{series}{{GRAPHITE} '04})}. \bibinfo{publisher}{Association for Computing Machinery}, \bibinfo{address}{New York, NY, USA}, \bibinfo{pages}{113--117}.
\newblock
\showISBNx{978-1-58113-883-2}
\urldef\tempurl%
\url{https://doi.org/10.1145/988834.988854}
\showDOI{\tempurl}


\bibitem[FAUCONNIER(2003)]%
        {fauconnier_description_2003}
\bibfield{author}{\bibinfo{person}{SANDRA FAUCONNIER}.} \bibinfo{year}{2003}\natexlab{}.
\newblock \bibinfo{booktitle}{\emph{Description models for unstable media art}}.
\newblock \bibinfo{type}{{T}echnical {R}eport}.
\newblock
\urldef\tempurl%
\url{https://v2.nl/wp-content/uploads/files/2003/publishing/articles/1_3_metadata.pdf}
\showURL{%
\tempurl}


\bibitem[Guljajeva(2018)]%
        {guljajeva_interaction_2018}
\bibfield{author}{\bibinfo{person}{Varvara Guljajeva}.} \bibinfo{year}{2018}\natexlab{}.
\newblock \emph{\bibinfo{title}{From {Interaction} to {Post}-participation: {The} {Disappearing} {Role} of the {Active} {Participant}}}.
\newblock \bibinfo{thesistype}{Ph.\,D. Dissertation}.
\newblock


\bibitem[Gwilt(1997)]%
        {gwilt_towards_1997}
\bibfield{author}{\bibinfo{person}{Ian Gwilt}.} \bibinfo{year}{1997}\natexlab{}.
\newblock \showarticletitle{Towards a visual taxonomy in {New} {Media} {Art}}. In \bibinfo{booktitle}{\emph{Proceedings of {ENGAGE}}}. \bibinfo{address}{Sydney}, \bibinfo{pages}{90--98}.
\newblock


\bibitem[Hannington and Reed(2002)]%
        {hannington_towards_2002}
\bibfield{author}{\bibinfo{person}{A. Hannington} {and} \bibinfo{person}{K. Reed}.} \bibinfo{year}{2002}\natexlab{}.
\newblock \showarticletitle{Towards a taxonomy for guiding multimedia application development}. In \bibinfo{booktitle}{\emph{Ninth {Asia}-{Pacific} {Software} {Engineering} {Conference}, 2002.}} \bibinfo{publisher}{IEEE Comput. Soc}, \bibinfo{address}{Gold Coast, Qld., Australia}, \bibinfo{pages}{97--106}.
\newblock
\showISBNx{978-0-7695-1850-3}
\urldef\tempurl%
\url{https://doi.org/10.1109/APSEC.2002.1182979}
\showDOI{\tempurl}


\bibitem[Huhtamo(2007)]%
        {huhtamo_twintouchtestredux_2007}
\bibfield{author}{\bibinfo{person}{Erkki Huhtamo}.} \bibinfo{year}{2007}\natexlab{}.
\newblock \showarticletitle{Twin–{Touch}–{Test}–{Redux}: {Media} {Archaeological} {Approach} to {Art}, {Interactivity}, and {Tactility}}.
\newblock  (\bibinfo{date}{Jan.} \bibinfo{year}{2007}).
\newblock
\urldef\tempurl%
\url{https://doi.org/10.7551/mitpress/4530.003.0008}
\showDOI{\tempurl}


\bibitem[Ihde(2002)]%
        {ihde_bodies_2002}
\bibfield{author}{\bibinfo{person}{Don Ihde}.} \bibinfo{year}{2002}\natexlab{}.
\newblock \bibinfo{booktitle}{\emph{Bodies in {Technology}}}.
\newblock \bibinfo{publisher}{U of Minnesota Press}.
\newblock
\showISBNx{978-0-8166-3846-8}
\newblock
\shownote{Google-Books-ID: kLM9gfnPcFAC}.


\bibitem[Ihde(2010)]%
        {ihde_embodied_2010}
\bibfield{author}{\bibinfo{person}{Don Ihde}.} \bibinfo{year}{2010}\natexlab{}.
\newblock \bibinfo{booktitle}{\emph{Embodied {Technics}}}.
\newblock \bibinfo{publisher}{Automatic Press / VIP}.
\newblock
\showISBNx{978-87-92130-27-3}
\newblock
\shownote{Google-Books-ID: tnb5QwAACAAJ}.


\bibitem[In et~al\mbox{.}({[n.\,d.]})]%
        {in_roy_nodate}
\bibfield{author}{\bibinfo{person}{In}, \bibinfo{person}{Randall Packer}, {and} \bibinfo{person}{Ken Jordan}.} \bibinfo{year}{[n.\,d.]}\natexlab{}.
\newblock \showarticletitle{Roy {Ascott}: {Behaviourist} {Art} and the {Cybernetic} {Vision}}.
\newblock
\urldef\tempurl%
\url{https://www.semanticscholar.org/paper/Roy-Ascott%3A-Behaviourist-Art-and-the-Cybernetic-In-Packer/a984eac0c55817c4d05276d2956c0aa0d4f263bf}
\showURL{%
\tempurl}


\bibitem[Institute({[n.\,d.]})]%
        {ludwig_boltzmann_institute_visualization_nodate}
\bibfield{author}{\bibinfo{person}{Ludwig~Boltzmann Institute}.} \bibinfo{year}{[n.\,d.]}\natexlab{}.
\newblock \bibinfo{booktitle}{\emph{Visualization {Showcase} - {Taxonomy} for {Interactive} {Art}}}.
\newblock \bibinfo{type}{{T}echnical {R}eport}.
\newblock
\urldef\tempurl%
\url{https://www.mediaartresearch.at/vis_subdomain/webarchive/public/view/mid_24.html}
\showURL{%
\tempurl}


\bibitem[Jones(2006)]%
        {jones_sensorium_2006}
\bibfield{editor}{\bibinfo{person}{Caroline~A. Jones}} (Ed.). \bibinfo{year}{2006}\natexlab{}.
\newblock \bibinfo{booktitle}{\emph{Sensorium: {Embodied} {Experience}, {Technology}, and {Contemporary} {Art}} (\bibinfo{edition}{illustrated edition} ed.)}.
\newblock \bibinfo{publisher}{The MIT Press}, \bibinfo{address}{Cambridge, Mass}.
\newblock
\showISBNx{978-0-262-10117-2}


\bibitem[Kluszczynski(2010)]%
        {kluszczynski_strategies_2010}
\bibfield{author}{\bibinfo{person}{RyszardW. Kluszczynski}.} \bibinfo{year}{2010}\natexlab{}.
\newblock \showarticletitle{Strategies of {Interactive} {Art}}.
\newblock \bibinfo{journal}{\emph{Journal of Aesthetics \& Culture}} \bibinfo{volume}{2}, \bibinfo{number}{1} (\bibinfo{date}{Jan.} \bibinfo{year}{2010}), \bibinfo{pages}{5525}.
\newblock
\showISSN{null}
\urldef\tempurl%
\url{https://doi.org/10.3402/jac.v2i0.5525}
\showDOI{\tempurl}
\newblock
\shownote{Publisher: Routledge \_eprint: https://doi.org/10.3402/jac.v2i0.5525}.


\bibitem[Koukal(2002)]%
        {koukal_bodies_2002}
\bibfield{author}{\bibinfo{person}{D.~R Koukal}.} \bibinfo{year}{2002}\natexlab{}.
\newblock \showarticletitle{Bodies in {Technology} (review)}.
\newblock \bibinfo{journal}{\emph{Technology and Culture}} \bibinfo{volume}{43}, \bibinfo{number}{4} (\bibinfo{date}{Oct.} \bibinfo{year}{2002}), \bibinfo{pages}{837--838}.
\newblock
\showISSN{1097-3729}
\urldef\tempurl%
\url{https://doi.org/10.1353/tech.2002.0169}
\showDOI{\tempurl}


\bibitem[Kwastek(2007)]%
        {kwastek_research_2007}
\bibfield{author}{\bibinfo{person}{Katja Kwastek}.} \bibinfo{year}{2007}\natexlab{}.
\newblock \bibinfo{booktitle}{\emph{Research {Project}: {A} {Taxonomy} of “{Interactive} {Art}”}}.
\newblock \bibinfo{type}{{T}echnical {R}eport}.
\newblock
\urldef\tempurl%
\url{https://www.kwastek.de/pdf/taxonomy_IA_200706.pdf}
\showURL{%
\tempurl}


\bibitem[Kwastek(2013)]%
        {kwastek_aesthetics_2013}
\bibfield{author}{\bibinfo{person}{Katja Kwastek}.} \bibinfo{year}{2013}\natexlab{}.
\newblock \bibinfo{booktitle}{\emph{Aesthetics of {Interaction} in {Digital} {Art}}}.
\newblock \bibinfo{publisher}{MIT Press}.
\newblock
\showISBNx{978-0-262-01932-3}
\newblock
\shownote{Google-Books-ID: 6Q7bAAAAQBAJ}.


\bibitem[Kwastek and Spörl(2009)]%
        {kwastek_research_2009}
\bibfield{author}{\bibinfo{person}{Katja Kwastek} {and} \bibinfo{person}{Ingrid Spörl}.} \bibinfo{year}{2009}\natexlab{}.
\newblock \bibinfo{booktitle}{\emph{Research {Report}: {Taxonomy} '{Interactive} {Art}' {II}. {Phase}}}.
\newblock \bibinfo{type}{{T}echnical {R}eport}.
\newblock
\urldef\tempurl%
\url{https://www.kwastek.de/pdf/taxonomy_IA_200911.pdf}
\showURL{%
\tempurl}


\bibitem[Leopoldseder(1990)]%
        {leopoldseder_prix_1990}
\bibfield{author}{\bibinfo{person}{Hannes Leopoldseder}.} \bibinfo{year}{1990}\natexlab{}.
\newblock \bibinfo{booktitle}{\emph{Der {Prix} {Ars} {Electronica} 90. {Internationales} {Kompendium} der {Computerkünste}}}.
\newblock \bibinfo{publisher}{Veritas Verlag}.
\newblock
\showISBNx{978-3-85329-833-6}


\bibitem[Leopoldseder et~al\mbox{.}(2014)]%
        {leopoldseder_cyberarts_2014}
\bibfield{editor}{\bibinfo{person}{Hannes Leopoldseder}, \bibinfo{person}{Christine Schöpf}, {and} \bibinfo{person}{Gerfried Stocker}} (Eds.). \bibinfo{year}{2014}\natexlab{}.
\newblock \bibinfo{booktitle}{\emph{{CyberArts} 2013: {International} {Compendum} {Prix} {Ars} {Electronica}} (\bibinfo{edition}{2013th edition} ed.)}.
\newblock \bibinfo{publisher}{Hatje Cantz}, \bibinfo{address}{Ostfildern}.
\newblock
\showISBNx{978-3-7757-3631-2}


\bibitem[Massumi(2002)]%
        {massumi_parables_2002}
\bibfield{author}{\bibinfo{person}{Brian Massumi}.} \bibinfo{year}{2002}\natexlab{}.
\newblock \bibinfo{booktitle}{\emph{Parables for the {Virtual}: {Movement}, {Affect}, {Sensation}}}.
\newblock \bibinfo{publisher}{Duke University Press}.
\newblock
\showISBNx{978-0-8223-8357-4}
\urldef\tempurl%
\url{https://doi.org/10.1215/9780822383574}
\showDOI{\tempurl}


\bibitem[Muller et~al\mbox{.}(2006)]%
        {muller_living_2006}
\bibfield{author}{\bibinfo{person}{L. Muller}, \bibinfo{person}{E. Edmonds}, {and} \bibinfo{person}{M. Connell}.} \bibinfo{year}{2006}\natexlab{}.
\newblock \showarticletitle{Living laboratories for interactive art}.
\newblock \bibinfo{journal}{\emph{CoDesign}} \bibinfo{volume}{2}, \bibinfo{number}{4} (\bibinfo{date}{Dec.} \bibinfo{year}{2006}), \bibinfo{pages}{195--207}.
\newblock
\showISSN{1571-0882, 1745-3755}
\urldef\tempurl%
\url{https://doi.org/10.1080/15710880601008109}
\showDOI{\tempurl}


\bibitem[Nappi(2004)]%
        {nappi_bodies_2004}
\bibfield{author}{\bibinfo{person}{Maureen Nappi}.} \bibinfo{year}{2004}\natexlab{}.
\newblock \showarticletitle{Bodies in {Technology} (review)}.
\newblock \bibinfo{journal}{\emph{Leonardo}}  \bibinfo{volume}{37} (\bibinfo{year}{2004}), \bibinfo{pages}{77--78}.
\newblock


\bibitem[Nardelli(2014)]%
        {nardelli_viewpoint_2014}
\bibfield{author}{\bibinfo{person}{Enrico Nardelli}.} \bibinfo{year}{2014}\natexlab{}.
\newblock \showarticletitle{A {Viewpoint} on the {Computing}-{Art} {Dialogue}: {The} {Classification} of {Interactive} {Digital} {Artworks}}.
\newblock \bibinfo{journal}{\emph{Leonardo}} \bibinfo{volume}{47}, \bibinfo{number}{1} (\bibinfo{date}{Feb.} \bibinfo{year}{2014}), \bibinfo{pages}{43--49}.
\newblock
\showISSN{0024-094X, 1530-9282}
\urldef\tempurl%
\url{https://doi.org/10.1162/LEON_a_00700}
\showDOI{\tempurl}


\bibitem[Paul(2011)]%
        {paul_new_2011}
\bibfield{author}{\bibinfo{person}{Christiane Paul}.} \bibinfo{year}{2011}\natexlab{}.
\newblock \showarticletitle{New {Media} in the {Mainstream}}.
\newblock \bibinfo{journal}{\emph{https://artnodes.uoc.edu/}} (\bibinfo{date}{Nov.} \bibinfo{year}{2011}).
\newblock
\showISSN{1695-5951}
\urldef\tempurl%
\url{https://openaccess.uoc.edu/handle/10609/10101}
\showURL{%
\tempurl}
\newblock
\shownote{Accepted: 2011-11-25T11:39:17Z Publisher: Universitat Oberta de Catalunya}.


\bibitem[Pepperell(2003)]%
        {pepperell_where_2003}
\bibfield{author}{\bibinfo{person}{Robert Pepperell}.} \bibinfo{year}{2003}\natexlab{}.
\newblock \showarticletitle{Where the {Action} {Is}: {The} {Foundations} of {Embodied} {Interaction} ({Review})}.
\newblock   \bibinfo{volume}{36} (\bibinfo{year}{2003}), \bibinfo{pages}{412--413}.
\newblock
\urldef\tempurl%
\url{https://doi.org/10.7551/mitpress/7221.001.0001}
\showDOI{\tempurl}


\bibitem[Schilhab(2010)]%
        {schilhab_embodiment_2010}
\bibfield{author}{\bibinfo{person}{Theresa S.~S. Schilhab}.} \bibinfo{year}{2010}\natexlab{}.
\newblock \showarticletitle{Embodiment and {Technics}—{At} the {Brink} of {Biology}: {Ihde}, {Don}, 2010. {Embodied} {Technics}, {New} {York}: {Automatic} {Press}}.
\newblock \bibinfo{journal}{\emph{Biosemiotics}} \bibinfo{volume}{3}, \bibinfo{number}{2} (\bibinfo{date}{Aug.} \bibinfo{year}{2010}), \bibinfo{pages}{253--255}.
\newblock
\showISSN{1875-1342, 1875-1350}
\urldef\tempurl%
\url{https://doi.org/10.1007/s12304-010-9091-z}
\showDOI{\tempurl}


\bibitem[Schraffenberger and van~der Heide(2012)]%
        {schraffenberger_interaction_2012}
\bibfield{author}{\bibinfo{person}{Hanna Schraffenberger} {and} \bibinfo{person}{Edwin van~der Heide}.} \bibinfo{year}{2012}\natexlab{}.
\newblock \showarticletitle{Interaction {Models} for {Audience}-{Artwork} {Interaction}: {Current} {State} and {Future} {Directions}}. In \bibinfo{booktitle}{\emph{Arts and {Technology}}}, \bibfield{editor}{\bibinfo{person}{Anthony~L. Brooks}} (Ed.). \bibinfo{publisher}{Springer}, \bibinfo{address}{Berlin, Heidelberg}, \bibinfo{pages}{127--135}.
\newblock
\showISBNx{978-3-642-33329-3}
\urldef\tempurl%
\url{https://doi.org/10.1007/978-3-642-33329-3_15}
\showDOI{\tempurl}


\bibitem[Schraffenberger and Heide(2015)]%
        {schraffenberger_audience-artwork_2015}
\bibfield{author}{\bibinfo{person}{Hanna~K. Schraffenberger} {and} \bibinfo{person}{Edwin Van~Der Heide}.} \bibinfo{year}{2015}\natexlab{}.
\newblock \showarticletitle{Audience-artwork interaction}.
\newblock \bibinfo{journal}{\emph{International Journal of Arts and Technology}} \bibinfo{volume}{8}, \bibinfo{number}{2} (\bibinfo{year}{2015}), \bibinfo{pages}{91}.
\newblock
\showISSN{1754-8853, 1754-8861}
\urldef\tempurl%
\url{https://doi.org/10.1504/IJART.2015.069550}
\showDOI{\tempurl}


\bibitem[Sparacino et~al\mbox{.}(2000)]%
        {sparacino_media_2000}
\bibfield{author}{\bibinfo{person}{F. Sparacino}, \bibinfo{person}{G. Davenport}, {and} \bibinfo{person}{A. Pentland}.} \bibinfo{year}{2000}\natexlab{}.
\newblock \showarticletitle{Media in performance: {Interactive} spaces for dance, theater, circus, and museum exhibits}.
\newblock \bibinfo{journal}{\emph{IBM Systems Journal}} \bibinfo{volume}{39}, \bibinfo{number}{3.4} (\bibinfo{year}{2000}), \bibinfo{pages}{479--510}.
\newblock
\showISSN{0018-8670}
\urldef\tempurl%
\url{https://doi.org/10.1147/sj.393.0479}
\showDOI{\tempurl}
\newblock
\shownote{Conference Name: IBM Systems Journal}.


\bibitem[Stern(2009)]%
        {stern_implicit_2009}
\bibfield{author}{\bibinfo{person}{Nathaniel Stern}.} \bibinfo{year}{2009}\natexlab{}.
\newblock \emph{\bibinfo{title}{The implicit body : {Understanding} interactive art through embodiment and embodiment through interactive art}}.
\newblock thesis. \bibinfo{school}{Trinity College (Dublin, Ireland). Department of Electronic \& Electrical Engineering}.
\newblock
\urldef\tempurl%
\url{http://www.tara.tcd.ie/handle/2262/76965}
\showURL{%
\tempurl}
\newblock
\shownote{Accepted: 2016-08-31T13:44:10Z}.


\bibitem[Trifonova et~al\mbox{.}(2008)]%
        {trifonova_software_2008}
\bibfield{author}{\bibinfo{person}{Anna Trifonova}, \bibinfo{person}{Letizia Jaccheri}, {and} \bibinfo{person}{Kristin Bergaust}.} \bibinfo{year}{2008}\natexlab{}.
\newblock \showarticletitle{Software engineering issues in interactive installation art}.
\newblock \bibinfo{journal}{\emph{International Journal of Arts and Technology}} \bibinfo{volume}{1}, \bibinfo{number}{1} (\bibinfo{year}{2008}), \bibinfo{pages}{43}.
\newblock
\showISSN{1754-8853, 1754-8861}
\urldef\tempurl%
\url{https://doi.org/10.1504/IJART.2008.019882}
\showDOI{\tempurl}


\bibitem[Xu et~al\mbox{.}(2023)]%
        {xu_describing_2023}
\bibfield{author}{\bibinfo{person}{Dan Xu}, \bibinfo{person}{Maarten~H Lamers}, {and} \bibinfo{person}{Edwin van~der Heide}.} \bibinfo{year}{2023}\natexlab{}.
\newblock \showarticletitle{Describing and {Comparing} {Co}-located {Interaction} in {Interactive} {Art} {Using} a {Relational} {Model}}. \bibinfo{publisher}{Springer}, \bibinfo{pages}{198--217}.
\newblock


\end{thebibliography}

\textbf{Note: }In this review, there are 49 artworks selected, but we choose not to create a reference entrance for each work and list all the information about each artwork in Tab.\ref{tab: corpus}.

\appendix
%
%
\begin{table}[h]
\centering
\captionof{table}{Literature Survey of Different Taxonomies I: Interactive Systems} \label{tab:taxonomies1} 
\begin{tabular}{lll}
\toprule
Literature                               & Attributes              & Values                                                                                                                         \\ \hline
Cornock and Edmonds (1973) \& Edmonds et al. (2004) & Art systems& \begin{tabular}[c]{@{}l@{}}• Static\\ • dynamic-passive\\ • dynamic-interactive \\ • (dynamic-interactive (varying))\end{tabular}\\ \hline
Sparacino et al. (2000)            & Interactive systems     & \begin{tabular}[c]{@{}l@{}}•  Scripted\\ • Responsive\\ • Behavioural\\ • Learning\\ • Intentional\end{tabular} \\ \hline
Trifonova et al. (2008) & Interaction rules       & \begin{tabular}[c]{@{}l@{}}• Static\\ • Dynamic \\ \hline \end{tabular} \\ 
                                   & Triggering parameters   & \begin{tabular}[c]{@{}l@{}}• Human presence\\ • Human actions\\ • Environment \\ \hline\end{tabular}   \\
                                   & Content origin          & \begin{tabular}[c]{@{}l@{}}• User input\\ • Predefined content by the artist\\ • Generated/algorithmic content\end{tabular} \\ \hline
Nardelli (2010)         & Content provider        & \begin{tabular}[c]{@{}l@{}}• Artist\\ • Audience\\ • Environment \\ \hline\end{tabular} \\
                                   & Processing dynamics     & \begin{tabular}[c]{@{}l@{}}• Static\\ • Predefined change\\ • Casual change\\ • Evolutionary change \\ \hline\end{tabular} \\
                                   & Processing contributors & \begin{tabular}[c]{@{}l@{}}• Self-audience\\ • Environment\end{tabular} \\ \hline
\end{tabular}
\end{table}

\begin{longtable}{lll}
\caption{Literature Survey of Different Taxonomies II: Audience-experience and Interaction}
\label{tab:taxonomies2}\\
\hline
Literature                                      & Attributes                             & Values                                                                                                                             \\ \hline
\endfirsthead
\multicolumn{3}{c}%
{{\bfseries Table \thetable\ continued from previous page}} \\
\hline
Literature                                      & Attributes                             & Values                                                                                                                             \\ \hline
\endhead
Costello et al. (2005)                          & Embodied experience                    & \begin{tabular}[c]{@{}l@{}}• Response\\ • Control\\ • Contemplation\\ • Belonging\\ • Disengagement\end{tabular}                   \\ \hline
Sommerer and Mignonneau (1999) & Interaction                            & \begin{tabular}[c]{@{}l@{}}• Non-linear\\ • Multi-layered\end{tabular}                                                             \\ \cline{2-3} 
                                                & Path of interaction                    & \begin{tabular}[c]{@{}l@{}}• Pre-designed/programmed \\ • Evolutionary processes\end{tabular}                                      \\ \hline
Dixon (2015)                                    & Level and depth of interactivity       & \begin{tabular}[c]{@{}l@{}}• Navigation\\ • Participation\\ • Conversation\\ • Collaboration\end{tabular}                          \\ \hline
Hannington and Reed(2002)                       & Level of interactivity (in multimedia) & \begin{tabular}[c]{@{}l@{}}• Passive\\ • Interactive\\ • Adaptive\end{tabular}                                                     \\ \hline
V2\_ Institute for the Unstable Media (2003)    & Interaction level                      & \begin{tabular}[c]{@{}l@{}}• Observational\\ • Navigational\\ • Participatory\\ • Co-authoring\\ • Intercommunication\end{tabular} \\ \hline
\end{longtable}

%
%
\begin{longtable}[c]{@{}lll@{}}
\caption{Categories from Ars Electronica and V2\_ Institute for the Unstable Media}
\label{tab:ars-v2}\\
\toprule
Literature                                                                                                             & Dimension                    & Option                                                                                                                                                                                                                                \\* \midrule
\endfirsthead
\multicolumn{3}{c}%
{{\bfseries Table \thetable\ continued from previous page}} \\
\toprule
Literature                                                                                                             & Dimension                    & Option                                                                                                                                                                                                                                \\* \midrule
\endhead
\begin{tabular}[c]{@{}l@{}}V2\_ Institute for the Unstable Media \\ (2003)\end{tabular}               & Time flexibility             & Any time, Scheduled                                                                                                                                                                                                                   \\* \cmidrule(l){2-3} 
                                                                                                                       & Interaction location         & Any location, Specific location                                                                                                                                                                                                       \\* \cmidrule(l){2-3} 
                                                                                                                       & User number                  & Single user, User group, Audience                                                                                                                                                                                                     \\* \cmidrule(l){2-3} 
                                                                                                                       & Interaction level            & \begin{tabular}[c]{@{}l@{}}Observational, Navigational, Participatory,\\ Co-authoring, Intercommunication\end{tabular}                                                                                                                \\* \cmidrule(l){2-3} 
                                                                                                                       & Sensor mode                  & Visual, Auditive, Olfactory, Tactile, Gustative                                                                                                                                                                                       \\* \midrule
\begin{tabular}[c]{@{}l@{}}Kwastek (2007, 2009) and \\ Ludwig Boltzmann Institute (2009)\end{tabular} & The performer (visitor) does & \begin{tabular}[c]{@{}l@{}}Observe, Explore, Activate, Control, Select, \\ Participate, Navigate, Leave traces, Co-author,\\ Collaborate, Exchange information, Create\end{tabular}                                                   \\* \cmidrule(l){2-3} 
                                                                                                                       & The work (project) does      & \begin{tabular}[c]{@{}l@{}}Monitor, Serve as an instrument, Document,\\ Enhance perception, Offer a game, \\ Enable communication, Visualise, Sonificate, \\ Transform, Store, Immerse, Process,\\ Mediate, Tell/narrate\end{tabular} \\* \bottomrule
\end{longtable}

\begin{longtable}{@{}lllllllllllll@{}}
\caption{Artwork Corpus: Selected Ars Electronica Interactive Art Awardee with Embodied Interaction (from 2009 to 2023)}
\label{tab: corpus}\\
\toprule
\textbf{Artwork}                                                                      & \textbf{Artist}                                                   & \textbf{Year} & \textbf{Body} & \multicolumn{9}{c}{\textbf{Somaesthetic}}      \\* \cmidrule(l){5-13} 
                                                                                                       &                                                                                    &                                &                                & VI  & AU  & TA  & VP & PA & TE & PR & JP & KI  \\* \midrule
\endfirsthead
\multicolumn{13}{c}%
{{\bfseries Table \thetable\ continued from previous page}} \\
\toprule
\textbf{Artwork}                                                                     \textbf{Artist}                                                   & \textbf{Year} & \textbf{Body} & \multicolumn{9}{c}{\textbf{Somaesthetic}}      \\* \cmidrule(l){5-13} 
                                                                                                       &                                                                                    &                                &                                & VI  & AU  & TA  & VP & PA & TE & PR & JP & KI  \\* \midrule
\endhead
\begin{tabular}[c]{@{}l@{}}BI0FILM.NET: \\ RESIST LIKE BACTERIA\end{tabular}                           & \begin{tabular}[c]{@{}l@{}}Jung Hsu \\ Natalia Rivera\end{tabular}                 & 2022                           & TS                             & S/R & S/R &     &    &    &    & S  &    &     \\* \midrule
MORPHECHORE                                                                                            & Daito Manabe                                                                       & 2022                           & TP                             & R   & R   &     &    &    &    &    &    & S   \\* \midrule
THE ZIZI SHOW                                                                                          & Jake Elwes                                                                         & 2022                           & TP                             & R   & R   &     &    &    &    &    &    & S   \\* \midrule
SHADOW STALKER                                                                                         & Lynn Leeson                                                                        & 2020                           & TP                             & R   & R   &     &    &    &    &    &    & S   \\* \midrule
GOOGLE MAPS HACKS                                                                                      & Simon Weckert                                                                      & 2020                           & TP                             & R   &     &     &    &    &    & S  &    &     \\* \midrule
MIND                                                                                                   & \begin{tabular}[c]{@{}l@{}}Shinseungback\\ Kimyonghun\end{tabular}                 & 2020                           & TS                             &     & R   &     &    &    &    &    & S  &     \\* \midrule
APPROPRIATE RESPONSE                                                                                   & Mario Klingemann                                                                   & 2020                           & TP                             & R   & R   &     &    &    &    &    &    & S   \\* \midrule
COMPRESSION CRADLE                                                                                     & Lucy McRae                                                                         & 2020                           & TS                             &     &     & R   &    &    &    &    &    & S/R \\* \midrule
THE OTHER IN YOU                                                                                       & \begin{tabular}[c]{@{}l@{}}Richi Owaki\\ YCAM\end{tabular}                         & 2018                           & VB                             & R   &     & R   &    &    & R  &    &    & S   \\* \midrule
TURNSTILE                                                                                              & Ursula Damm                                                                        & 2018                           & TS                             & R   &     &     &    &    &    & S  &    &     \\* \midrule
MONITOR MAN                                                                                            & Yassine Khaled                                                                     & 2018                           & HS                             & S/R & S/R &     &    &    &    &    &    & S   \\* \midrule
\begin{tabular}[c]{@{}l@{}}INSTITUTE OF HUMAN \\ OBSOLESCENCE\end{tabular}                             & Manuel Beltrán                                                                     & 2018                           & TS                             & R   &     &     &    &    & S  &    &    & S   \\* \midrule
ECHO                                                                                                   & \begin{tabular}[c]{@{}l@{}}Georgie Pinn\\ Kendyl Rossi\end{tabular}                & 2018                           & TS                             & R   & R   &     &    &    &    &    & S  &     \\* \midrule
AI DJ PROJECT                                                                                          & Shoya Dozono, et al.                                                               & 2018                           & TS                             & R   & R   &     &    &    &    & S  &    &     \\* \midrule
\begin{tabular}[c]{@{}l@{}}CONSPIRACY: \\ CONJOINING THE VIRTUAL\end{tabular}                          & Kristin McWharter                                                                  & 2018                           & VS                             & R   & R   &     &    &    &    &    & S  &     \\* \midrule
INFERNO                                                                                                & \begin{tabular}[c]{@{}l@{}}Louis-Philippe Demers\\ Bill Vorn\end{tabular}          & 2016                           & TS                             &     &     &     & R  &    &    &    &    & R   \\* \midrule
PATHFINDER                                                                                             & \begin{tabular}[c]{@{}l@{}}Christian Loclair\\ Onformative\end{tabular}            & 2016                           & TP                             & R   & R   &     &    &    &    &    &    & S   \\* \midrule
EPIPHYTE CHAMBER                                                                                       & Philip Beesley                                                                     & 2014                           & TP                             & R   & R   &     &    &    &    & S  & S  &     \\* \midrule
\begin{tabular}[c]{@{}l@{}}PEACE CAN BE REALIZED \\ EVEN WITHOUT ORDER\end{tabular}                    & teamLab                                                                            & 2014                           & TS                             & R   & R   &     &    &    &    & S  &    &     \\* \midrule
\begin{tabular}[c]{@{}l@{}}THE MACHINE TO BE \\ ANOTHER\end{tabular}                                   & BeAnotherLab                                                                       & 2014                           & HS                             & S/R & S/R & R   &    &    &    &    &    & S   \\* \midrule
SWARM                                                                                                  & James Coupe                                                                        & 2014                           & HS                             & R   &     &     &    &    &    &    &    & S   \\* \midrule
SPORTS TIME MACHINE                                                                                    & \begin{tabular}[c]{@{}l@{}}Ryoko Ando\\ Hiroshi Inukai\end{tabular}                & 2014                           & TP                             & R   &     &     &    &    &    &    &    & S   \\* \midrule
PENDULUM CHOIR                                                                                         & Cod.Act, et al.                                                                    & 2013                           & TS                             & R   & S   & R   & R  &    &    &    & S  & R   \\* \midrule
VOICES OF ALIVENESS                                                                                    & Masaki Fujihata                                                                    & 2013                           & TP                             & R   & S/R &     &    &    &    &    &    & S   \\* \midrule
RAIN ROOM                                                                                              & rAndom International                                                               & 2013                           & TP                             & R   & R   & R   &    &    & R  & S  &    &     \\* \midrule
AHORA                                                                                                  & \begin{tabular}[c]{@{}l@{}}Hernán Kerlleñevich \\ Mene Savasta Alsina\end{tabular} & 2013                           & TS                             &     & R   &     &    &    &    & S  &    &     \\* \midrule
ANGLES MIRROR                                                                                          & Daniel Rozin                                                                       & 2013                           & TP                             & R   &     &     &    &    &    &    &    & S   \\* \midrule
\begin{tabular}[c]{@{}l@{}}DOWN WITH WRESTLERS \\ WITH SYSTEMS AND\\  MENTAL NONADAPTERS!\end{tabular} & \begin{tabular}[c]{@{}l@{}}Dmitry Kawarga\\ Elena Kawarga\end{tabular}             & 2013                           & TP                             & R   & S   &     & R  &    &    &    &    & S   \\* \midrule
EXPLODED VIEWS 2.0                                                                                     & Marnix de Nijs                                                                     & 2013                           & TP                             & R   & R   &     &    &    &    &    & S  &     \\* \midrule
ISHIN-DEN-SHIN                                                                                         & \begin{tabular}[c]{@{}l@{}}Disney Research \\ Pittsburgh, et al.\end{tabular}      & 2013                           & TS                             &     & S/R & S/R & R  &    &    &    & S  &     \\* \midrule
\begin{tabular}[c]{@{}l@{}}PERFUME GLOBAL \\ SITE PROJECT\end{tabular}                                 & Satoshi Horii, e tal.                                                              & 2013                           & TP                             & R   & R   &     &    &    &    &    &    & S   \\* \midrule
\begin{tabular}[c]{@{}l@{}}THE KURAMOTO MODEL\\ (1,000 FIREFLIES)\end{tabular}                         & David Allan Rueter                                                                 & 2013                           & TS                             & R   &     &     &    &    &    & S  &    &     \\* \midrule
VOICE ARRAY                                                                                            & Rafael Lozano-Hemmer                                                               & 2013                           & TS                             & R   & S   &     &    &    &    &    &    &     \\* \midrule
IDEOGENETIC MACHINE                                                                                    & Nova Jiang                                                                         & 2012                           & TP                             & R   &     &     &    &    &    &    &    & S   \\* \midrule
CROSS COORDINATES                                                                                      & Iván Abreu                                                                         & 2012                           & TS                             & R   &     &     &    &    &    &    &    & S   \\* \midrule
ANDROID THEATER                                                                                        & \begin{tabular}[c]{@{}l@{}}Oriza Hirata\\ Hiroshi Ishiguro\end{tabular}            & 2011                           & TP                             & R   & S/R &     &    &    &    &    &    & S   \\* \midrule
\begin{tabular}[c]{@{}l@{}}REDES VESTÍVEIS \\ WEARABLE NETS\end{tabular}                               & Claudio Bueno                                                                      & 2011                           & TS                             & R   &     &     &    &    &    & S  &    &     \\* \midrule
EMPATHETIC HEARTBEAT                                                                                   & Hideyuki Ando, et al.                                                              & 2011                           & TP                             & R   & R   &     &    &    &    &    &    &     \\* \midrule
SENSITIVE TO PLEASURE                                                                                  & Sonia Cillari                                                                      & 2011                           & TS                             & R   &     & S/R &    & R  & S  &    & S  &     \\* \midrule
TUNNEL                                                                                                 & \begin{tabular}[c]{@{}l@{}}Rejane Canton\\ Leonardo Crescenti\end{tabular}         & 2011                           & TS                             &     & R   &     & R  &    &    & S  &    &     \\* \midrule
\begin{tabular}[c]{@{}l@{}}SIX-FORTY BY \\ FOUR-EIGHTY\end{tabular}                                    & \begin{tabular}[c]{@{}l@{}}Marcelo Coelho\\ Jamie Zigelbaum\end{tabular}           & 2011                           & TP                             & R   &     & S   &    &    &    &    & S  &     \\* \midrule
THE EYEWRITER                                                                                          & Zach Lieberman, et al.                                                             & 2010                           & TP                             & S/R &     &     &    &    &    &    & S  &     \\* \midrule
TAXILINK                                                                                               & Tal Chalozin, et al.                                                               & 2010                           & HS                             & S/R & S/R &     &    &    &    & S  &    &     \\* \midrule
MOBILE CRASH                                                                                           & Lucas Bambozzi                                                                     & 2010                           & TS                             & R   & R   &     &    &    &    &    & S  &     \\* \midrule
\begin{tabular}[c]{@{}l@{}}WHEN LAUGHTER TRIPS \\ AT THE THRESHOLD OF \\ THE DIVINE\end{tabular}       & \begin{tabular}[c]{@{}l@{}}Kim Beck\\ Osman Khan\end{tabular}                      & 2009                           & TP                             & R   & R   &     &    &    &    & S  &    &     \\* \midrule
AUDIENCE                                                                                               & Hannes Koch, et al.                                                                & 2009                           & TP                             & R   &     &     &    &    &    & S  &    &     \\* \midrule
DOUBLE-TAKER (SNOUT)                                                                                   & Steven Benders, et al.                                                             & 2009                           & TP                             & R   &     &     &    &    &    & S  &    &     \\* \midrule
FUTURE KISS                                                                                            & Lenka Klimesova                                                                    & 2009                           & TS                             &     &     &     & R  &    &    &    & S  &     \\* \midrule
IN THE LINE OF SIGHT                                                                                   & \begin{tabular}[c]{@{}l@{}}Daniel Sauter\\ Fabian Winkler\end{tabular}             & 2009                           & TP                             & R   &     &     &    &    &    &    &    & S   \\* \bottomrule
\end{longtable}
\end{document}